\documentclass[jacsat]{achemso}

\setkeys{acs}{
abbreviations=true,
articletitle=true,
biblabel=plain,
chaptertitle=true,
doi=true,
email=true,
etalmode=truncate,
keywords=true,
maxauthors=100,
super=true
}

\setcitestyle{super,open={},close={}}

\usepackage[utf8]{inputenc}
\usepackage[T1]{fontenc}
\usepackage[english]{babel}
\usepackage[version=4]{mhchem}
\usepackage{siunitx}
\usepackage{graphicx}
\usepackage{longtable}

\usepackage{geometry}
\usepackage{caption}
\usepackage{subcaption}
\usepackage{float}
\usepackage{natbib}
\usepackage{setspace}
\usepackage{xkeyval}
\usepackage{array}
\usepackage{booktabs}
\usepackage{listings}
\usepackage{lmodern}
\usepackage{mathpazo}
\usepackage{microtype}
\usepackage{hyperref}
\hypersetup{breaklinks,colorlinks=true,linkcolor=black,filecolor=black,urlcolor=black,citecolor=black}

\usepackage{amsmath}
\usepackage{amssymb}
\usepackage{amsfonts}
\usepackage{latexsym}
\usepackage[compact]{titlesec}
\usepackage[usenames,dvipsnames]{xcolor}
\usepackage{mathptmx}
\usepackage{calc}
\usepackage{titletoc}
\usepackage{lipsum}
\usepackage{multirow}
\usepackage{amstext}
\usepackage{tabularx}
\usepackage{adjustbox}
\usepackage{epsfig,color}
\usepackage[section]{placeins}
\usepackage{natmove}
\usepackage{titlesec}
\usepackage{tablefootnote}
\usepackage{threeparttable}
\usepackage[d]{esvect}
\usepackage{mathtools}
\usepackage{physics}
\usepackage{textcomp}

\usepackage[none]{hyphenat}
\usepackage{microtype}
\usepackage{xcolor}

\vfuzz \hfuzz
\SectionsOn
\SectionNumbersOn
\AbstractOn

\title{Interplay between structural, electronic and topological properties in low-dimensional tellurium} 

\author{Gabriel Elyas Gama Araújo}
\email{gabrielelyas@proton.me}
\author{Andreia Luisa da Rosa}
\email{andreialuisa@ufg.br}
\affiliation{Federal University of Goi\'as, Institute of Physics, Campus Samambaia, 74960600 Goi\^ania, Brazil}


\date{\today}

\begin{document}

\maketitle

\begin{abstract}
 We present a comprehensive first-principles investigation of the structural, electronic, vibrational, and topological properties of tellurium across its dimensional hierarchy, including bulk trigonal Te-I, two-dimensional tellurene polymorphs, and one-dimensional helical nanowires. Using density functional theory with full inclusion of spin–orbit coupling, we confirm that bulk Te-I is a narrow-gap semiconductor hosting Weyl nodes arising from broken inversion symmetry and degenerate phonon modes suggestive of chiral phonon behavior. In contrast, two-dimensional $\alpha$- and $\beta$-tellurene are found to be topologically trivial ($\mathbb{Z}_2 = 0$), with no spin–orbit-driven band inversion in the occupied manifold. Beyond these established phases, we find that buckled kagome and buckled square tellurene lattices exhibit a nontrivial two-dimensional $\mathbb{Z}_2 = 1$ topology of the occupied electronic bands, indicating incipient quantum spin Hall character in metallic systems. In contrast, one-side hydrogen-passivated hexagonal tellurene realizes a fully gapped quantum spin Hall phase with a robust $\mathbb{Z}_2 = 1$ invariant, preserved under applied strain and chemical functionalization. In the one-dimensional limit, helical tellurium nanowires preserve chirality and host edge-localized states accompanied by pronounced anisotropy in carrier effective masses. These results establish tellurium as a highly tunable platform for engineering topological phenomena across dimensionality, bridging three-dimensional Weyl physics, two-dimensional quantum spin Hall and incipient $\mathbb{Z}_2$ phases, and one-dimensional helical systems.  
\end{abstract}

\section{Introduction}

Tellurium (Te) is a rare, silvery metalloid found only in trace quantities in Earth’s crust and seawater. Despite its scarcity, Te has attracted increasing attention owing to its distinctive physical and chemical properties, particularly as a semiconductor with emerging technological relevance. As a group-16 chalcogen, Te exhibits remarkable crystalline and electronic characteristics, including non-trivial topological behavior driven by strong spin–orbit coupling (SOC).\cite{Medina-Cruz2020,zhang2019evidenceweylfermionselemental}

In its bulk phase, Te crystallizes in a trigonal structure composed of
one-dimensional (1D) chiral helical chains aligned along the
$c$-axis. This non-centrosymmetric motif breaks inversion symmetry and
gives rise to a rich landscape of topological phenomena. Trigonal
tellurium is a narrow-gap semiconductor widely recognized as hosting
Weyl nodes in its bulk band
structure.\cite{Hirayama2015,PhysRevB.97.201402,PhysRevB.103.235421}
The characteristic hedgehog spin texture in momentum space further
emphasizes the critical role of inversion-symmetry breaking and SOC in
shaping Te electronic
topology.\cite{zhang2019evidenceweylfermionselemental}

While bulk trigonal Te is now well understood, the evolution of
topology in its low-dimensional derivatives remains less
systematically understood across dimensionality. The two-dimensional
$\alpha$-phase (P$\bar{3}$m1) and monoclinic $\beta$-phase (P2/m) were
first predicted theoretically\cite{multi} and later synthesized on
GaAs\,\cite{alphon,10.1002/adma.202309023} and
graphene/6H-SiC(0001)\,\cite{beton} substrates,
respectively. Additional 2D allotropes have since been proposed and,
in some cases, realized, including a rectangular phase on
Ni(111)\cite{10.1021/acsami.2c20400} and a honeycomb tellurene lattice
exhibiting an SOC-induced Dirac
gap\,\cite{acs.nanolett.4c02171}. These reduced-dimensional phases
provide fertile ground for exploring tunable topological states
through strain, substrate interaction, or chemical functionalization.

Tellurium can also form one-dimensional nanostructures such as
nanowires and nanoribbons. First synthesized via hydrothermal
methods,\cite{fio} these systems preserve the helical atomic backbone
of bulk Te, offering an ideal platform to probe chirality-driven
physics under quantum confinement. The recent fabrication of
ultrathin, high-crystallinity Te
nanowires\,\cite{10.1021/acsanm.5c01278} underscores the growing
interest in identifying low-dimensional electronic and boundary-state
phenomena in 1D Te architectures.

The interplay of structural flexibility, chirality, strong SOC, and
symmetry breaking thus positions tellurium as a uniquely versatile
platform for realizing and engineering emergent topological
phases.\cite{PhysRevResearch.5.023142,acsaem.4c02722}

Despite these advances, a unified understanding of how topology
evolves across Te dimensionalities remains incomplete. Here, we
address this gap through first-principles calculations of tellurium in
its 3D, 2D, and 1D forms, elucidating how chirality, symmetry
breaking, and SOC govern its electronic topology. We confirm the
existence of Weyl nodes in bulk Te, reveal that 2D allotropes exhibit
symmetry-tunable electronic structures, and demonstrate that helical
Te nanowires retain bulk chirality while hosting edge-localized states
associated with quantum confinement.

This work establishes symmetry-driven connections between the Weyl
physics of bulk Te and its low-dimensional derivatives. Beyond
advancing the fundamental understanding of symmetry-protected states
in reduced dimensions, our findings highlight tellurium as a promising
candidate for next-generation spintronic, optoelectronic, and quantum
devices, where topological phases can be engineered via strain,
confinement, and chemical modification.

\section{Computational details}

Our investigation employs first-principles calculations performed with
the Vienna \textit{Ab initio} Simulation Package
(VASP)\cite{vasp1,vasp2}. The exchange–correlation potential is
treated using both the generalized gradient approximation (GGA),
parameterized by Perdew, Burke, and Ernzerhof (PBE)\cite{pbe}, and the
hybrid Heyd–Scuseria–Ernzerhof (HSE06) functional.\cite{hse06}
Long-range van der Waals (vdW) interactions are included via the
DFT-D3 correction method proposed by Grimme.\cite{dftd3} Core–valence
electron interactions are described using the projector augmented-wave
(PAW) method,\cite{paw} and the Kohn–Sham single-electron
wavefunctions are expanded in a plane-wave basis set with a kinetic
energy cutoff of 520\,eV.

All atomic structures are fully relaxed until the residual forces on
each atom are smaller than $1 \times 10^{-6}$~eV/\AA. To avoid
spurious interactions arising from periodic boundary conditions, a
vacuum spacing of 12~\AA\ is introduced—along the $z$-direction for
monolayers and along the transverse ($x$ and $y$) directions for
nanowires. The Brillouin zone is sampled using \textbf{k}-point meshes
generated by the Monkhorst–Pack scheme,\cite{monkhorst1976special}
with grids of ($15 \times 15 \times 8$) for bulk Te, ($13 \times 17
\times 1$) for $\alpha$-tellurene, ($9 \times 12 \times 1$) for
$\beta$-tellurene, ($13 \times 17 \times 1$) for hexagonal planar,
($24 \times 24 \times 1$) for hexagonal buckled, ($13 \times 13 \times
1$) for pentagonal, ($17 \times 17 \times 1$) for Lieb-like, ($21
\times 21 \times 1$) for planar kagome, ($19 \times 19 \times 1$) for
buckled kagome, ($23 \times 23 \times 1$) for square planar, ($23
\times 23 \times 1$) for buckled square, and ($1 \times 1 \times 18$)
for nanowires.

Vibrational and thermodynamic properties are evaluated from phonon
dispersion calculations using the finite-displacement
method. Supercells of sizes ($4 \times 4 \times 4$) for bulk, ($4
\times 4 \times 1$) for monolayers, and ($1 \times 1 \times 15$) for
nanowires are employed. The same plane-wave cutoff energy (520~eV) and
\textbf{k}-point densities as in the electronic structure calculations
are used, with a force convergence threshold of
$10^{-6}$~eV/\AA. Phonon dispersion relations and corresponding
density of states (DOS) are computed using the Phonopy
package.\cite{phonopy-phono3py-JPCM, phonopy-phono3py-JPSJ}

To evaluate the Chern number and the $\mathbb{Z}_2$ topological
invariant, maximally localized Wannier functions (MLWFs) are
constructed using the \textsc{Wannier90} package.\cite{wantool} The
corresponding topological invariants are subsequently computed with
the \textsc{WannierTools} post-processing code.\cite{WU2017} The topological properties of the one-dimensional nanowire were analyzed using the modern theory of polarization. 
Atomic structures are visualized using the \textsc{VESTA}
software,\cite{Momma:db5098} and all plots and graphical analyses are
produced with the \textsc{Matplotlib} library.\cite{Hunter:2007}

\section{Results and Discussions}

Tellurium crystallizes in a stable trigonal phase composed
of one-dimensional helical chains of Te atoms along the $c$-axis. Each
atom has a $5s^2 5p^4$ configuration, where two $5p$ electrons form
covalent bonds within the chains, while the $5s$ electrons remain
core-like. The remaining $5p$ electrons form lone pairs oriented
between chains, leading to interchain vdW interactions. This results
in strong intrachain bonding and a quasi-layered, anisotropic
structure~\cite{core,core2}.

\begin{figure}[H]
	\centering
	\begin{subfigure}[b]{0.32\columnwidth}
    \subcaption[]{}
		\includegraphics[width=\columnwidth]{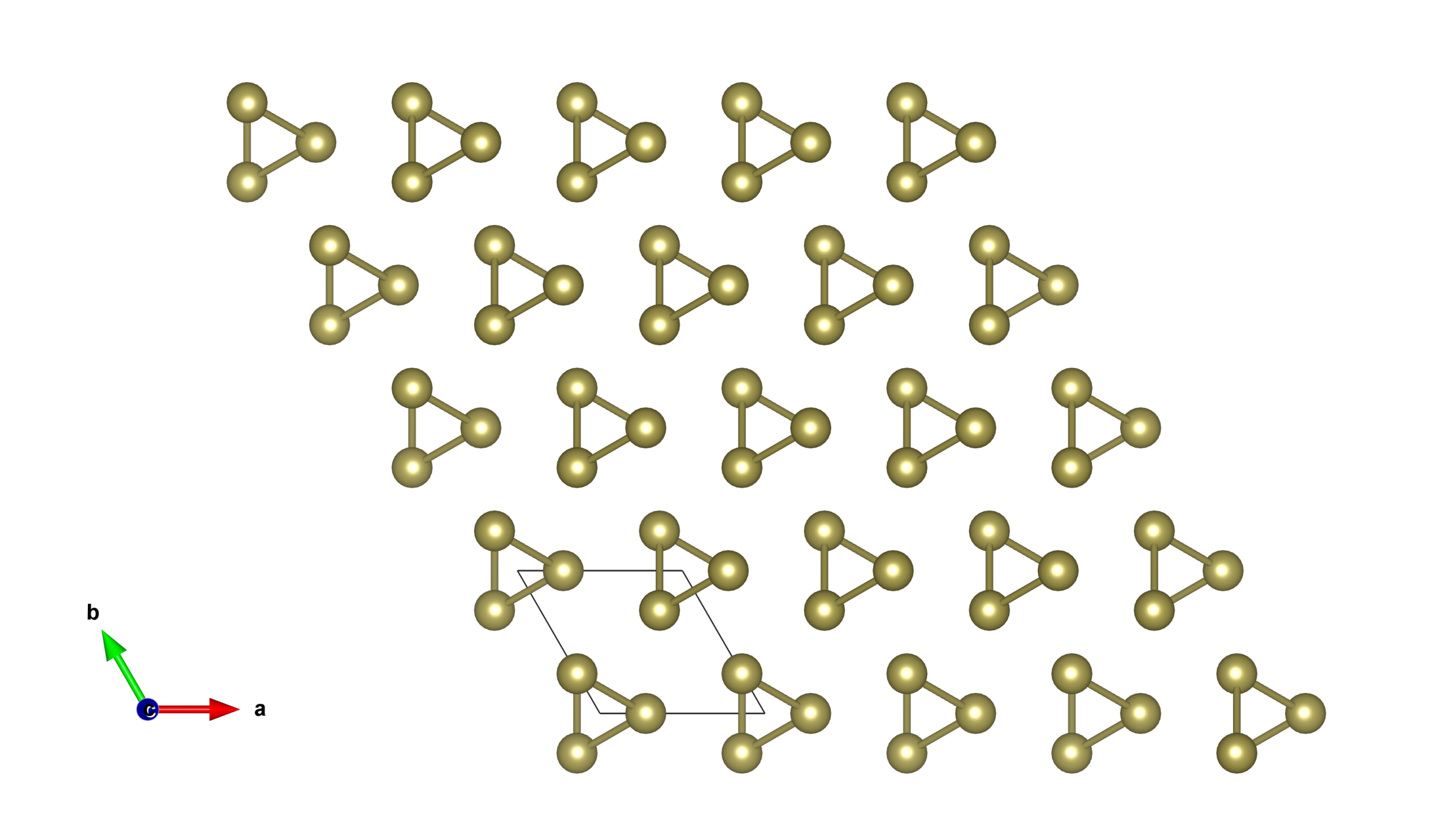}
	\end{subfigure}
	\hspace{0.1cm}
	\begin{subfigure}[b]{0.32\columnwidth}
    \subcaption[]{}
		\includegraphics[width=\columnwidth]{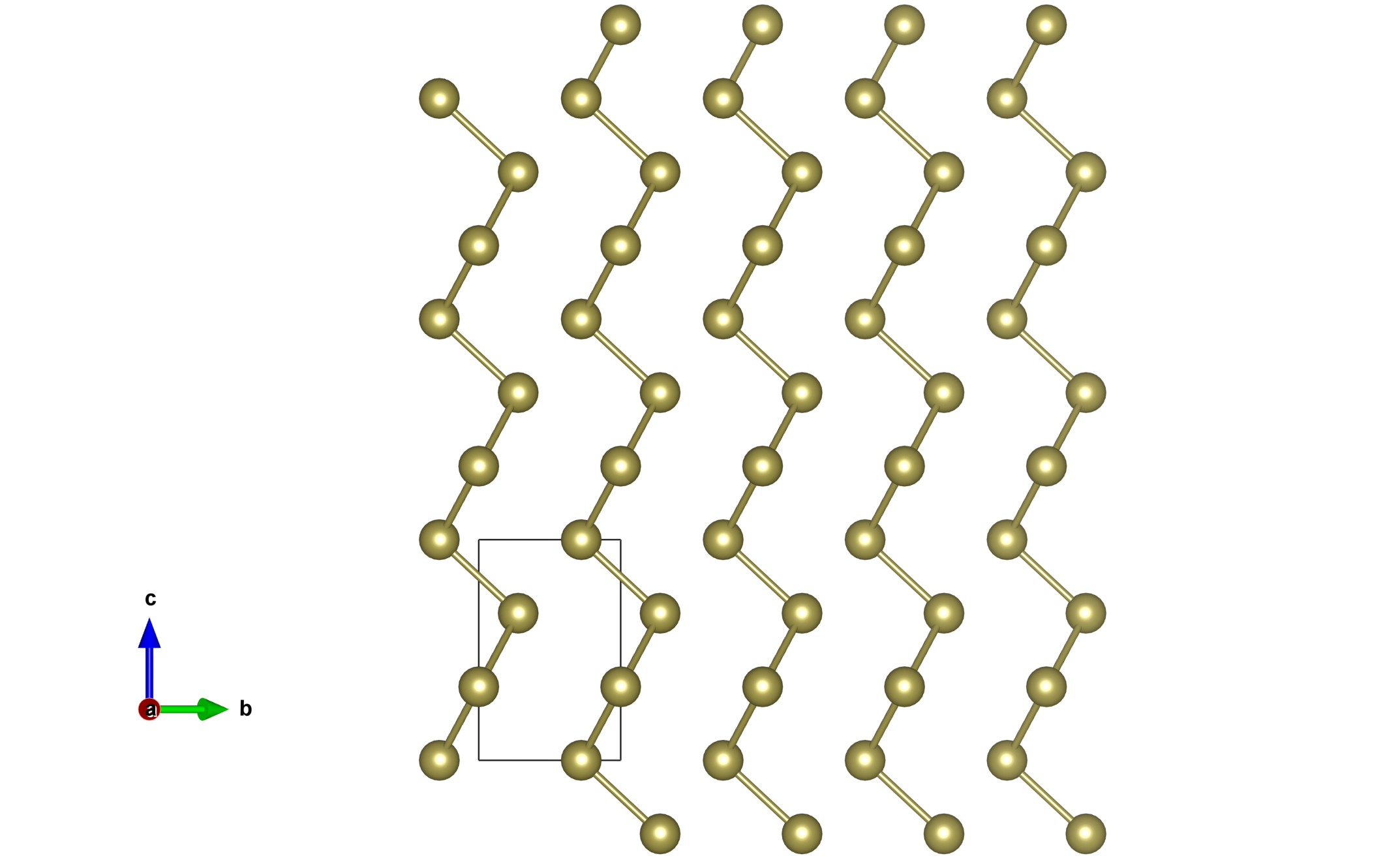}
	\end{subfigure}
	\hspace{0.1cm}
	\begin{subfigure}[b]{0.32\columnwidth}
    \subcaption{}
		\includegraphics[width=\columnwidth]{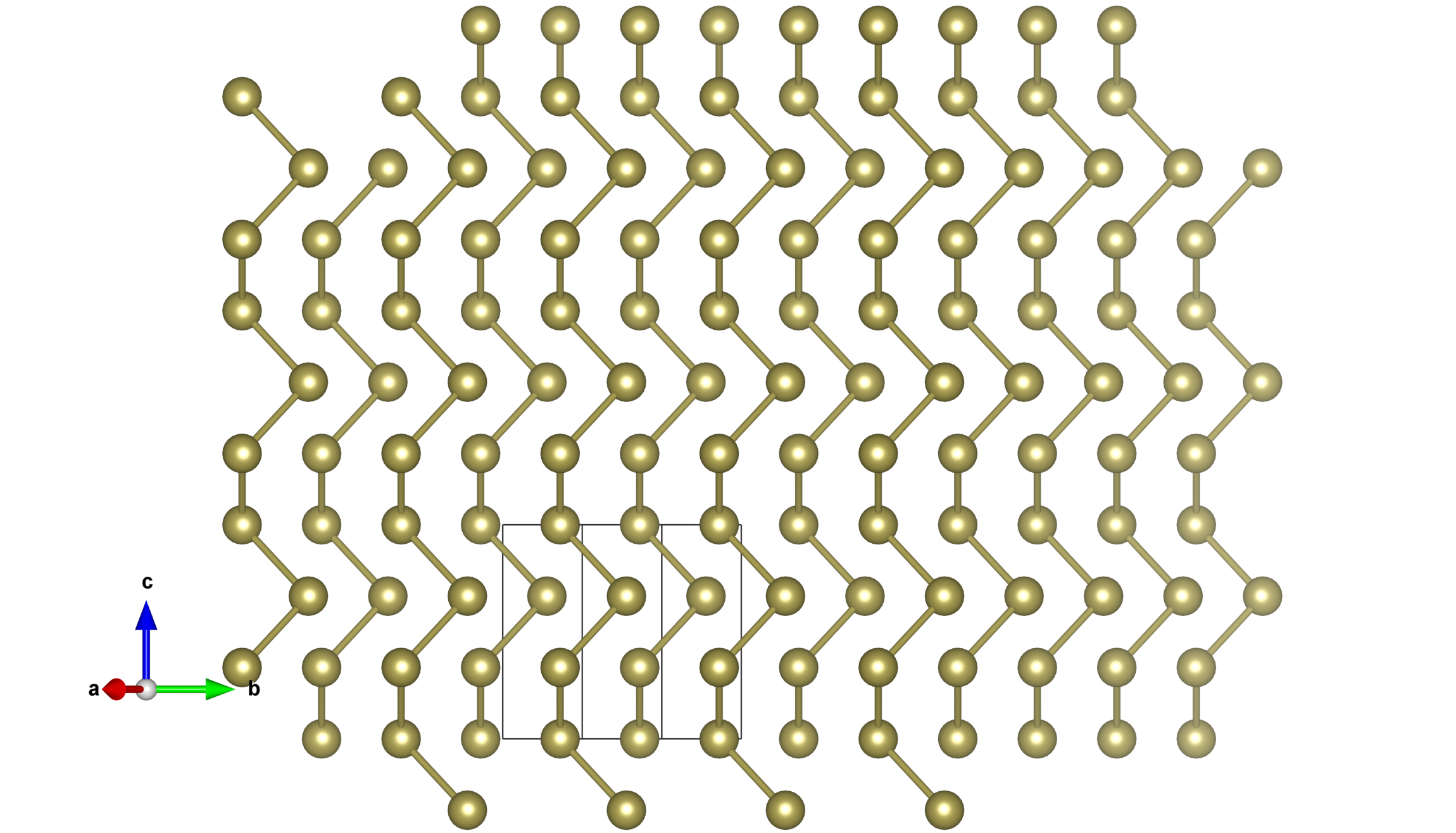}
	\end{subfigure}
	\caption{~\label{fig:te1_geometry}~Relaxed geometry of trigonal Te-I: (a) top view projected along  $\vec{c}$; (b) along $\vec{a}$; (c) perpendicular to $\vec{b}$.}
\end{figure}

The helical structure exhibits structural chirality, adopting either the
right-handed $P3_121$ ($D_3^4$) or the left-handed $P3_221$ ($D_3^6$)
space group, making them
non-centrosymmetric systems.  After structure optimization shown in
Fig. \ref{fig:te1_geometry}, the obtained lattice parameters within
GGA-PBE are $a = b = 4.41$~\AA\ and $c = 5.93$~\AA, with a Te-Te bond
length of $d_h = 2.90$~\AA. These values are very close to previous
theoretical calculations
\cite{lattice_te,Yi_Zhu_Cai_Jia_Cho_2017,Kramer} and close to the
experimental values of $a = b = 4.45$~\AA\ and $c = 5.93$~\AA\ \cite{EXP_TE_BULK}.

The dynamic stability of trigonal tellurium (Te-I) is investigated
through phonon dispersion calculations within the harmonic
approximation, employing the finite displacement method.  The
calculated phonon band structure of trigonal tellurium (Te-I) along with thermal properties are shown in Fig.~\ref{fig:S1}. A closer inspection of the optical phonon modes at the $\Gamma$ point reveals doubly degenerate branches along the $\Gamma$-A direction at approximately 2.44 THz and 3.89 THz. These modes involve coupled atomic displacements combining bond stretching and angular distortions, and reflect the intrinsic chiral symmetry of the trigonal structure. While such degeneracies are compatible with chiral lattice dynamics, we emphasize that a definitive identification of Weyl phonons \cite{weyl_phon} would require an explicit evaluation of phonon Berry curvature and topological charges, which is beyond the scope of the present work. The observed phonon features therefore suggest, but do not by themselves establish, topologically nontrivial phononic behavior.

\begin{figure}[H]
    \centering
    \begin{subfigure}[b]{0.3\textwidth}
        \subcaption{}
        \includegraphics[width=\textwidth]{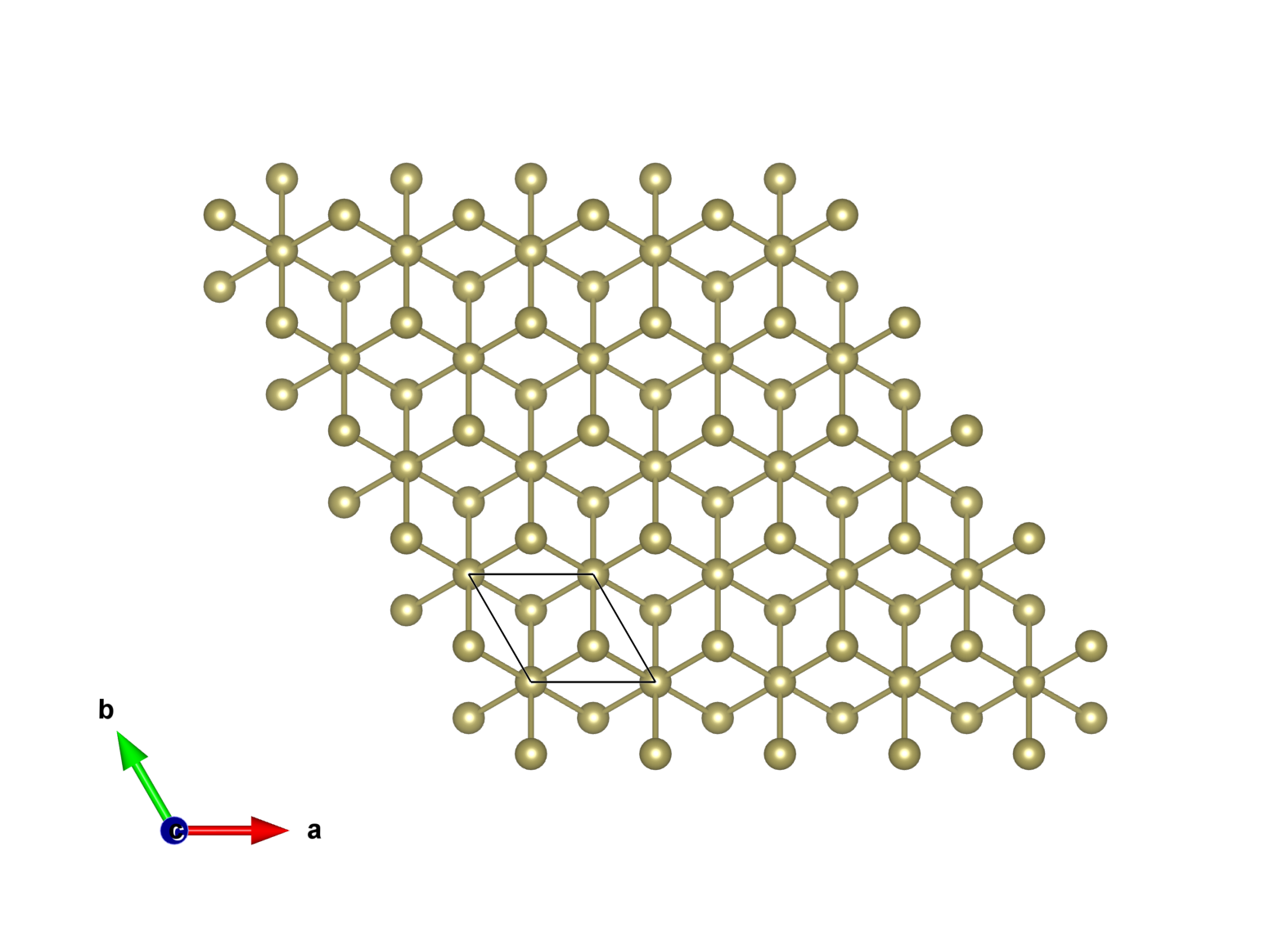}
    \end{subfigure}
    \begin{subfigure}[b]{0.3\textwidth}
        \subcaption{}
        \includegraphics[width=\textwidth]{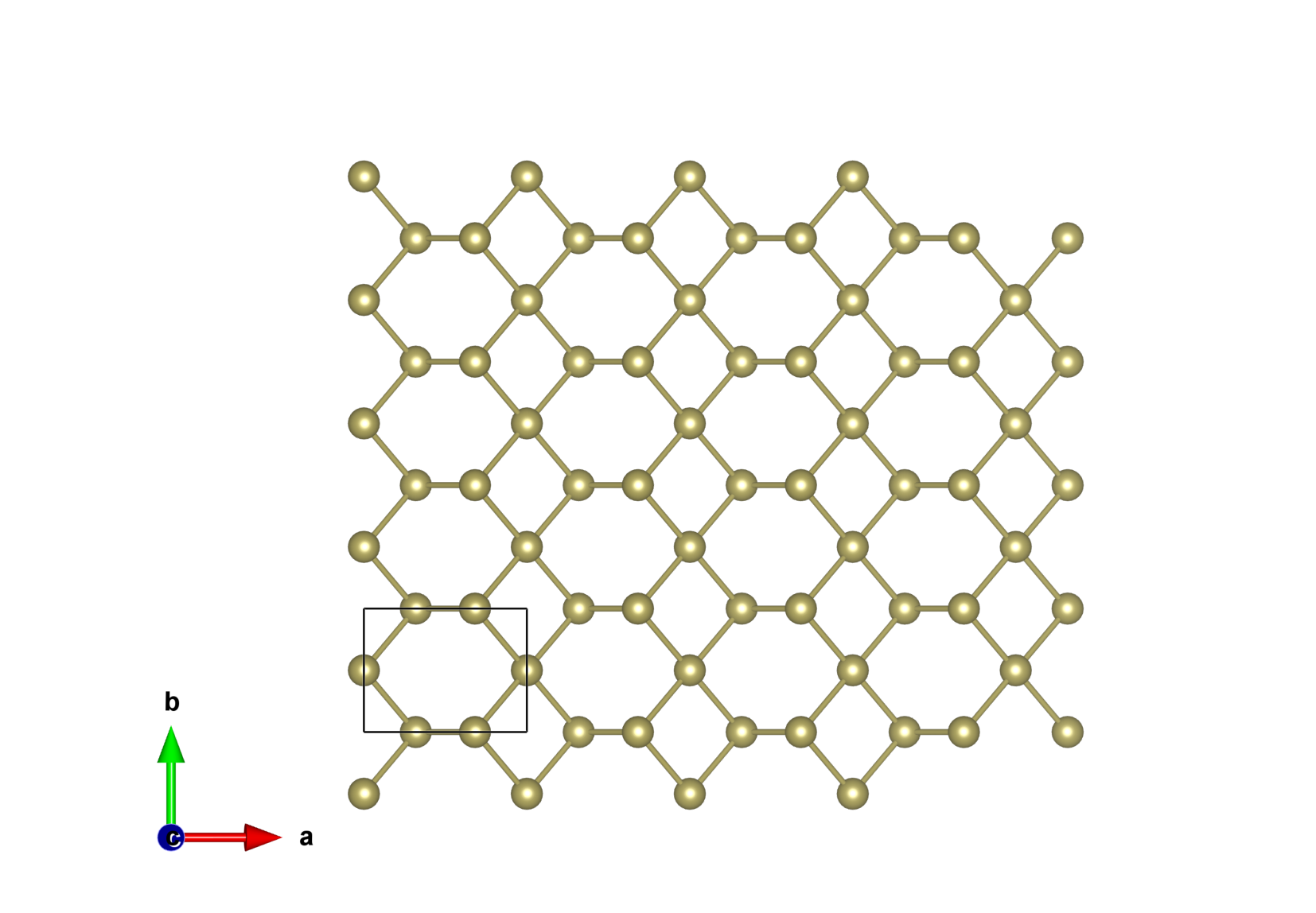}
    \end{subfigure}
    \begin{subfigure}[b]{0.3\textwidth}
        \subcaption{}
        \includegraphics[width=\textwidth]{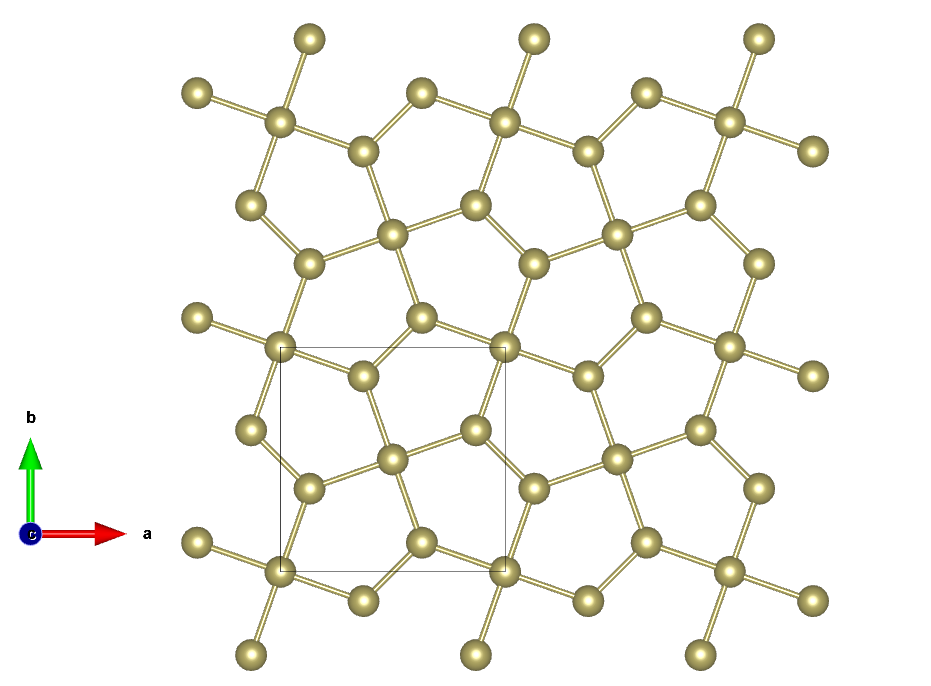}
    \end{subfigure}   
    \begin{subfigure}[b]{0.3\textwidth}
        \includegraphics[width=\textwidth]{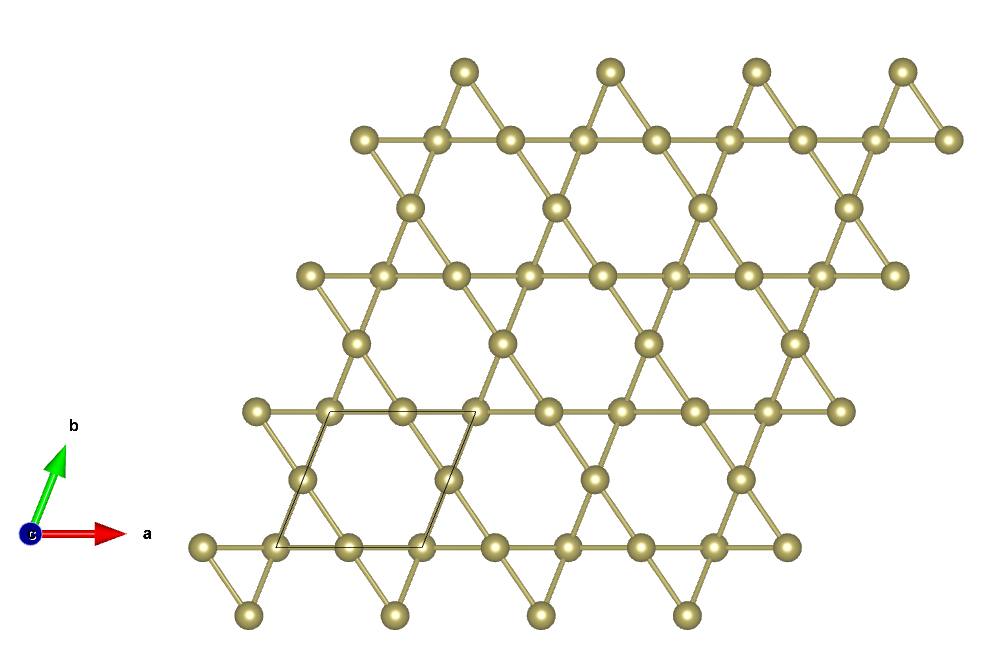}
        \subcaption{}
    \end{subfigure}
    \begin{subfigure}[b]{0.3\textwidth}
         \subcaption{}
        \includegraphics[width=\textwidth]{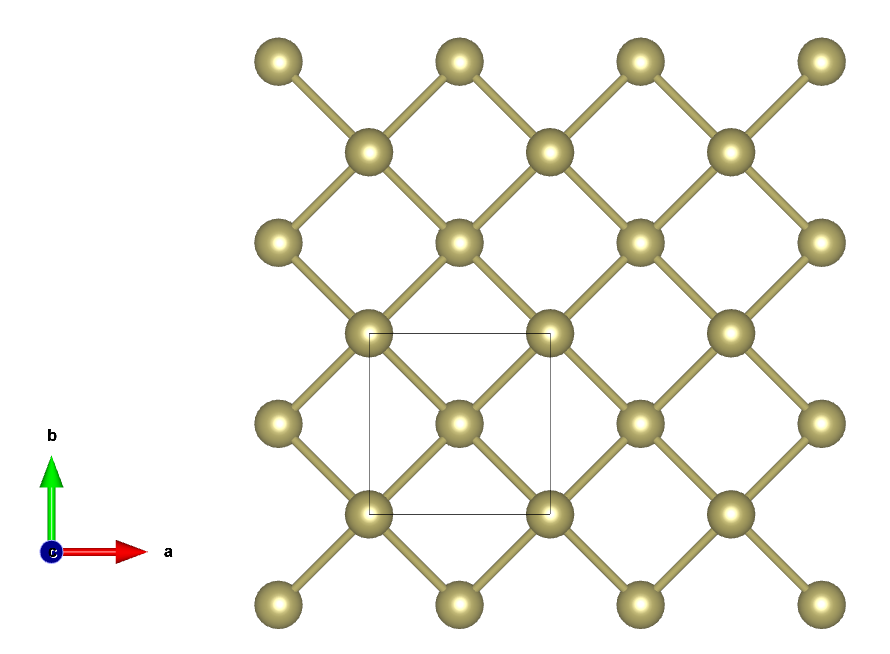}
    \end{subfigure}
\begin{subfigure}[b]{0.30\textwidth}
\centering
\subcaption{}
\includegraphics[width=\textwidth]{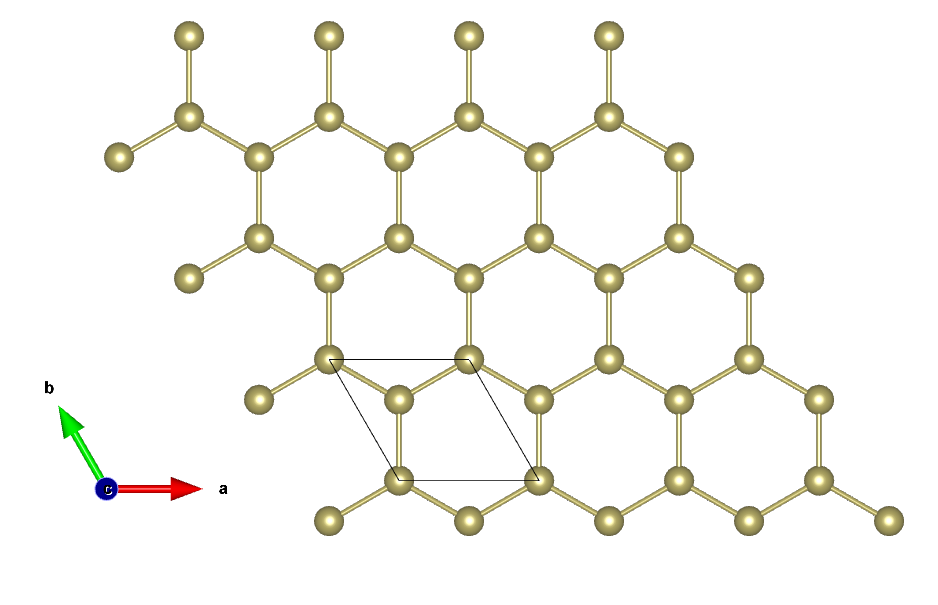}
\end{subfigure}
\begin{subfigure}[b]{0.30\textwidth}
\subcaption{}
\includegraphics[width=\textwidth]{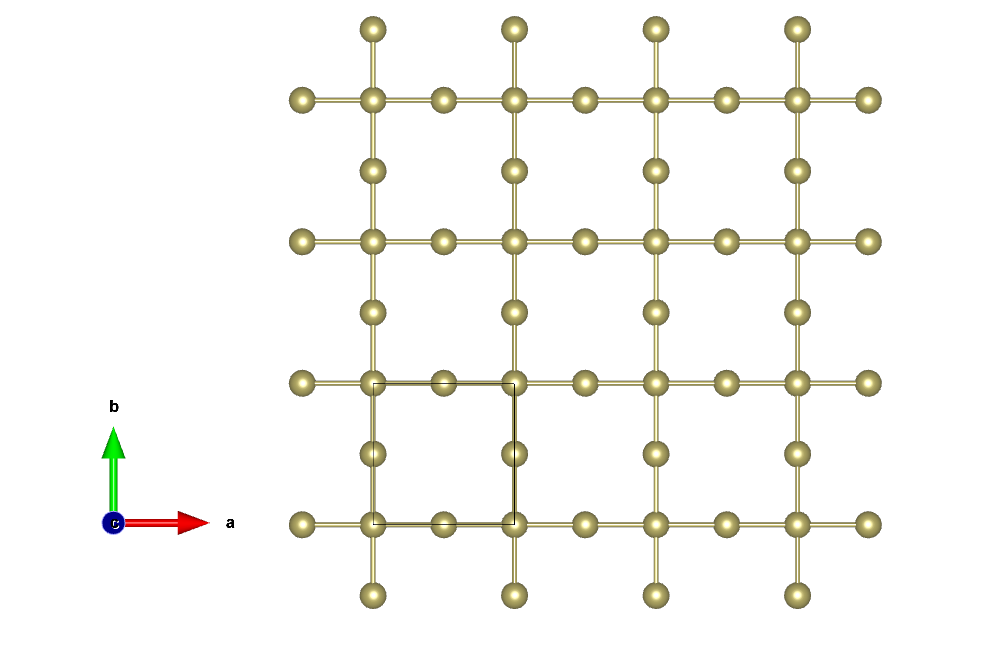}
\end{subfigure}
\begin{subfigure}[b]{0.30\textwidth}
\subcaption{}
\includegraphics[width=\textwidth]{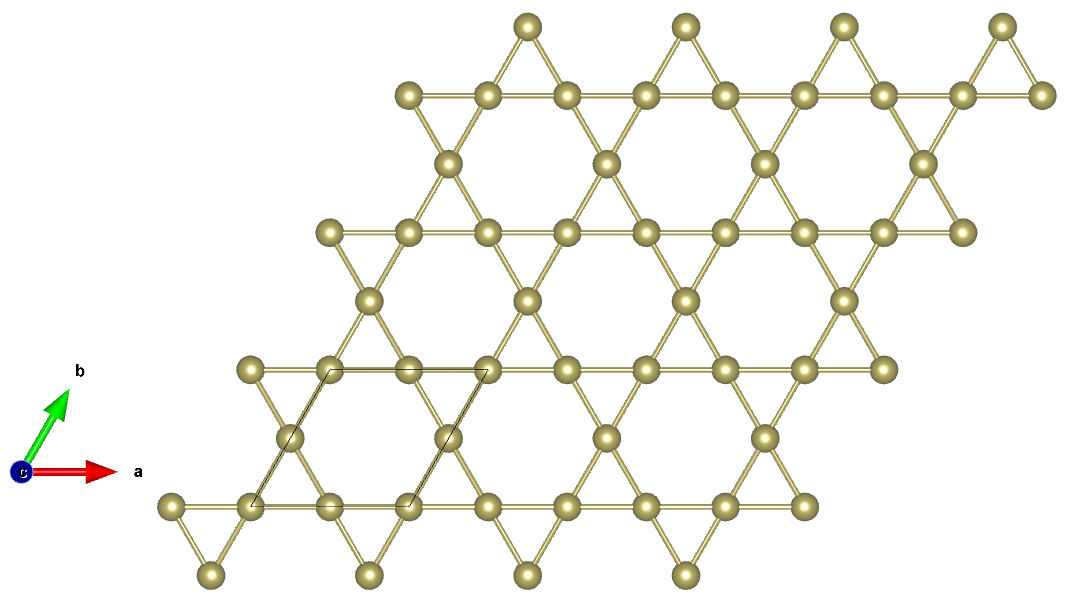}
\end{subfigure}
\begin{subfigure}[b]{0.30\textwidth}
\centering
\subcaption{}
\includegraphics[width=\textwidth]{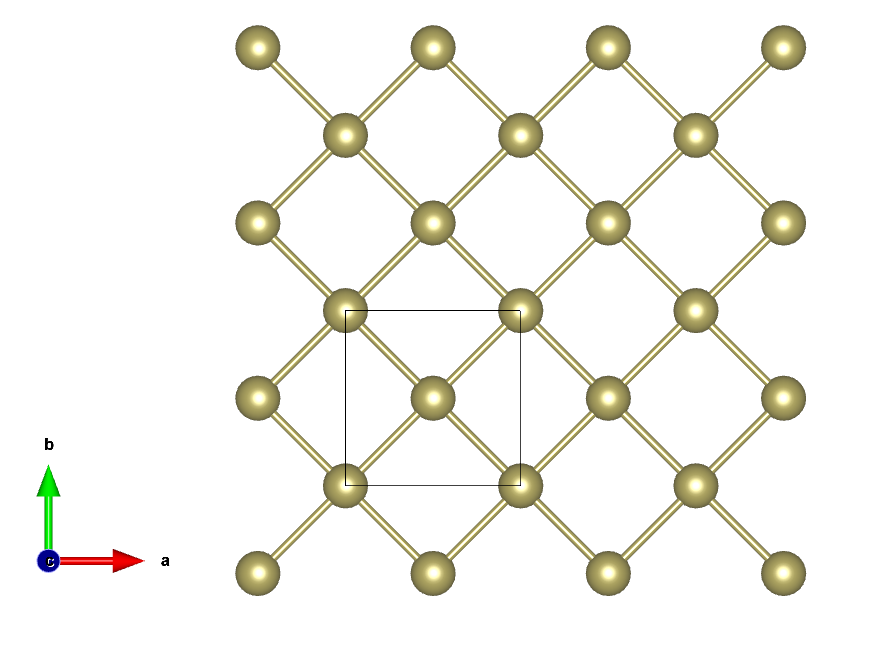}
\end{subfigure}
\caption{\label{fig:2d_structures} Relaxed crystal structures of 2D  tellurium phases: a) $\alpha$-Te, b) $\beta$-Te, c) buckled pentagonal, d) buckled kagome, e) buckled square, f) planar hexagonal, g) Lieb-like, h) planar kagome and i) planar square.}
\end{figure}

Much less explored are the 2D allotropes of tellurium predominantly
found in  the trigonal $\alpha$-phase ($\alpha$-Te),
characterized by the $P\bar{3}m1$ space group, and the monoclinic $\beta$-phase
($\beta$-Te), belonging to the $P2/m$ space group with distinct
zigzag and armchair directions. Tellurium propensity to form these
2D monolayers is attributed to its outer
valence electron configuration and a potential Peierls instability, a
distortion of the periodic lattice in a one-dimensional crystal that
breaks its perfect translational symmetry, in this case, tellurium
helical chains~\cite{peil,alpha_exp_peil}. This instability can drive
a spontaneous structural transition toward energetically more
favorable 2D configurations, resulting in the formation
of $\alpha$-phase, show in
Fig.~\ref{fig:2d_structures}(a). This structure exhibits a lower total energy, than the
$\beta$-phase shown in Fig.~\ref{fig:2d_structures}(b). Additionally, we investigate the following 2D-tellurium structures: Fig.~\ref{fig:2d_structures}(c) (buckled pentagonal), Fig.~\ref{fig:2d_structures}(d) (buckled kagome), Fig.~\ref{fig:2d_structures}(e) (buckled square), Fig.~\ref{fig:2d_structures}(f) (planar hexagonal), Fig.~\ref{fig:2d_structures}(g) (Lieb-like), Fig.~\ref{fig:2d_structures}(h) (planar kagome) and Fig.~\ref{fig:2d_structures}(i) (planar square).
These forms of tellurium are proposed here inspired by novel newly synthesized 2D square/rectangular\,\cite{10.1021/acsami.2c20400} and hexagonal\,\cite{acs.nanolett.4c02171} phases, kagome metals\,\cite{cdjy-6jgs,NWAOGBO2025114125}, pentagonal bismuthene\,\cite{LU2024565}, graphene\cite{graphene_penta,10.1073/pnas.1520402112} and phosphorene\,\cite{10.1002/adma.202411182} just to cite a few.

For \(\alpha\)-Te, the calculated lattice parameters are \(a = b =
4.22\,\text{\AA}\), with a Te-Te bond length of \(d =
3.03\,\text{\AA}\), very similar to previous theoretical~\cite{lattice_te,zhu2016tellureneamonolayertelluriumfirstprinciples,adatom}
and experimental~\cite{alpha_exp_peil,alpha_exp}values. The \(\beta\)-tellurene
phase exhibits lattice parameters \(a = 5.61\,\text{\AA}\) and \(b =
4.22\,\text{\AA}\), featuring Te-Te bond lengths of \(d =
3.03\,\text{\AA}\) and \(d = 2.76\,\text{\AA}\), similar to previous theoretical~\cite{beta_theo,lattice_te,adatom} and experimental
data~\cite{beta_exp}. The buckled pentagonal structure has lattice
parameters of 7.71{\AA}, the buckled kagome of 5.51{\AA} and the
buckled square of 4.10{\AA}. This buckling leads to Te-Te bond lengths
of 3.02 {\AA}, 2.96{\AA} and 3.03{\AA}, respectively.

Topology is conditional on stabilization.  The buckled kagome and buckled square tellurene lattices are found to host nontrivial $Z_2$ = 1 topology arising from SOC-gapped near-crossings in the electronic structure. Phonon calculations indicate that these free-standing configurations exhibit soft modes, suggesting that their stabilization likely requires interaction with a substrate or external constraints. Importantly, similar kagome- and square-based tellurium phases have been experimentally realized on metallic substrates, supporting the physical relevance of these structures. Our results therefore establish the intrinsic topological character of these lattices, conditional on structural stabilization.
Applying 5\% strain modifies
the band dispersions and shifts the band edges, but SOC continues to
gap the near-crossings along $\Gamma$-X-M-$\Gamma$ [Fig.~\ref{fig:S6}(l)].
For the buckled
structures, our results are comparable to previous ones for
$\alpha$-Te, $\beta$-Te~\cite{alpha_beta_phon},
pentagonal\,\cite{ZHANG2021149851},
hexagonal\,\cite{acs.nanolett.4c02171}. We emphasize the importance of the substrate: the DFT
ground state is a square lattice, while in the experiment a
rectangular reconstruction is found on Ni(111)
substrate\,\cite{10.1021/acsami.2c20400}. 

Fig.~\ref{fig:S2} shows the phonon dispersion curves of mechanically
stable structures, with exception of the pentagonal structure that
shows a couple of imaginary frequencies. The planar hexagonal, kagome,
Lieb-like and square lattices in the absence of buckling are not
mechanically stable (phonons not shown here), but we suggest
nevertheless that these may be substrate-stabilized candidate phases.

In $\alpha$-Te, one of the acoustic branches exhibits a soft mode
[Fig.~\ref{fig:S2}(a)], indicating a possible dynamic instability under small
perturbations. In $\beta$-Te, all three acoustic modes are softened,
particularly along the armchair direction ($\Gamma$–X), and several
optical branches show a noticeable reduction in energy
[Fig.~\ref{fig:S2}(b)]. This softening implies a higher susceptibility of these
2D structures to structural distortions or phase transitions when
subjected to external perturbations. Although the buckled pentagonal,
[Fig.~\ref{fig:S2}(c)], and buckled kagome phases show small imaginary
frequencies, as seen in Fig.~S\ref{fig:S2}(d) and (e), it is be possible to
stabilize these phases on an appropriate substrate. Strain induced by the substrate possibly introduce some electrostatic interaction and change the corrugation. This is consistent with experimental results for
pentagonal\cite{ZHANG2021149851},
hexagonal,\cite{acs.nanolett.4c02171}, and rectangular
lattices\cite{10.1021/acsami.2c20400} which are reported to be topological systems.

In $\beta$-Te, the isolated and nearly flat optical mode above 5 THz
exhibits a noticeable hardening. This behavior can be attributed to
modifications in interatomic bonding induced by reduced
dimensionality. Additionally, several optical branches in the 1–2 THz
range intersect with higher-energy acoustic phonons, suggesting
enhanced phonon–phonon interactions and possible anharmonic effects in
this phase. The modes in $\alpha$-Te (a)–(e) presented in
Fig.~\ref{fig:S3} correspond to contraction and expansion motions
within the $yz$-plane. In $\beta$-Te, the modes (f)–(i) shown in
produce expansion and contraction within the
$xy$-plane, reflecting in-plane lattice vibrations. The stiffer mode
corresponds to an out-of-phase torsional motion of the
atoms. Crossover points in the phonon dispersion are also observed at
the high-symmetry K point.

We now turn our discussion to single-helix tellurium nanowires
(Te-h). These one-dimensional structures can be obtained, for example,
by decoupling the helical atomic chains that constitute bulk trigonal
tellurium (Te-I),\cite{NAK2024,teexp} thereby preserving the intrinsic
geometrical chirality and screw symmetry of Te-I at the single-chain
level while removing the interchain packing present in the bulk
crystal. The resulting quantum-confined geometry, combined with the
strong spin–orbit coupling of tellurium, gives rise to distinctive
one-dimensional electronic properties. We emphasize that the
structural chirality of the helical chain should not be confused with
chiral (sublattice) symmetry of the electronic Hamiltonian; in the
present nanowires the latter is absent (see Fig.~S7 in SI and the
details for Zak phase calculations.

\begin{figure}[H]
  \centering
  \begin{subfigure}[b]{0.4\textwidth}
\subcaption{}
\includegraphics[width=\textwidth]{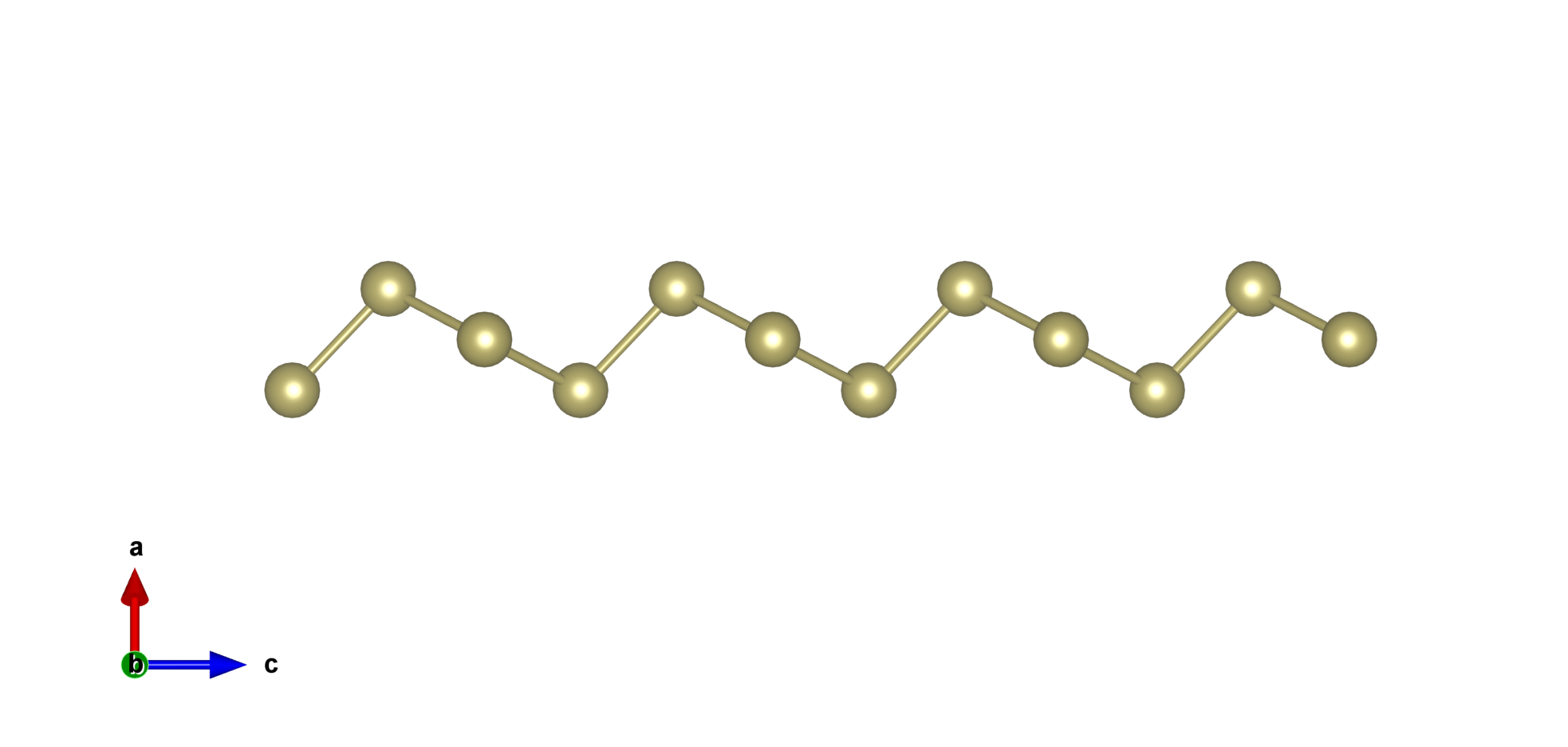}
  \end{subfigure}
  \begin{subfigure}[b]{0.4\textwidth}
\subcaption{}
        \includegraphics[width=\textwidth]{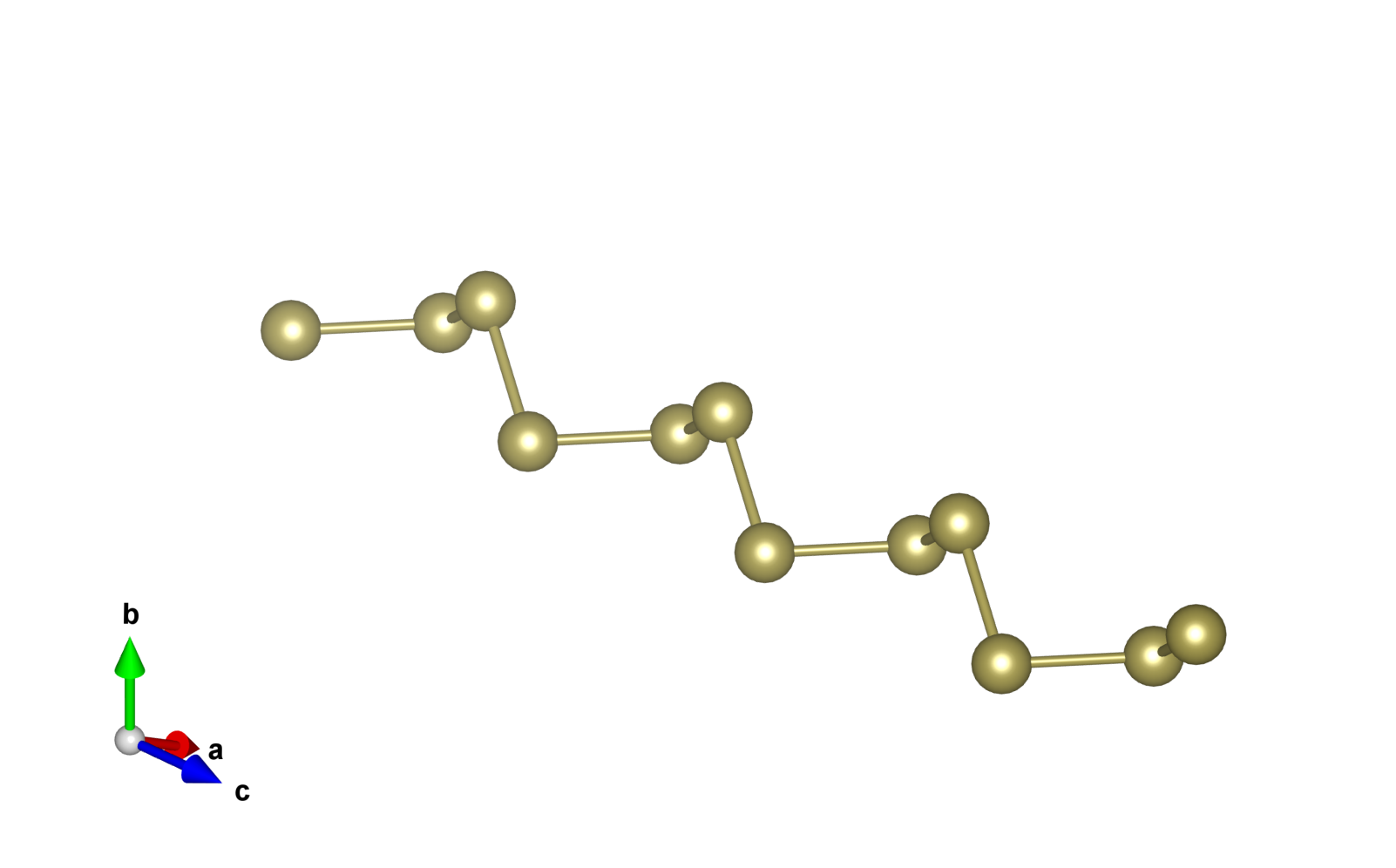}
 \end{subfigure}
        \caption{Helical geometry of the tellurium nanowire (Te-h).}
	\label{fig:teh_geometry}
\end{figure}

Fig.~\ref{fig:teh_geometry} shows the optimized structures of an ultra-thin tellurium
nanowire. The obtained lattice parameter is \(c = 5.67\,\text{\AA}\)
with a Te-Te bond length of $d = 2.74$~\AA\, in good
agreement with previous theoretical
results~\cite{teh_stab,Kramer}.  From the phonon dispersion curves
depicted in Fig.~\ref{fig:S5}, the absence of
imaginary frequencies in Te-h along the high symmetry paths,
indicating the possibility of local stability at low temperature, in agreement  with
previously reported theoretical data~\cite{teh_stab}. Also, we see that the acoustic
branches and one optical branch have softened, since their energies
decrease compared to Te-I ($\Gamma$ - A), suggesting potential
structural phase transitions under small perturbations. On the other
hand, the energy of some optical branches increased significantly
(hardening). For example, the mode at 0.14 THz [Fig.~\ref{fig:S5}(a)] 
corresponds to torsional oscillations, similarly to the Te-I phase,
but with a significantly lower eigenfrequency. This reduction is
attributed to the absence of inter-chain interactions in the
system. The Te-h phase hardens the modes that involve a combination of
bond stretching and bending. The shorter Te–Te bond length in Te-h
(2.74~\AA) as opposed to bulk Te (2.90~\AA) is consistent with this
behavior.

The cohesive energy serves as a key
metric for evaluating the relative stability of various Te-based
phases. Accordingly, the cohesive energy, $E_{\text{coh}}$, is defined
as $E_{\text{coh}} = \frac{n E_{\text{atom}} - E_{\text{tot}}}{n}$, where $E_{\text{atom}}$ is the total energy of a single, isolated Te atom, $E_{\text{tot}}$ is the total energy of the fully relaxed system, and $n$ is the number of Te atoms in the structure.

\begin{table}[H]
\centering
\caption{\label{tab:param_coh}Lattice parameters $a$, $b$ and $c$, interatomic distances d$_{\rm Te-Te}$, coh\
esive energies ${\rm E_{coh}}$ and C$_v$ for tellurium phases calculated within GGA. E$_g$ calculated within HSE06.}
\begin{tabular}{cccccccc}
\toprule 
&  \multicolumn{3}{c}{lattice constant (Å)} & d$_{\rm Te-Te}$ ({\AA}) & ${\rm E_{coh}}$ (eV) & C$_v$ ({J/K$\cdot$mol})  & E$_g$ (eV) \\
\midrule 
Phase & a & b & c  &  & &  \\
\midrule
Te-I        &   4.41 & 4.41 & 5.93 &      2.90       & -2.75 &    73.68  & 0.30 \\ 
$\alpha$-Te &   4.22 & 4.22 & -    &      3.03       & -2.61  &   73.50 & 0.75\\
$\beta$-Te  &   5.61 & 4.22 & -    &     3.03, 2.76  & -2.55   &  73.53 &  1.44 \\
buckled pentagonal      & 7.71 & 7.71 & -    &       3.02  & -2.21  &  72.07 & -\\
buckled kagome &   5.51 & 5.51 & -    &        2.96  & -2.30   &   72.70 &  -\\
buckled square & 4.10 & 4.10 & -    &        3.03  & -2.39 & 48.06 & - \\
Te-h     &   -    &   -   & 5.67    &         2.74  & -2.38 & 73.07 &  2.23 \\
\bottomrule
\end{tabular}
\end{table}

According to Table~\ref{tab:param_coh}, $\alpha$-Te is identified as
the most stable 2D phase, while the remaining phases are metastable
with respect to $\alpha$-Te. The smaller cohesive energy of
$\alpha$-Te compared to $\beta$-Te confirms its higher thermodynamic
stability. Although $\beta$-Te is thermodynamically less stable than
$\alpha$-Te, the monoclinic structure may still be experimentally
accessible under specific synthesis conditions,\cite{beton}. The
buckled pentagonal, kagome, and square structures exhibit comparable
cohesive energies, in agreement with experimental observations for
these phases obtained under different substrates and growth
conditions,\cite{ZHANG2021149851,acs.nanolett.4c02171,10.1021/acsami.2c20400}. This
finding suggests that alternative tellurium phases can be stabilized
depending on the experimental environment and synthesis parameters.

For Te-h, the blue curve exhibits a steep slope, indicating a rapid
increase in entropy with temperature [Fig.~\ref{fig:S5}(b)]. This behavior arises from the
enhanced vibrational degrees of freedom intrinsic to its
one-dimensional structure. The results indicate that the primary contributions to the
thermodynamic properties arise from in-plane atomic interactions as
the dimensionality decreases from bulk to monolayer. Consequently,
interlayer interactions—absent in 2D systems—appear to play only a
minor role in determining the thermal stability of these
structures. 

The specific heat ($C_v$) at $T = 300$~K is calculated to be around
73~J/(K·mol) for 3D, 2D and 1D phases (with exception of buckled
square) shown in Figs.~\ref{fig:S1},\ref{fig:S4},\ref{fig:S5} and Table~\ref{tab:S1}. In comparison,
graphene exhibits a much lower value of approximately
7~J/(K·mol),\cite{graph_cv}. For phosphorene, the specific heat is not
constant across its allotropes; for black phosphorene, it has been
reported as 12.39~J/(K$\cdot$mol) at room
temperature,\cite{PhysRevB.92.081408}. Similarly, for monolayer
2H-MoS$_2$, the reported value is 61.12~J/(K$\cdot$mol) at
300~K,\cite{BANO20216464}. These thermodynamic results, cohesive
energy values, and dynamical stability analyses exhibit a consistent
trend, reinforcing the conclusion that Te-I represents the most stable
phase, followed by the 2D phases and the 1D Te-h nanowire.

The MLWF-HSE06 band structures calculated with and without spin–orbit
coupling (SOC), shown in Fig.~S\ref{fig:S6}(j) for Te-I, reveal an
indirect electronic band gap located at the high-symmetry point H. The computed
band gaps are 0.49 (0.30)~eV with (without) SOC, in good agreement
with the experimental value of 0.33~eV,\cite{tei_gap}. These results
classify Te-I as a narrow-gap semiconductor. Within the energy range from $-1$~eV to $0$~eV, the valence band at
the H point is fourfold degenerate when SOC is neglected. The
inclusion of SOC lifts this degeneracy, resulting in two
non-degenerate states and one doubly degenerate state at lower
energy. The region between the
high-symmetry points L–H–A highlights features of the conduction band
where doubly degenerate states exhibit linear dispersion—indicative of
Weyl nodes located close to the Fermi level.

\begin{figure}[H]
\centering
\begin{subfigure}[b]{0.4\columnwidth}
\subcaption{}
\includegraphics[width=\columnwidth,clip=true,keepaspectratio]{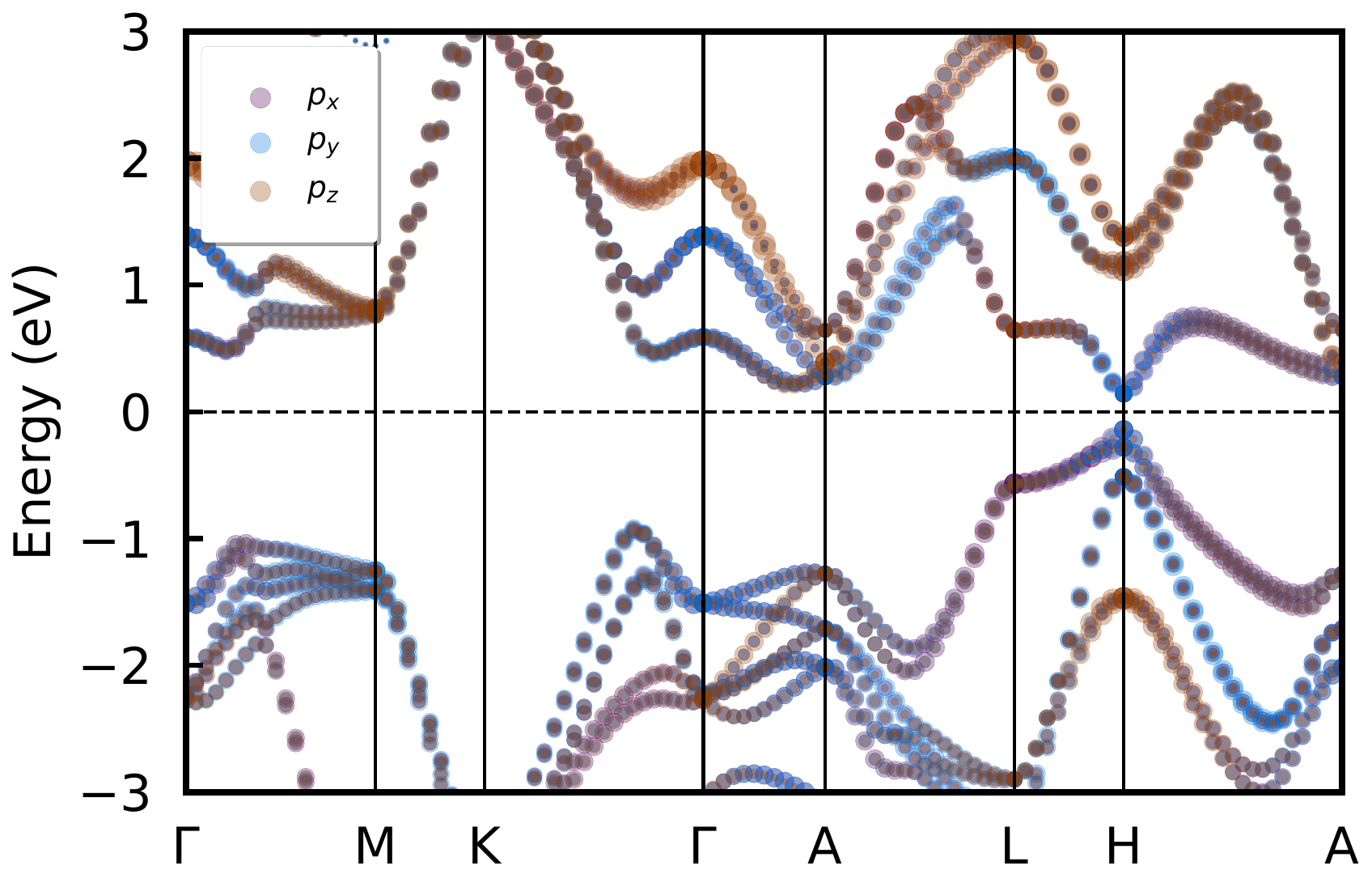}
\end{subfigure}
\begin{subfigure}[b]{0.4\columnwidth}
\subcaption{}
\includegraphics[height=4cm,clip=true,keepaspectratio]{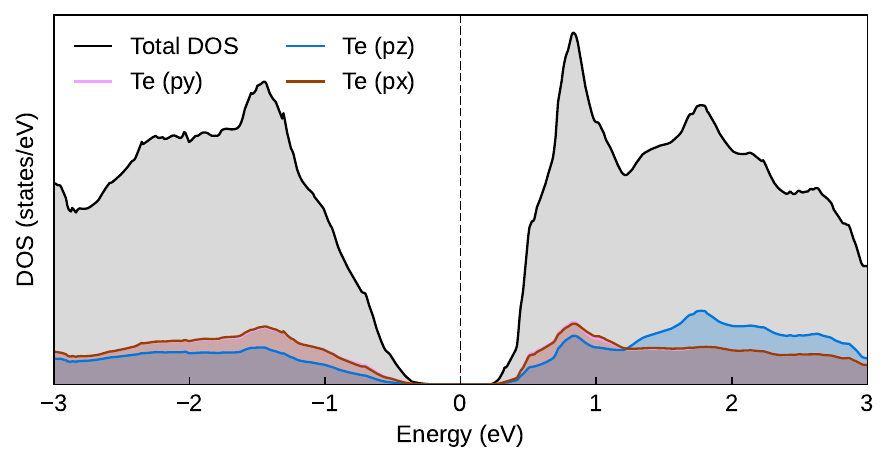}
\end{subfigure}
\caption{Orbital-projected a) electronic band structure and b) density of states of Te-I calculated within MLWF-HSE06+SOC.}
\label{fig:te_band_trigonal}
\end{figure}

The states in this energy range primarily originate from lone-pair
electrons derived from $p_x$ orbitals. The six unoccupied states
correspond to anti-bonding configurations dominated by $p_z$–$p_y$
orbital interactions. The projected band structure and partial density
of states (PDOS), shown in Figs.~\ref{fig:te_band_trigonal}(a) and
(b), confirm that the $p_x$ and $p_y$ orbitals dominate near the Fermi
level, while contributions from $p_z$ orbitals remain negligible.

Within the valence band, the PDOS reveals a substantial overlap
between the $p_x$ and $p_y$ orbitals, indicating hybridization and
strong orbital mixing. Moreover, the $p_z$ orbital in this energy
range exhibits a PDOS profile similar in shape to those of the $p_x$
and $p_y$ orbitals. A comparable trend is observed in the conduction
bands within the 0–1~eV range, suggesting analogous orbital
interactions at higher energies.

Given the identification of possible Weyl nodes in the band structure,
we now turn to the analysis of spin textures. This investigation is
crucial to confirm the topological nature of the material, since Weyl
nodes are associated with characteristic spin–momentum locking, where
the electron spin orientation is intrinsically coupled to its momentum
direction.

\begin{figure}[H]
\centering
\begin{subfigure}[b]{0.4\columnwidth}
\subcaption{}
\includegraphics[width=\columnwidth,clip=true]{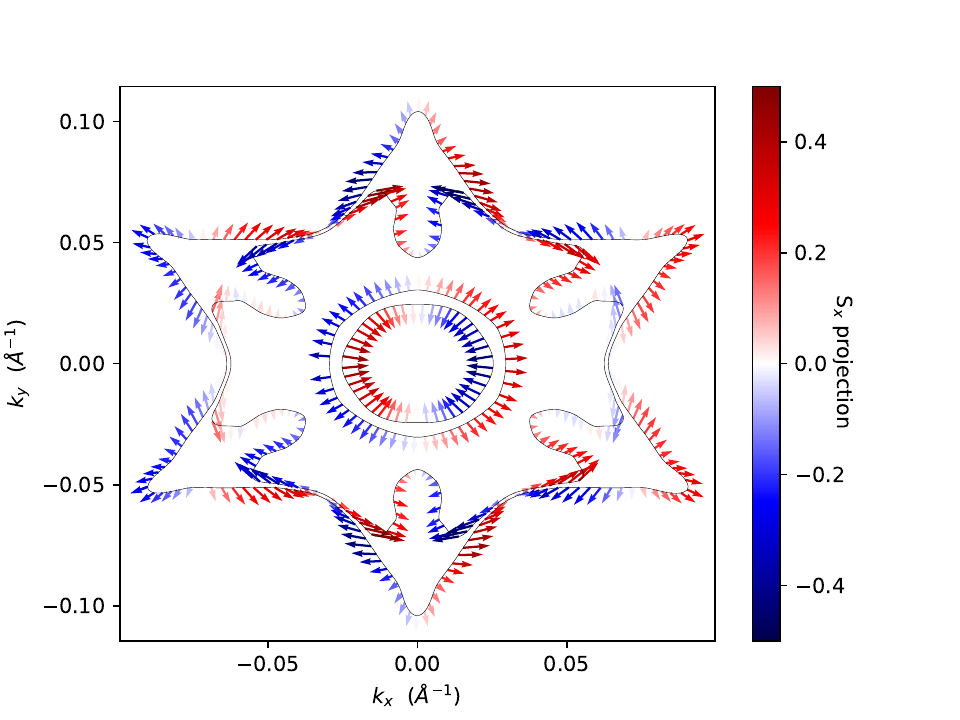} 
\end{subfigure}
\begin{subfigure}[b]{0.4\columnwidth}
\subcaption{}
\includegraphics[width=\columnwidth,clip=true]{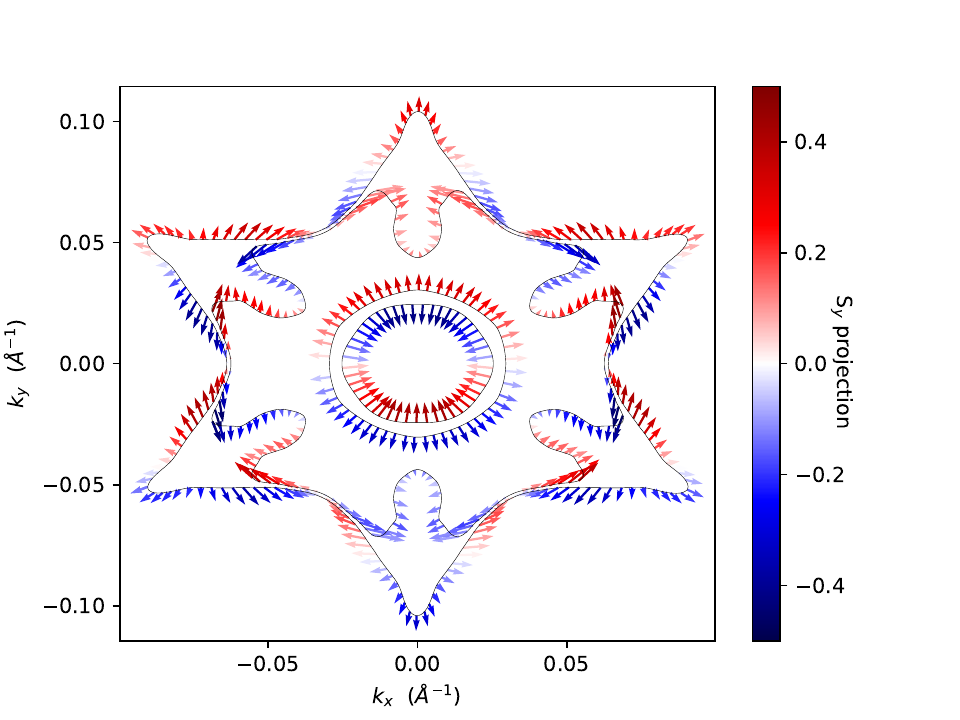}
\end{subfigure}
\caption{\label{fig:te_spin_trigonal}  Spin textures of Te-I. a)  $\langle S_x \rangle$ and b) $\langle S_y \rangle$ components of the trigonal phase at 0.9 eV. The color scale denotes the expectation values of the spin components.}	
\end{figure}

Fig.\,\ref{fig:te_spin_trigonal} shows indeed the
two crossing points at the high-symmetry point H correspond to Weyl
nodes. This identification is supported by the characteristic
hedgehog-like spin texture in momentum space, where the spins align
radially, creating spin patterns that act as “Berry monopoles” in
momentum space and are connected with a defined chirality. A chirality
charge of positive sign corresponds to a positive Chern number, while
a Weyl node with a negative chirality charge has a negative Chern
number. The magnitude of the spin components, as illustrated in the
color bars in Fig.~\ref{fig:te_spin_trigonal}, results directly from
the effects of SOC. Consequently, in regions where SOC exerts a more
significant influence on the electronic band structure, the spin
components tend to display larger expectation values, indicating a
stronger spin polarization.

Figs.~\ref{fig:S6}(a) and (b) present the electronic band structures of $\alpha$-Te and $\beta$-Te, respectively, calculated using the MLWF-HSE06 exchange–correlation functional, both with and without spin–orbit coupling (SOC). The inclusion of SOC leads to a reduction in the band gap for both phases. Specifically, for $\alpha$-Te, the band gap decreases from $1.04$~eV to $0.75$~eV upon inclusion of SOC, in excellent agreement with previous theoretical reports,\cite{multi,zhu2016,Cai_2020}. Similarly, for $\beta$-Te, the band gap decreases from $1.77$~eV to $1.44$~eV with SOC, consistent with earlier theoretical studies,\cite{multi,Cai_2020,betao,betoso}.

Upon inclusion of spin–orbit coupling (SOC), $\beta$-Te undergoes a
transition from an indirect to a direct band gap at the $\Gamma$
point, whereas $\alpha$-Te retains its indirect gap. This band-gap
transition in $\beta$-Te has the potential to enhance the material
optical absorption efficiency. SOC induces a significant reshaping of
the valence bands in both monolayers, notably lifting the degeneracy
at linearly dispersive crossing points. In the conduction band,
$\beta$-Te exhibits a smaller degree of band splitting compared with
$\alpha$-Te. Furthermore, $\alpha$-Te presents a quasi-flat band along
the M–$\Gamma$–K direction, while $\beta$-Te displays a similar
feature in the conduction band along the $\Gamma$–X–M path. The
presence of such flat bands implies an accumulation of electronic
states within a narrow energy range, typically leading to strong
electronic correlations and enhanced carrier localization—properties
of great interest for optical applications and potentially for
superconductivity.

\begin{figure}[H]
    \centering
\begin{subfigure}[b]{0.9\columnwidth}
\subcaption{}
\includegraphics[height= 4cm,clip=true,keepaspectratio]{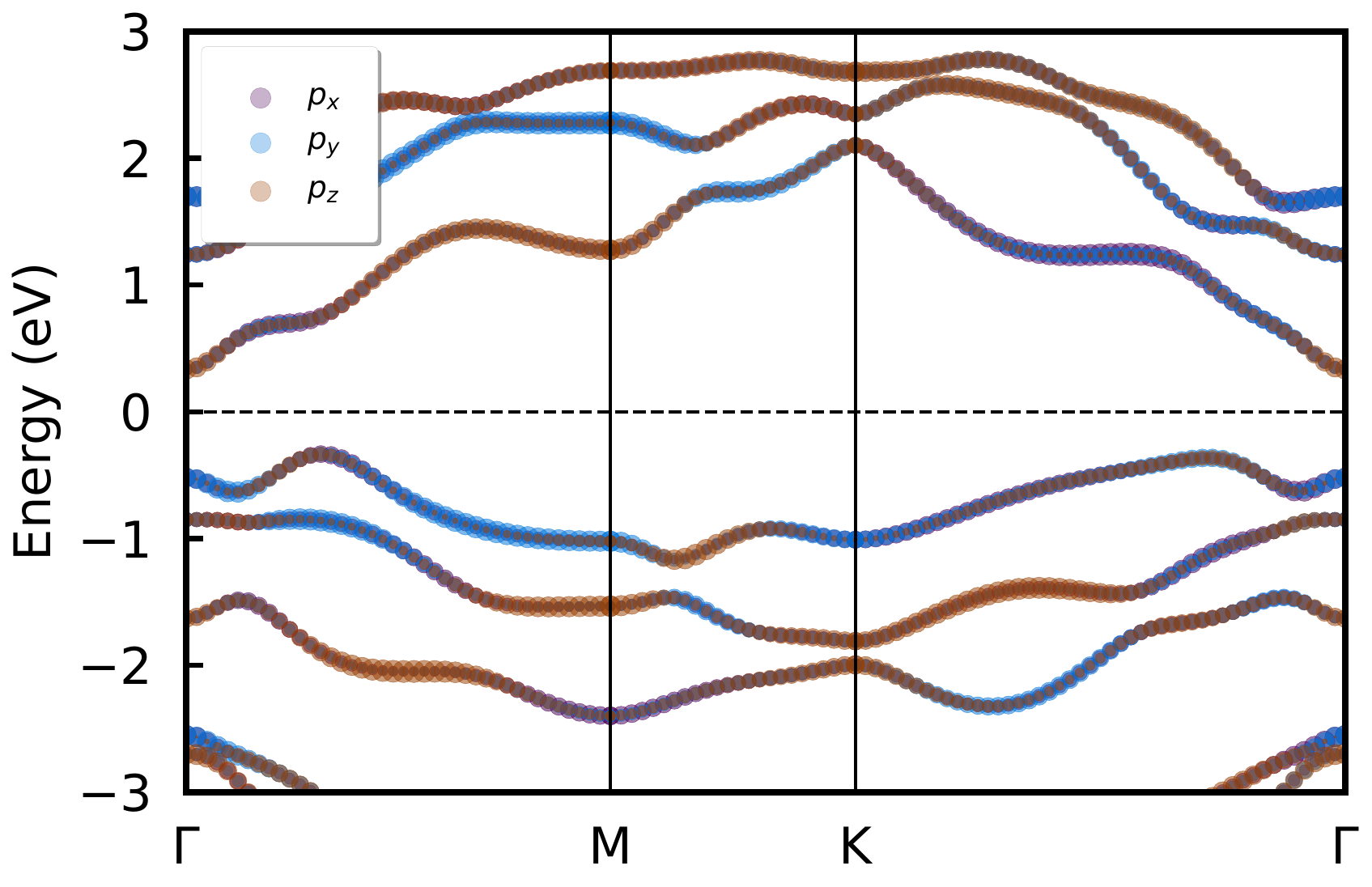}
\includegraphics[height= 4cm,clip=true,keepaspectratio]{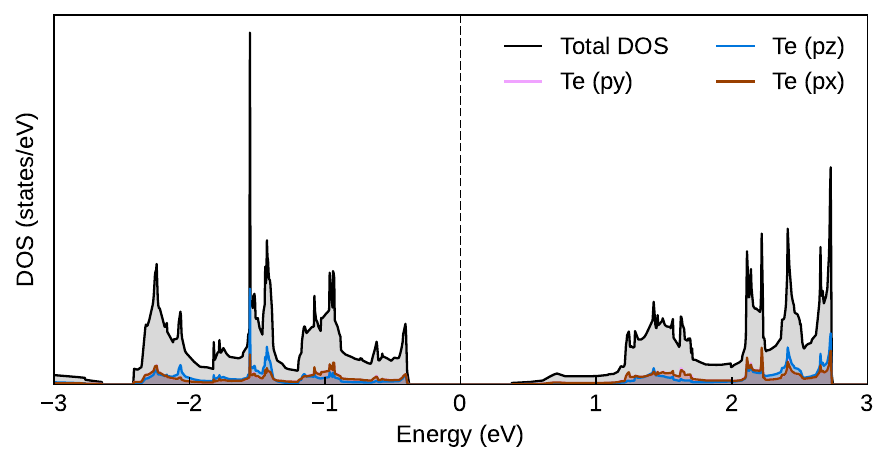}
\end{subfigure}
\begin{subfigure}[b]{0.9\columnwidth}
\subcaption{}
\includegraphics[height= 4cm,clip=true,keepaspectratio]{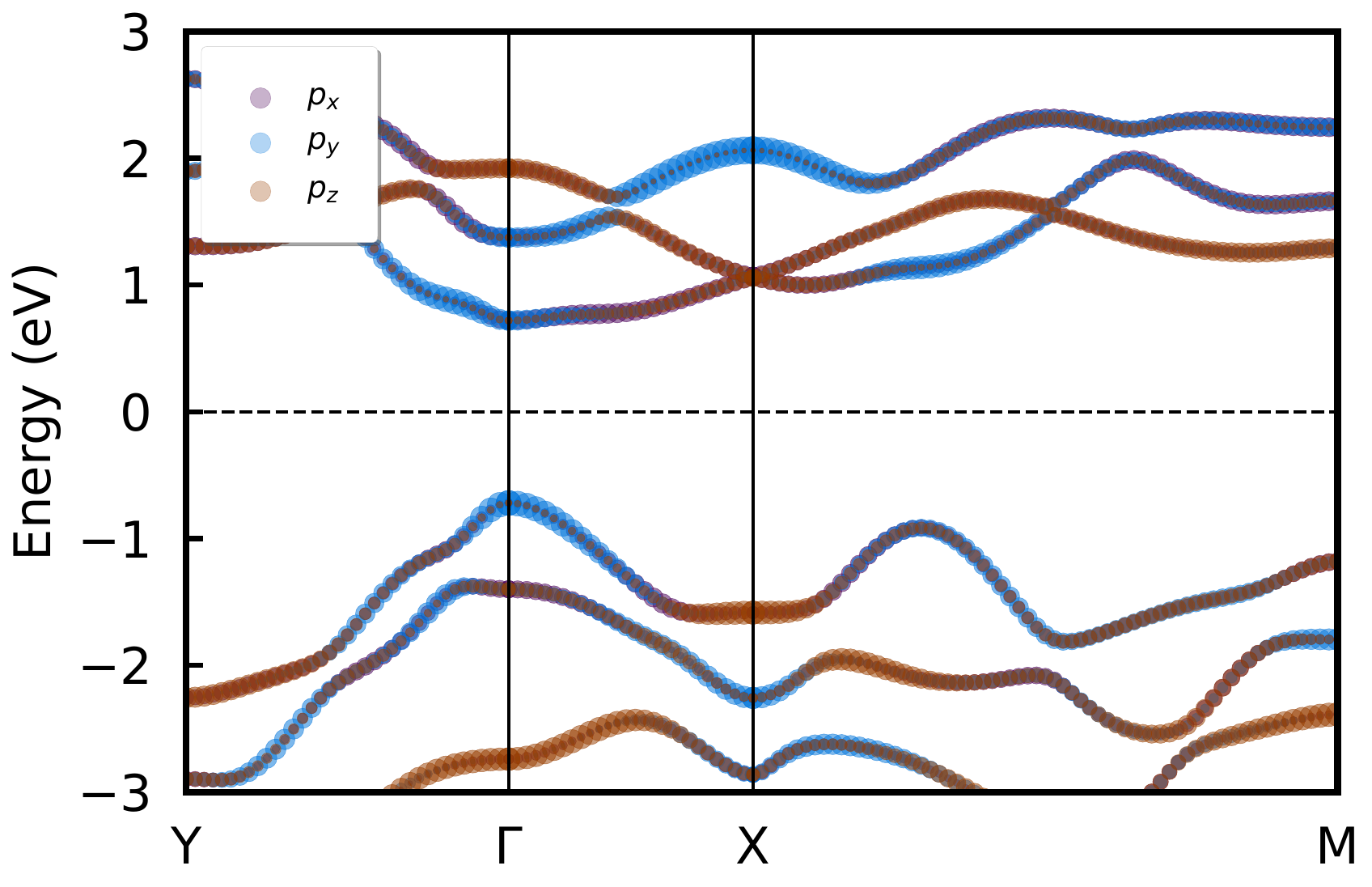}
\includegraphics[height= 4cm,clip=true,keepaspectratio]{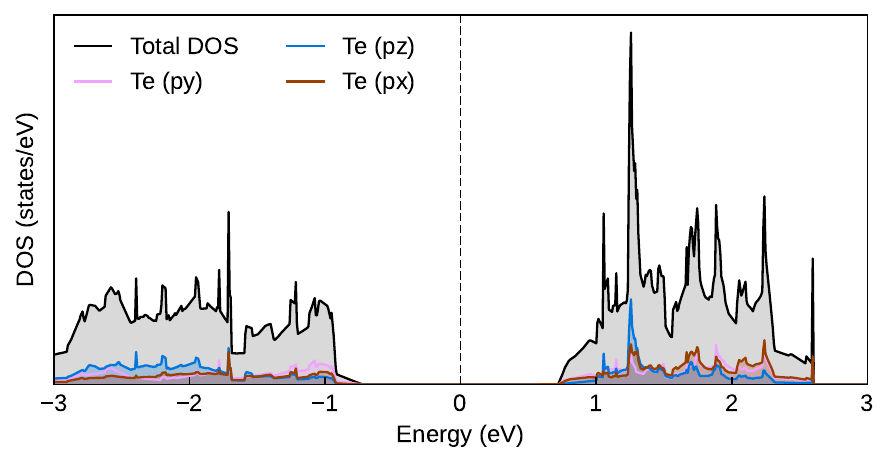}
\end{subfigure}
\begin{subfigure}[b]{0.9\columnwidth}
 \subcaption{}
\includegraphics[height= 4cm,clip=true,keepaspectratio]{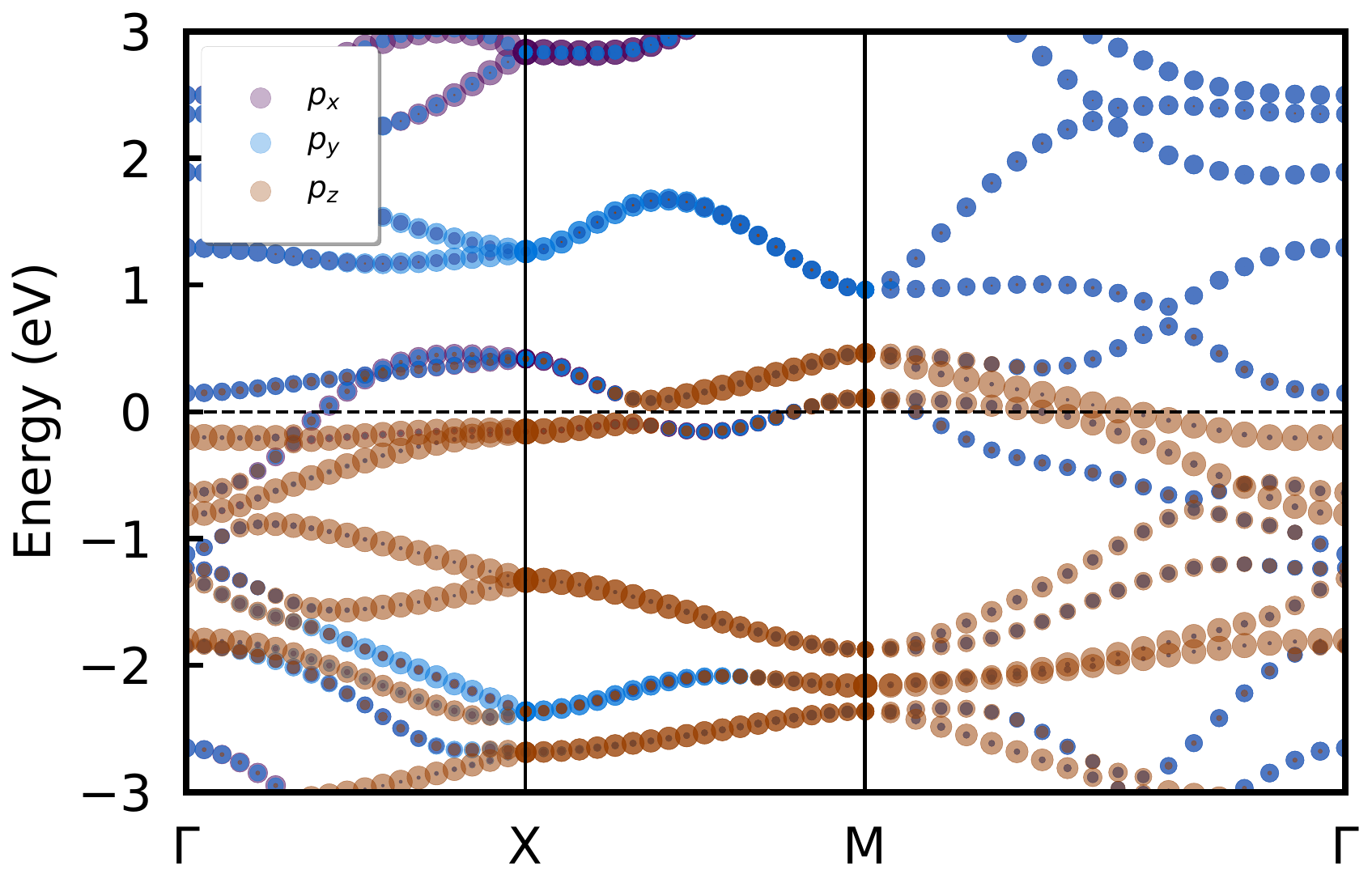}
\includegraphics[height= 4cm,clip=true,keepaspectratio]{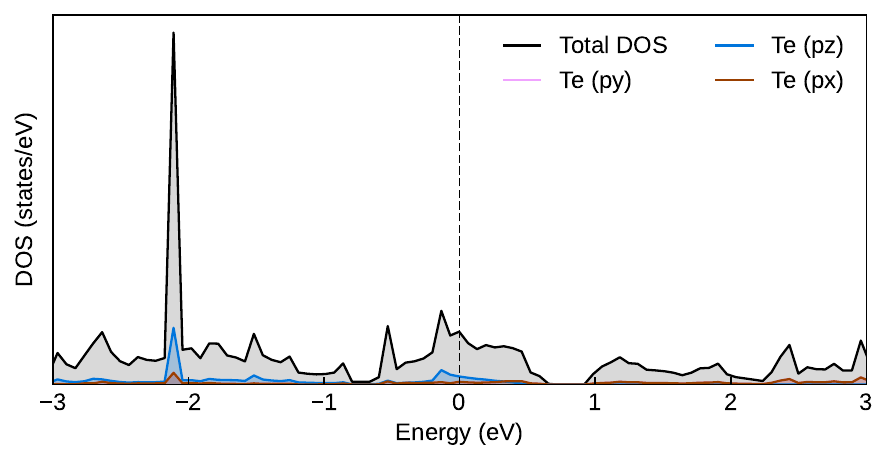}
\end{subfigure}
\begin{subfigure}[b]{0.9\columnwidth}
\subcaption{}
\includegraphics[height= 4cm,clip=true,keepaspectratio]{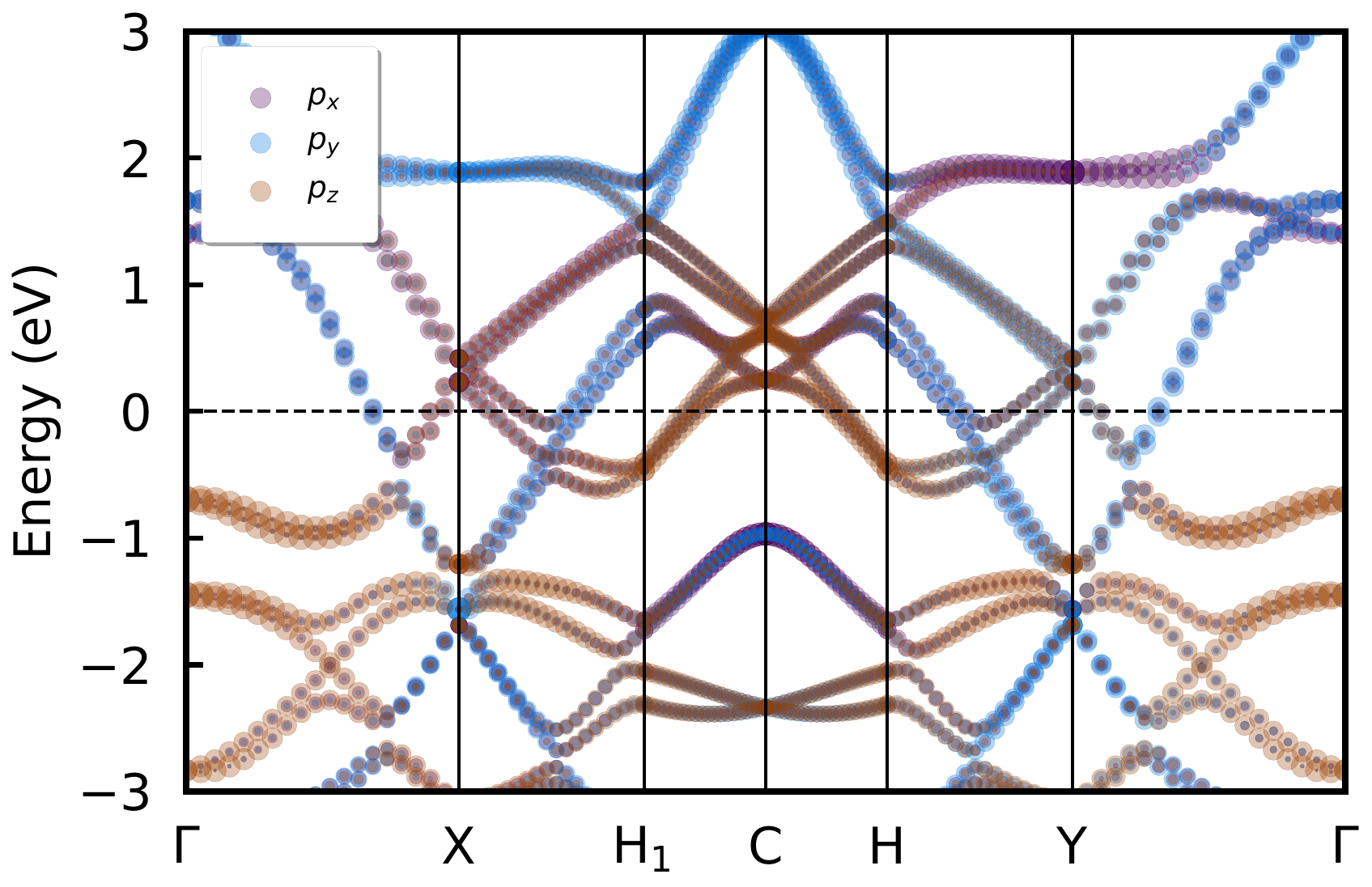}
\includegraphics[height= 4cm,clip=true,keepaspectratio]{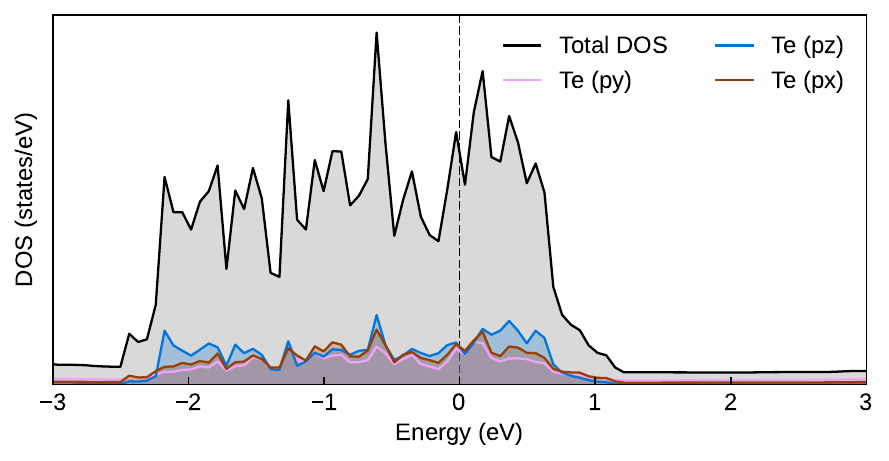}
\end{subfigure}
\begin{subfigure}[b]{0.9\columnwidth}
  \subcaption{}
\includegraphics[height=4cm,clip=true,keepaspectratio]{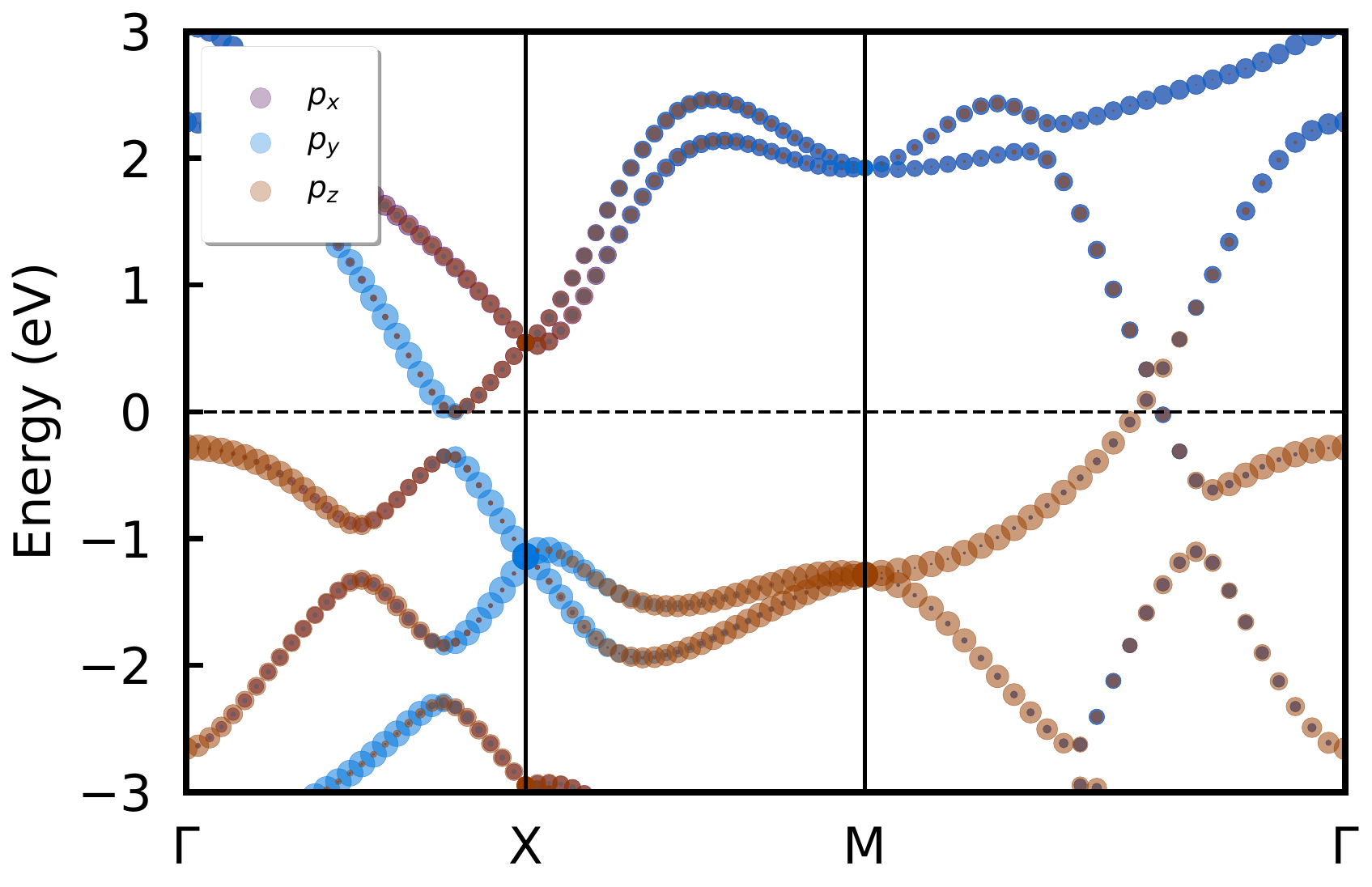}
\includegraphics[height= 4cm,clip=true,keepaspectratio]{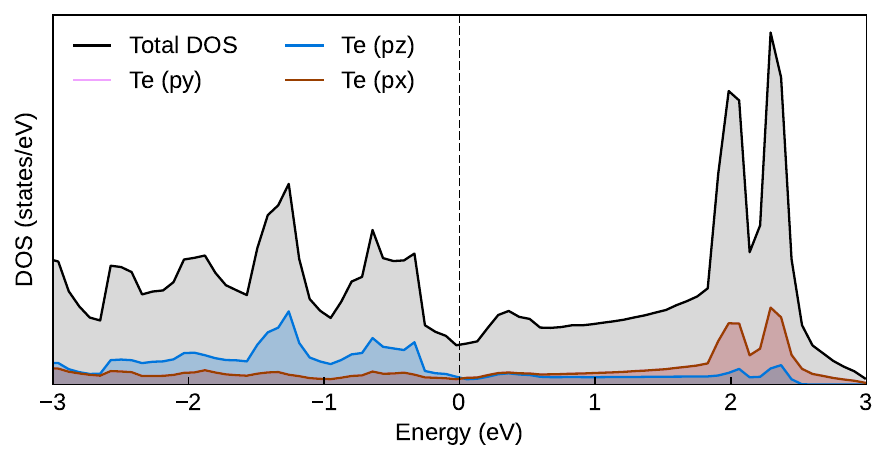}
\end{subfigure}
\caption{Orbital-projected electronic band structures and DOS for 2D tellurium phases calculated within MLWF-HSE06+SOC: (a) $\alpha$-Te, (b) $\beta$-Te, (c) buckled pentagonal,  (d) buckled kagome and (e) buckled square. The contributions from $p_x$, $p_y$, and $p_z$ orbitals are indicated by the thickness of the dots.}
\label{fig:all_fat_bands}
\end{figure}

For $\beta$-Te, the $p_y$ orbital provides the dominant contribution
to the valence band maximum (VBM) near the Fermi level. In contrast,
the conduction band minimum (CBM) exhibits a mixed contribution from
both $p_y$ and $p_x$ orbitals, with the latter being slightly more
prominent, as shown by the projected band structure in
Fig.~\ref{fig:all_fat_bands}. The total DOS indicates the presence of
highly localized states in both monolayers: $\beta$-Te
[Fig.~\ref{fig:all_fat_bands}(b)] shows a slightly lower degree of
delocalization compared to $\alpha$-Te
[Fig.~\ref{fig:all_fat_bands}(a)]. This behavior arises from quantum
confinement when the dimensionality is reduced from three to two. The
reduction in available electron degrees of freedom perpendicular to
the monolayer forces the particles to occupy discrete energy levels
within this confined region. The projected band structures of buckled pentagonal
[Fig.~\ref{fig:all_fat_bands}(c)], buckled kagome
[Fig.~\ref{fig:all_fat_bands}(d)], and buckled square
[Fig.~\ref{fig:all_fat_bands}(e)] lattices all exhibit metallic
character. Among them, the buckled kagome lattice shows the highest
DOS at the Fermi level, primarily originating from Te-$p$ states.

\begin{figure}[H]
	\centering
	\begin{subfigure}[b]{0.20\columnwidth}
     \subcaption{}
\includegraphics[width=\columnwidth]{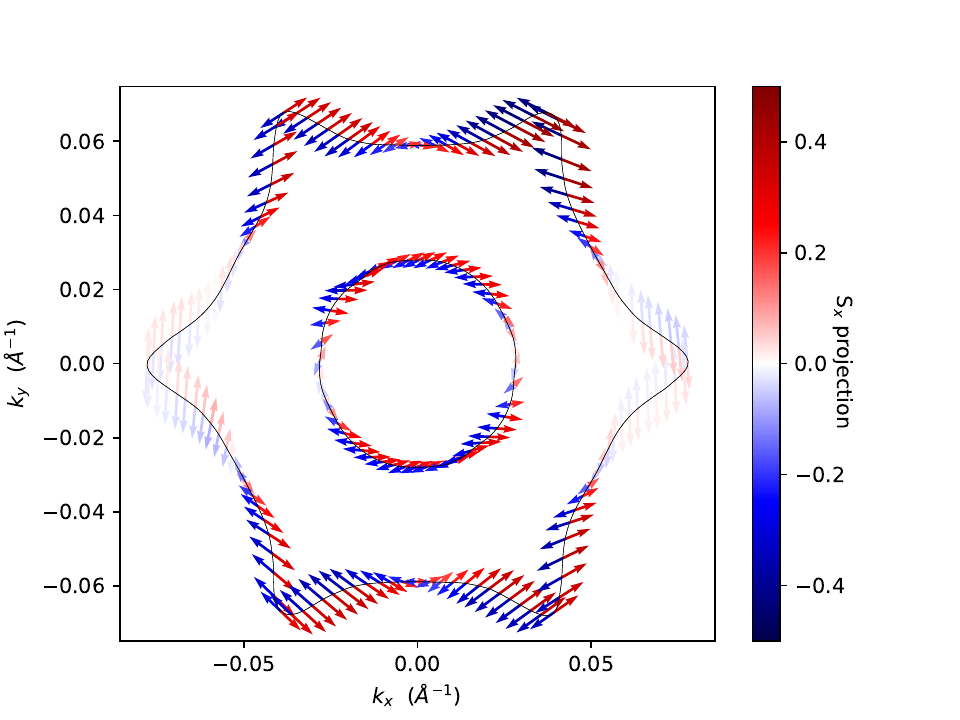} 
        \end{subfigure}
        \begin{subfigure}[b]{0.20\columnwidth}
        \subcaption{}
\includegraphics[width=\columnwidth]{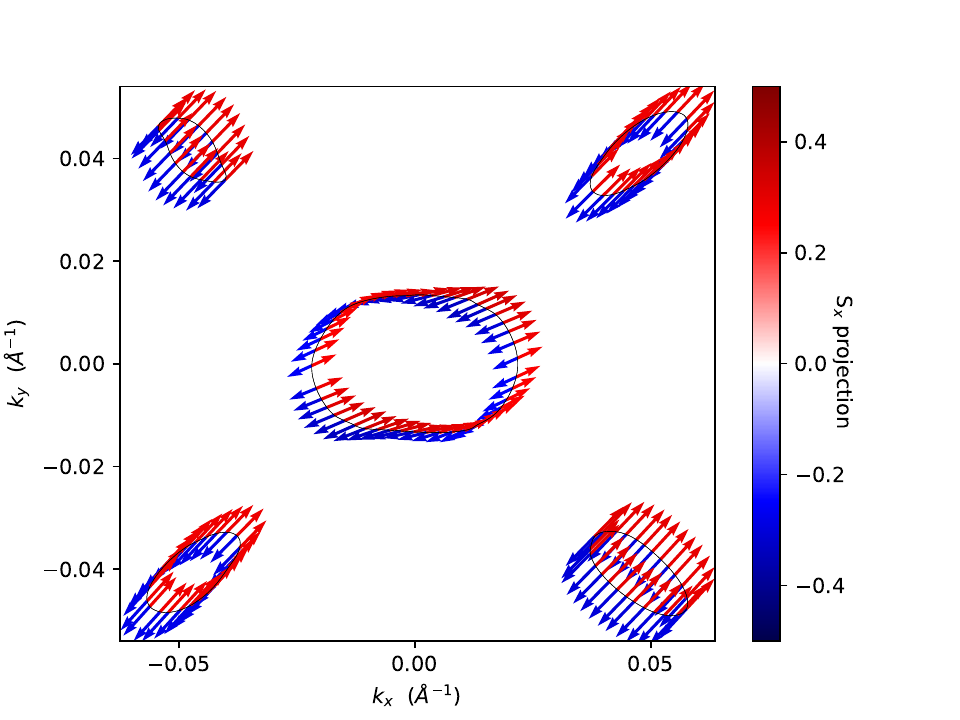} 
	\end{subfigure}
\begin{subfigure}[b]{0.20\columnwidth}
        \subcaption{}
        \includegraphics[width=\columnwidth,clip=true]{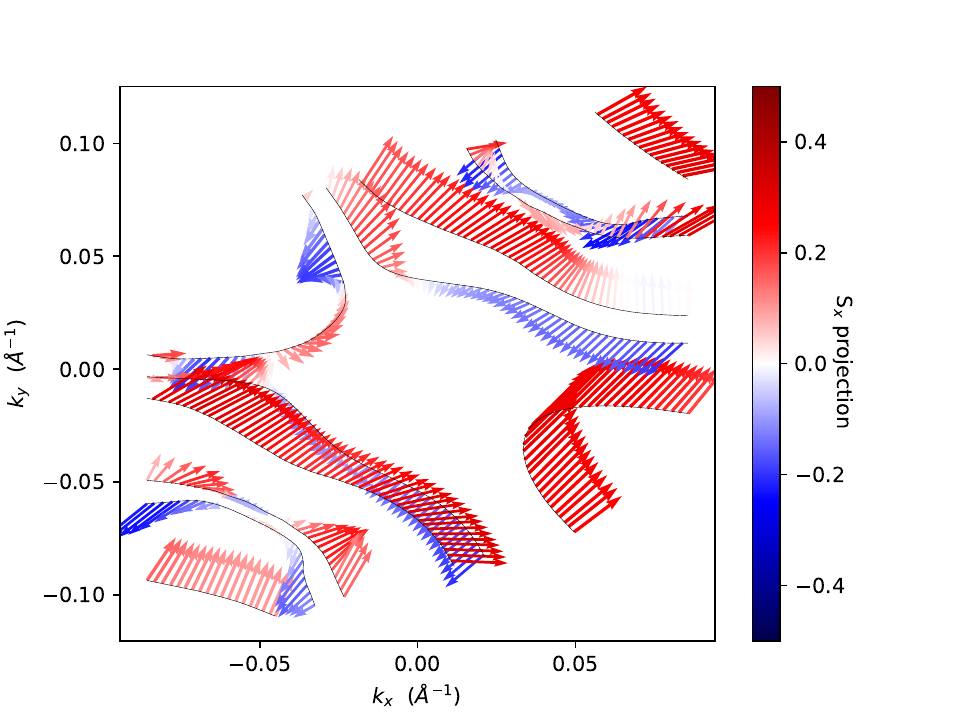}
\end{subfigure}
\begin{subfigure}[b]{0.20\columnwidth}
        \subcaption{}
        \includegraphics[width=\columnwidth,clip=true]{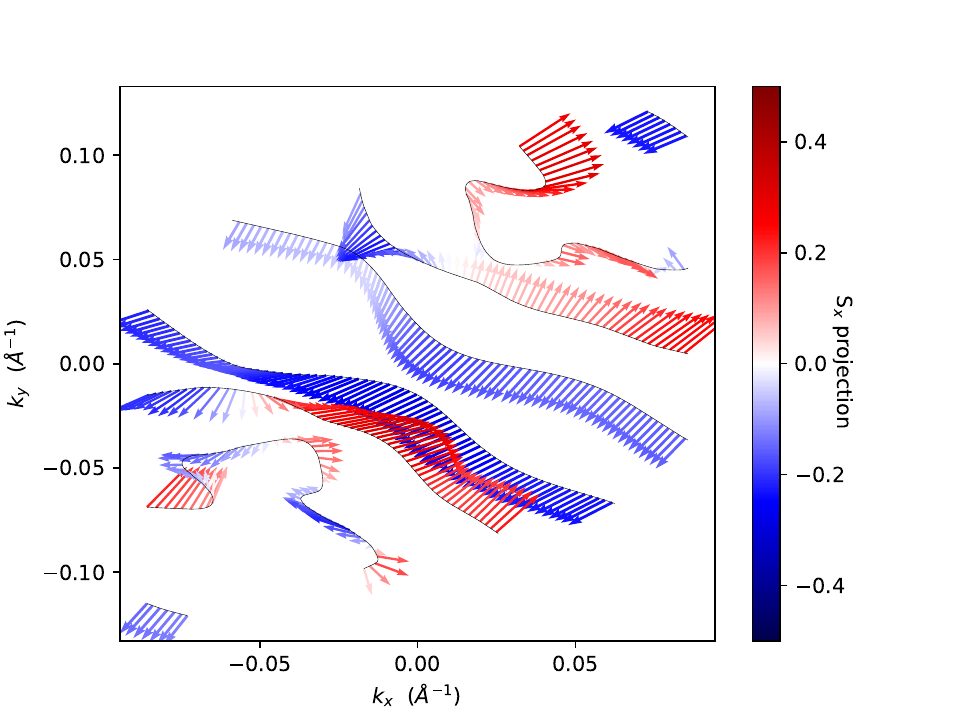}
\end{subfigure}\\
\begin{subfigure}[b]{0.20\columnwidth}
        \subcaption{}
        \includegraphics[width=\columnwidth,clip=true]{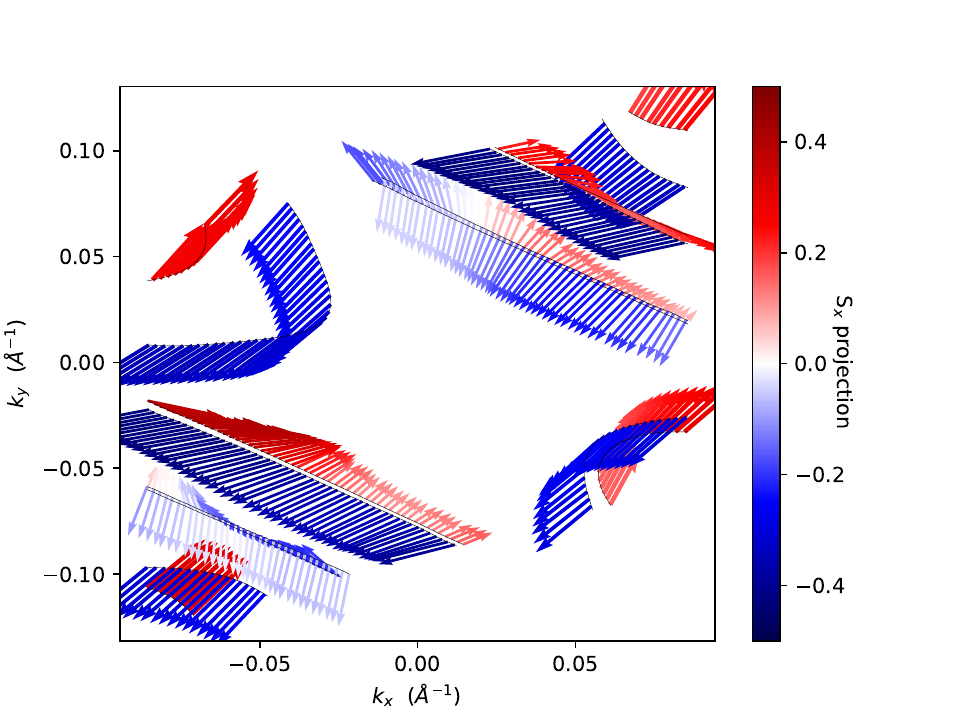}
\end{subfigure}
\begin{subfigure}[b]{0.20\columnwidth}
        \subcaption{}
        \includegraphics[width=\columnwidth,clip=true]{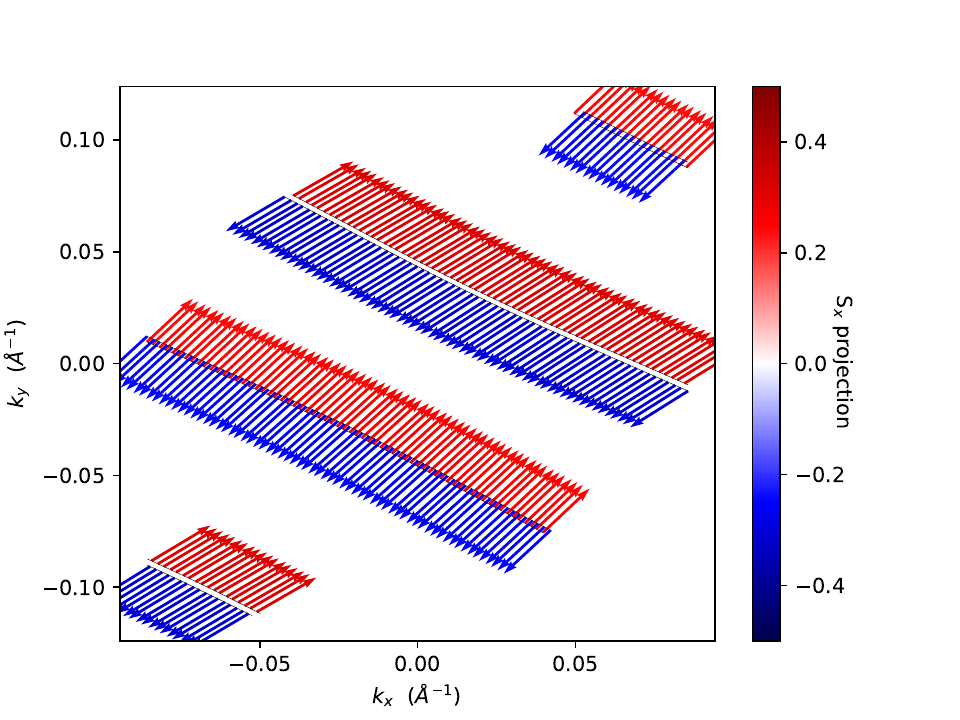}
\end{subfigure}
\begin{subfigure}[b]{0.20\columnwidth}
        \subcaption{}
        \includegraphics[width=\columnwidth,clip=true]{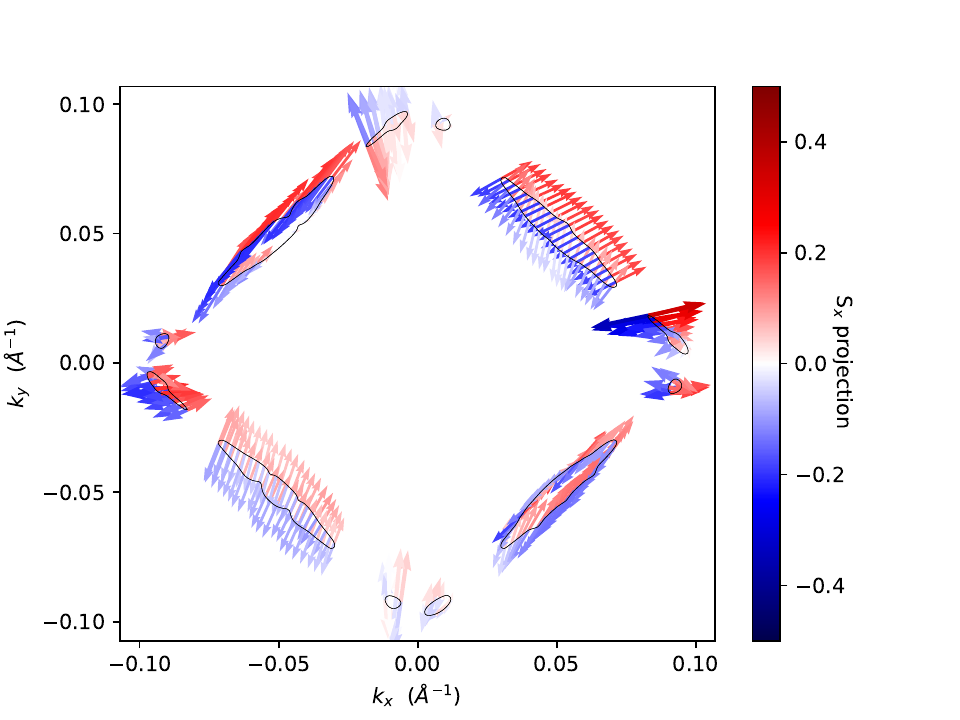}
\end{subfigure}
\begin{subfigure}[b]{0.20\columnwidth}
        \subcaption{}
        \includegraphics[width=\columnwidth,clip=true]{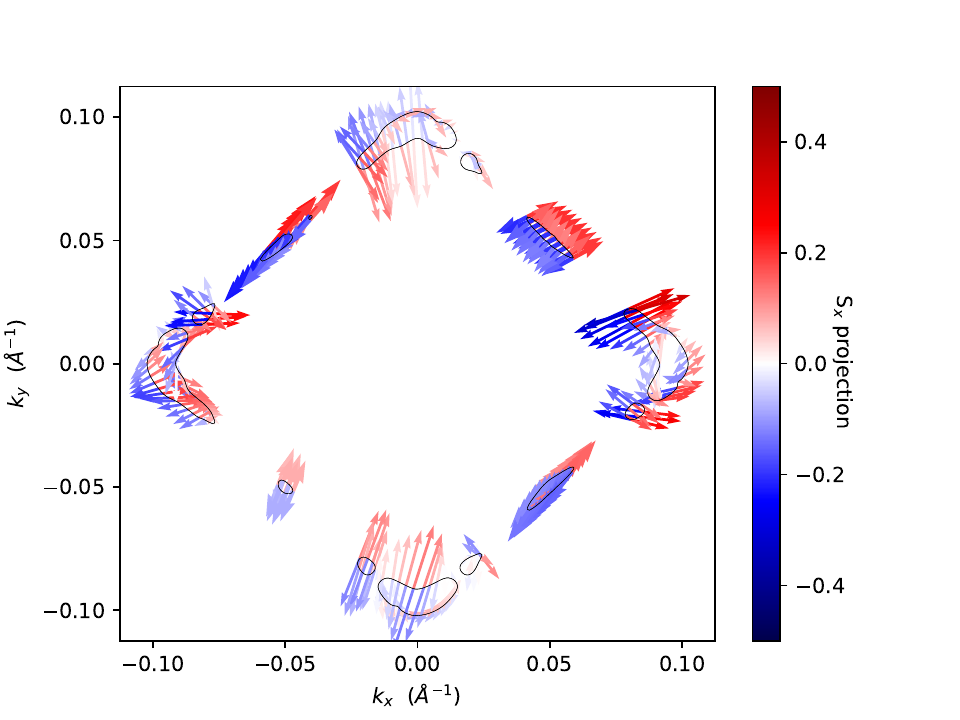}
\end{subfigure}
        \caption{\label{fig:2dte_spin} Spin textures of tellurium phases. $\langle S_x \rangle$ component for a) $\alpha$-Te, b) $\beta$-Te, (c,d) buckled pentagonal, (e,f)  buckled kagome, (g,h) buckled square phase. The color scale denotes the expectation values of the spin components. All calculation performed within the MLWF-SE06+SOC.}
\end{figure}

In order to confirm the presence or absence of non-trivial electronic
states, a spin-texture analysis will be carried out to identify
signatures of non-trivial spin behavior that may not be apparent from
the band structures alone. Fig.~\ref{fig:2dte_spin} reveals an intriguing behavior: despite the
distinct spin patterns—tangential in the inner band and radial in the
outer band for $\alpha$-Te (Fig.~\ref{fig:2dte_spin}(a)), and
predominantly radial for $\beta$-Te (Fig.~\ref{fig:2dte_spin}(b))—no
clear spin splitting is observed. This indicates that the bands remain
degenerate even in the presence of SOC. Although spin degeneracy
breaking is often associated with non-trivial electronic states,
particularly in Weyl semimetals and topological insulators, it is
important to note that topological properties may arise from
mechanisms other than spin splitting. To conclusively determine the
topological character, we compute both the Chern number and the
$\mathbb{Z}_2$ invariant. The results obtained using the
MLWF–HSE06+SOC approach confirm that both $\alpha$-Te and $\beta$-Te
monolayers are topologically trivial.

This behavior is attributed to the fact that 
spatial inversion symmetry—$\alpha$-Te belongs to the space group
$P\bar{3}m1$, which contains an inversion center. On the other hand, $\beta$-Te
belongs to the space group $P2/m$, which has both spatial and time-reversal symmetry inversion. This prevents the formation of Weyl
nodes in the band structure. Likewise, the lack of SOC-induced band
inversion precludes the characterization of these systems as
topological insulators.

Nevertheless, it is important to note that non-trivial topological
phases can arise under suitable external conditions. Topological phase
transitions may be induced by perturbations such as mechanical strain,
magnetic impurities, or doping. As discussed in
Ref.~\cite{Niu_Zhang_Graf_Lee_Wang_Wu_Low_Ye_2023}, for instance,
applying isotropic strain can drive tellurene from a trivial state
into a topological phase. For completeness, the $\langle S_x \rangle$ spin components of the:
pentagonal phase are shown in Figs.~\ref{fig:2dte_spin}(c,d), 
buckled kagome phase [Figs.~\ref{fig:2dte_spin}(e,f)], and  buckled square phase  [Figs.~\ref{fig:2dte_spin}(g,h)]. 

\begin{figure}[H]
\centering	
\begin{subfigure}[b]{0.4\columnwidth}
\subcaption{}
\includegraphics[height = 4cm,clip=true,keepaspectratio]{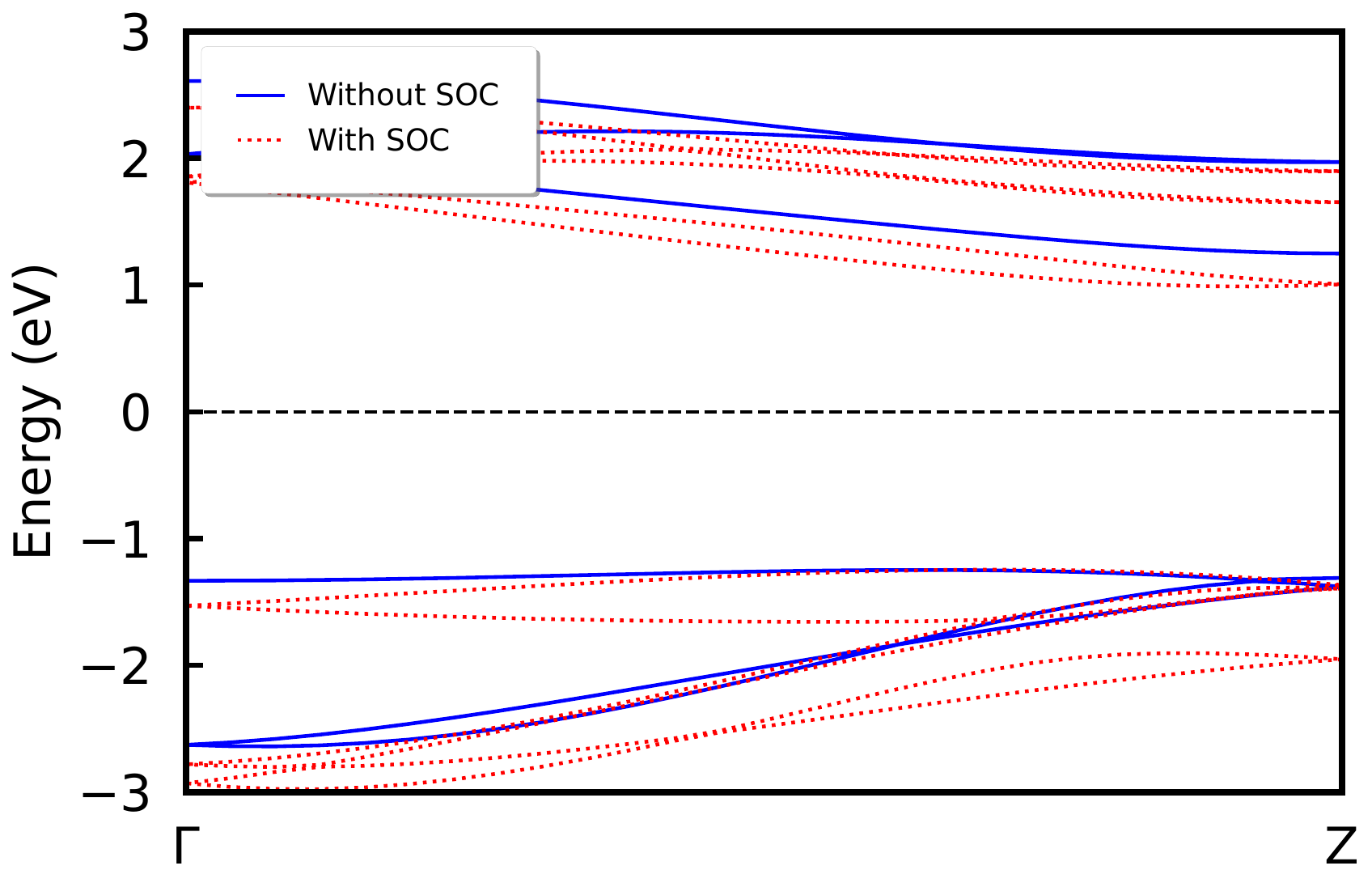}
\end{subfigure}
\begin{subfigure}[b]{0.4\columnwidth}
\subcaption{}
\includegraphics[height = 4cm,clip=true,keepaspectratio]{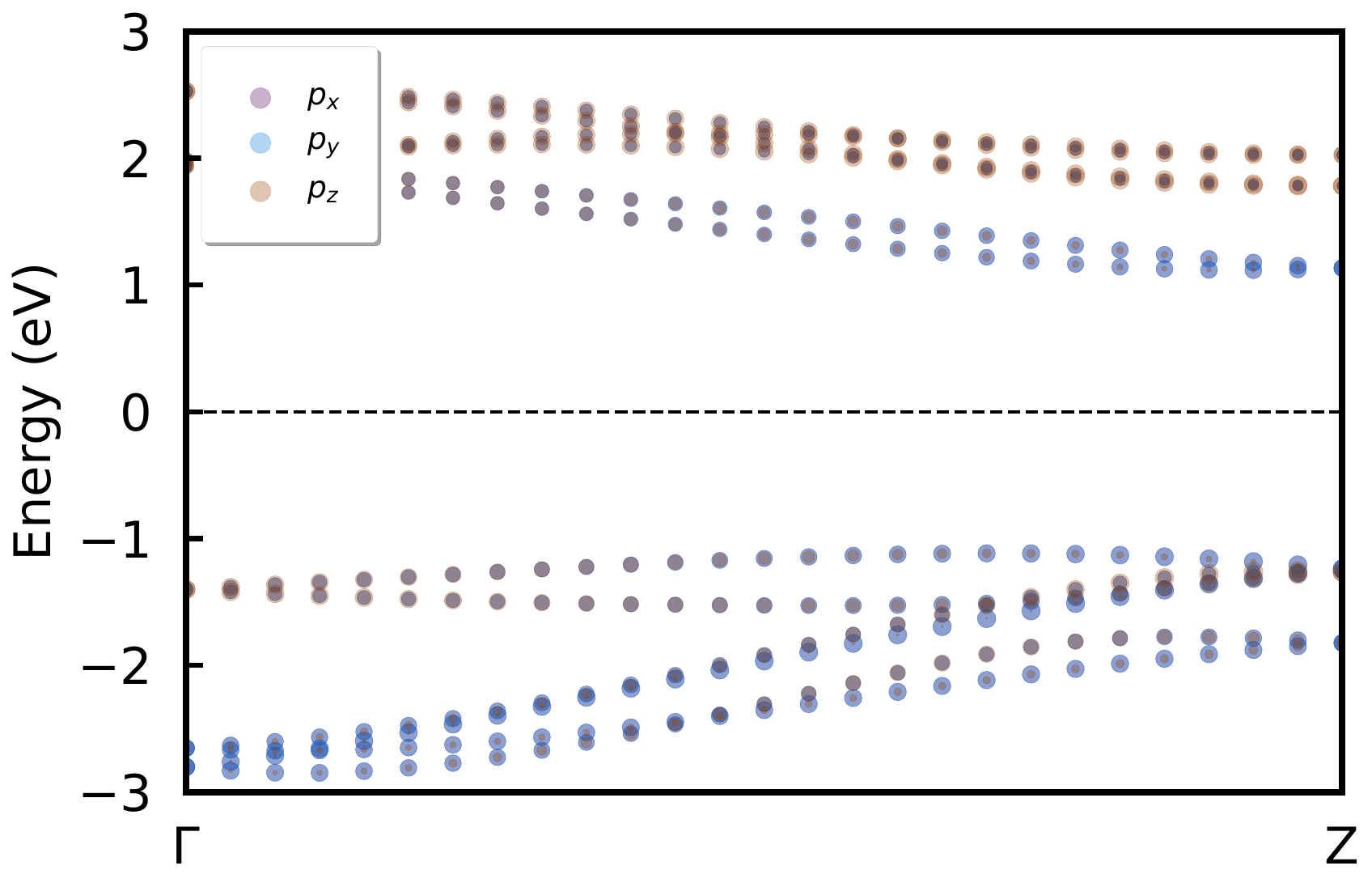}
\end{subfigure}
\begin{subfigure}[b]{0.40\columnwidth}
\subcaption{}
\includegraphics[width=\columnwidth,clip=true,keepaspectratio]{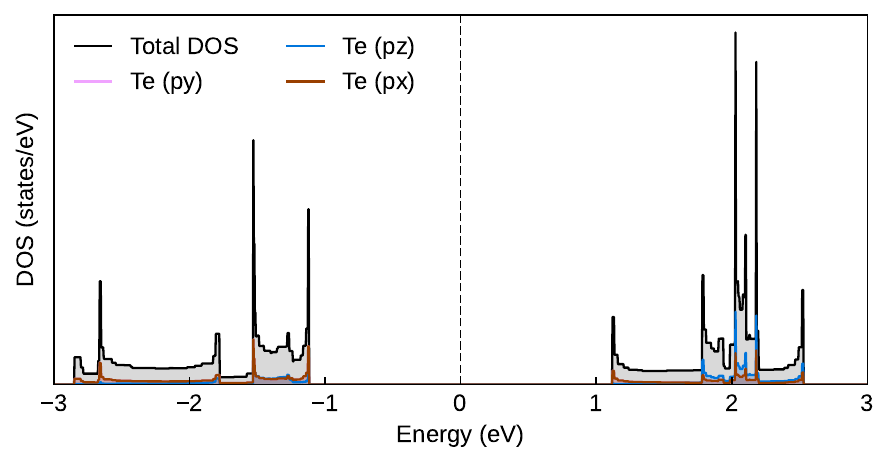}
\end{subfigure}
\begin{subfigure}[b]{0.40\columnwidth}
\subcaption{}
\includegraphics[width=\columnwidth,clip=true,keepaspectratio]{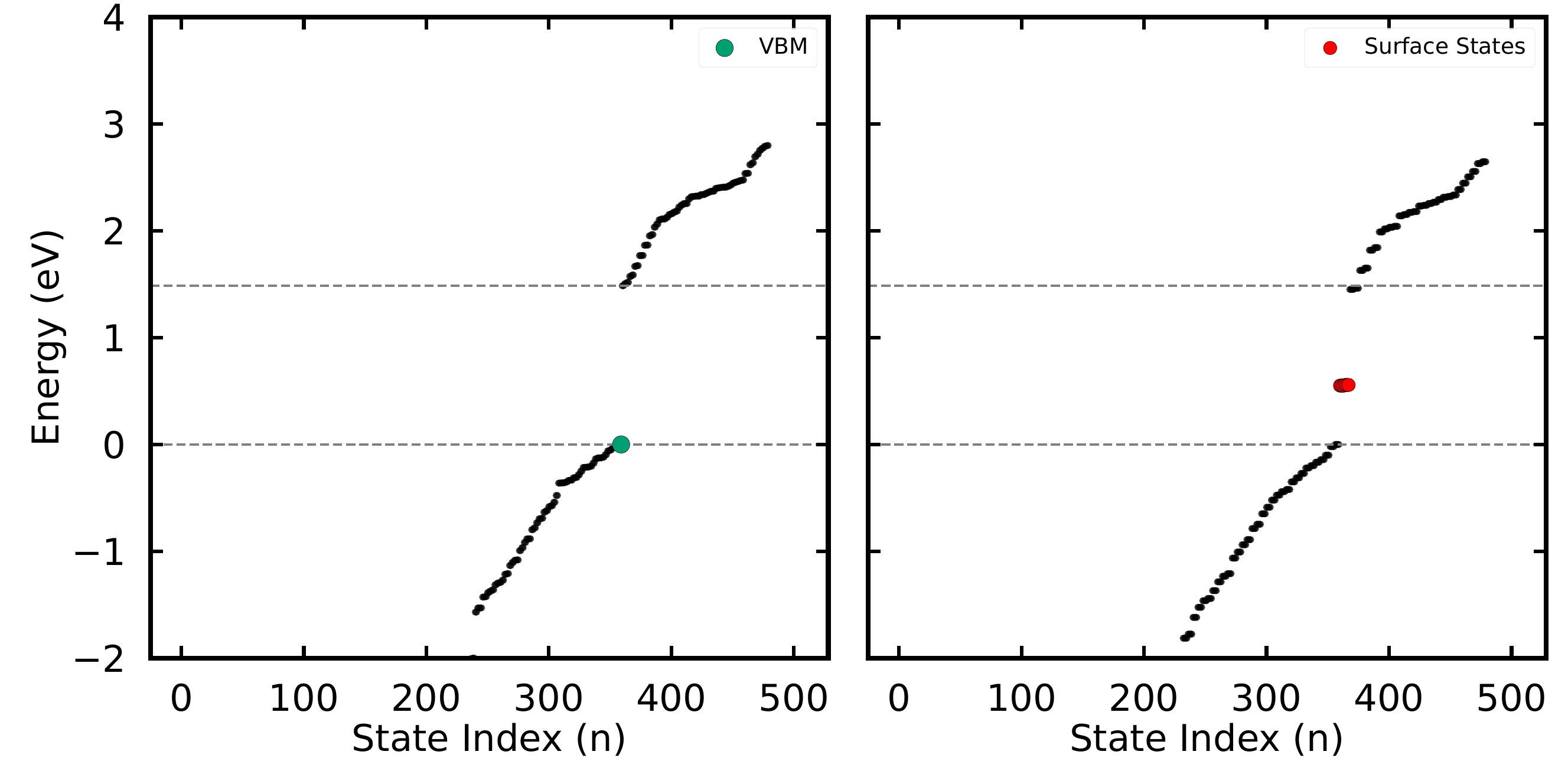}
\end{subfigure}\\
\begin{subfigure}[b]{0.40\columnwidth}
\subcaption{}
\includegraphics[width=\columnwidth,height = 3cm,clip=true,keepaspectratio]{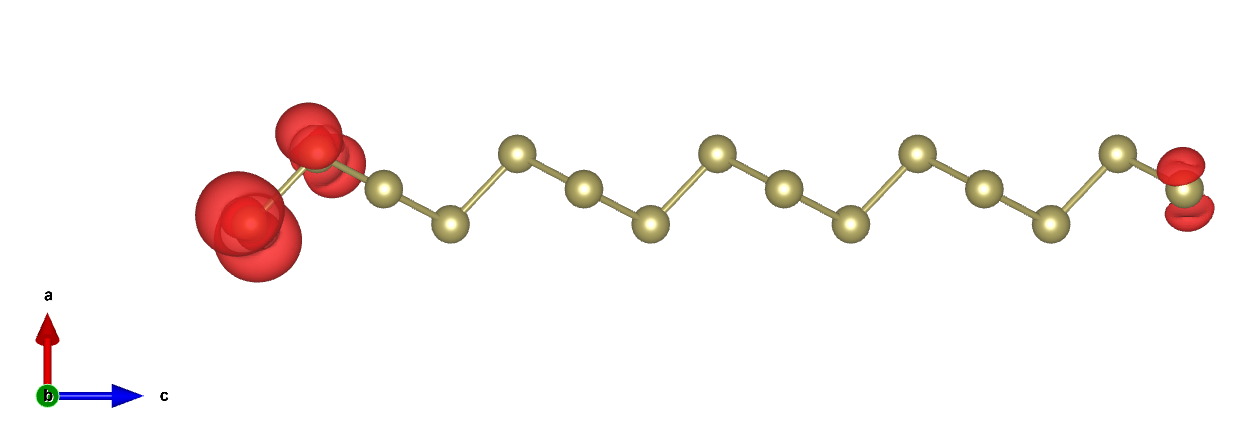}
\end{subfigure}
\caption{\label{fig:teh_band_surface_states} (a) Electronic band structure of the nanowire calculated without (gray) and with (colored) spin–orbit coupling (SOC), highlighting the pronounced SOC-induced modification of the confined spectrum.
(b) Orbital-resolved representation of the SOC-included band structure, indicating that the states near the Fermi level are dominated by Te $p$-orbitals. (c) Total electronic density of states (DOS) of the nanowire, confirming the presence of a finite energy gap in the quantum-confined geometry.
  (e) Real-space charge-density isosurfaces of representative d) in-gap states, revealing strong localization at the edge atoms of the nanowire.}
\end{figure}

Fig.~\ref{fig:teh_band_surface_states}(a) presents the band structure of the Te
helicoidal nanowire (Te-h), computed using the MLWF-HSE06 functional both
with and without SOC. The inclusion of SOC consistently narrows the
band gap, reducing it from 2.49 eV to 2.23 eV. These values align well
with previous theoretical reports,\cite{Kramer}. In both cases, the
band structure displays predominantly flat dispersion. This
combination of a finite band gap and quasi-flat bands makes Te-h an
appealing material for photonic applications,\cite{fast}. Fig.~\ref{fig:teh_band_surface_states}(b) shows the inclusion of SOC
lifts band degeneracies in a manner similar to that observed in Te-I
along the $\Gamma$–A direction. Along the high-symmetry path of Te-h,
four band crossings are identified: two occurring at the same
$K$-point—P1 in the valence band at $-2.7,\text{eV}$ and P2 in the
conduction band near $2,\text{eV}$—and two additional crossings near
the $\Gamma$-point, P3 at $-2.7,\text{eV}$ and P4 around
$2,\text{eV}$. Fig.~\ref{fig:teh_band_surface_states}(c) presents the PDOS for
Te-h, showing that the $p_x$, $p_y$, and $p_z$ orbitals contribute
comparably to both the VBM and CBM. However, the CBM exhibits a more
pronounced contribution from the $p_x$ orbital. Due to quantum
confinement, Te-h displays highly localized electronic states, which
is reflected in the discrete features of the band structure and in the
pronounced peaks observed in the DOS.

Fig.~\ref{fig:teh_band_surface_states}(d) illustrates the
  edge-derived states that lie within the nanowire band gap. The
  presence of highly localized states originating from the edge atoms
  is further highlighted in
  Fig.~\ref{fig:teh_band_surface_states}(e). The helical tellurium
  nanowire preserves the broken inversion symmetry and structural
  chirality of bulk trigonal Te, while introducing strong quantum
  confinement. In the presence of spin–orbit coupling, this reduced
  dimensionality leads to the emergence of in-gap electronic states
  that are strongly localized at the terminal atoms of the nanowire,
  as evidenced by the projected charge-density analysis. Although a
  strict $\mathbb{Z}_2$ topological invariant is not formally defined
  for isolated one-dimensional systems, these edge-localized states
  can be naturally interpreted as boundary manifestations inherited
  from the higher-dimensional Weyl semimetal parent phase. The
  coexistence of time-reversal symmetry, strong SOC, and chirality
  stabilizes these boundary modes, distinguishing them from trivial
  dangling-bond states. The topological properties of the tellurium
  nanowire were analyzed within the modern theory of
  polarization. Since the system is one-dimensional and preserves
  time-reversal symmetry, no $\mathbb{Z}_2$ topological invariant associated with
  quantum spin Hall phases can be defined. Instead, the relevant bulk
  quantity is the Berry (Zak) phase accumulated along the
  one-dimensional Brillouin zone. Because the nanowire lacks inversion
  and chiral symmetries, the Zak phase is not symmetry-quantized. Our
  calculations yield a Zak phase essentially equal to zero,
  corresponding to a vanishing bulk polarization (SI, Fig.~\ref{fig:berry_zak_nw}). This demonstrates
  that the periodic nanowire is topologically trivial in the normal
  (non-superconducting) state. Consistently, any end-localized states
  observed in finite nanowires originate from termination effects and
  are not protected by bulk topology.

The effective masses [Table~\ref{tab:S2}] were extracted from the band
structure near the inflection points at the VBM and CBM. Te-h
effective masses are suggestive of potentially high mobility\cite{gaas,si1,geh}. Moreover, the modest reduction in electron and hole
mobility in Te-h can help mitigate current leakage in nanoelectronic
devices~\cite{tetete}. In units of the free-electron mass, the
electron effective mass of Te-h is 0.484, while the hole effective
mass is 0.817 (see Table~\ref{fig:S2}).

In Table~\ref{tab:Z2}, the topological invariant $\mathbb{Z}_2$
for the 2D phases of tellurium is presented. The invariants are
calculated following the method described in Ref.,\cite{WU2017}. The
buckled kagome and buckled square phases exhibit non-trivial topology,
whereas $\alpha$-Te, $\beta$-Te, and the buckled pentagonal phase are
topologically trivial. These topological classifications are
consistent with the spin textures shown in Fig.~\ref{fig:2dte_spin}.

Using experimentally reported lattice parameters\cite{acs.nanolett.4c02171}, the planar hexagonal tellurene phase is found to exhibit semimetallic behavior and a nontrivial $\mathbb{Z}_2 = 1$ index when SOC is included, consistent with previous experimental observations. However, our phonon calculations indicate that the free-standing planar hexagonal lattice is dynamically unstable. We therefore investigated stabilization mechanisms through strain engineering and surface functionalization. Under 5\% isotropic in-plane strain, as well as under one-side hydrogen passivation, the system undergoes a transition to a gapped electronic state while preserving a nontrivial $\mathbb{Z}_2 = 1$ invariant. In these stabilized configurations, the SOC-induced gap and the winding of the Wilson loop unambiguously identify a quantum spin Hall phase.

Similarly, one-side hydrogen passivation of the hexagonal phase yields a  $\mathbb{Z}_2 = 1$ phase
[Fig.~\ref{fig:S6}(k)], depending on the Fermi-level position it can appear semimetallic, but the SOC-opened gap indicate QSH-type topology. These findings demonstrate that the topological features
remain robust under both strain and surface
functionalization. Therefore, controlling strain or chemical
termination provides viable strategies to engineer and stabilize
topological phases in two-dimensional tellurium.

We now turn our attention to the calculation of the tellurium
effective masses, obtained from the slopes of the band structure near
the VBM and CBM. The obtained values are shown in Table~\ref{tab:S2}. Since the effective mass is inversely related to
carrier mobility, the relatively low values
found for Te-I suggest anisotropic electron and hole mobilities. For
Te-I, the electron and hole effective masses are 0.614 and 0.335,
respectively. 

For the 2D phases, the effective masses for electrons (holes) in
$\alpha$-Te are 0.108 (0.135). In $\beta$-Te, the electron effective
masses are 1.009 (along X) and 0.203 (along Y), while the hole
effective masses are 0.368 (X) and 0.127 (Y), indicating strong
transport anisotropy. Our results suggest that both $\alpha$-Te and
$\beta$-Te could have higher electron and hole mobilities than
structurally or symmetrically similar materials such as
2H-MoS$_2$~\cite{mos1,mos2} and phosphorene~\cite{phos1,phos2}. While
$\alpha$-Te exhibits higher mobilities for both charge carriers
overall, $\beta$-Te shows pronounced anisotropy due to its geometry,
with significantly reduced mobility along the armchair direction
($\Gamma$–X) compared to the zigzag direction ($\Gamma$–Y).

For the other phases, the buckled pentagonal structure yields
effective masses of 0.220 (electron) and 0.172 (hole). The buckled
square lattice is anisotropic, with electron masses of 0.100
(CBM–$\Gamma$) and 0.148 (CBM–X), and hole masses of 0.459
(VBM–$\Gamma$) and 0.239 (VBM–M). Finally, the hydrogen-passivated
hexagonal phase is slightly asymmetric, with electron effective masses
of 2.400 ($\Gamma$–M) and 2.310 ($\Gamma$–K), and a hole effective
mass of 1.184 (VBM–$\Gamma$).

\begin{figure}
\begin{subfigure}[b]{0.4\columnwidth}
\subcaption{}
\includegraphics[width=\columnwidth,clip=true,keepaspectratio]{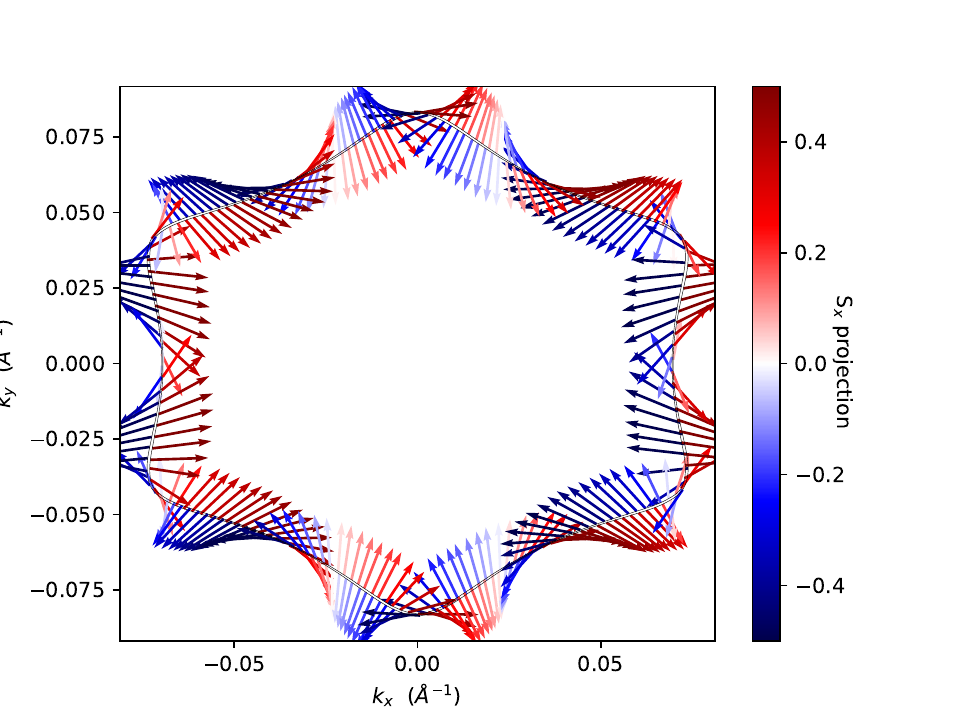}
\end{subfigure}
\begin{subfigure}[b]{0.4\columnwidth}
\subcaption{}
\includegraphics[width=\columnwidth,clip=true,keepaspectratio]{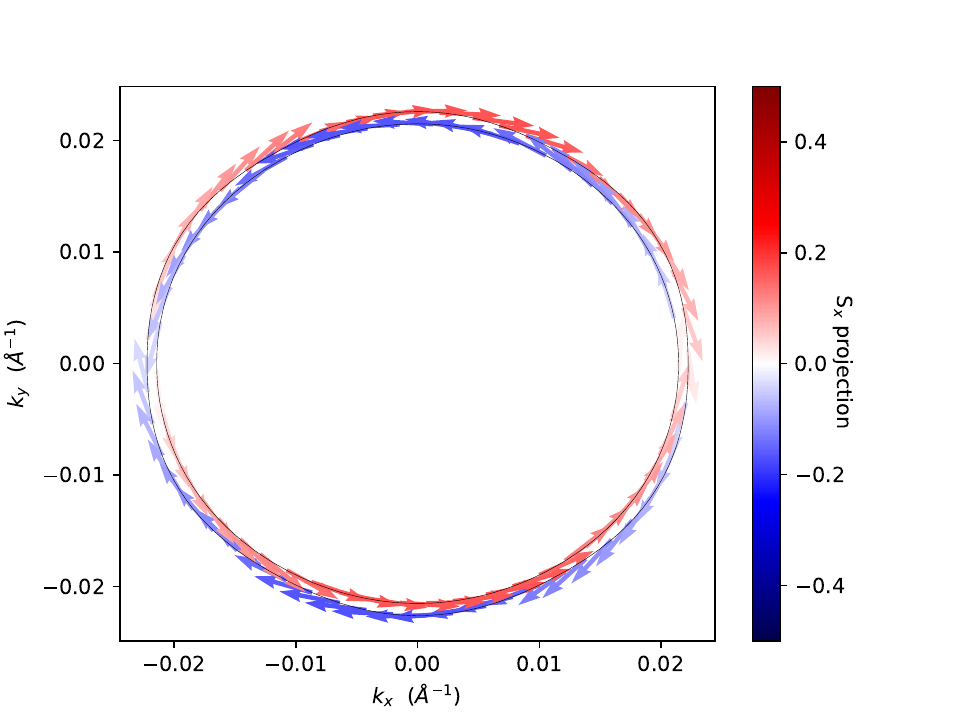}
\end{subfigure}
\caption{\label{fig:spin_hex_passiv} a) and b) spin texture of hydrogen passivated hexagonal  tellurium at ${\rm E_F}$-0.3 eV calculated within MLWF-HSE06+SOC.}
\end{figure}

Fig.\,\ref{fig:spin_hex_passiv} shows the spin texture of hydrogen
passivated hexagonal tellurium at an energy ${\rm E_F}$ - 0.3\,eV. The
one-side H-passivated hexagonal tellurene exhibits clear in-plane
spin–momentum locking and Rashba-type spin splitting. The
constant-energy spin textures show nearly circular spin-split contours
as well as pronounced hexagonal warping, indicating strong
inversion asymmetry induced SOC and anisotropic spin polarization
around $\Gamma$.

\begin{figure}
\begin{subfigure}[b]{0.4\columnwidth}
\subcaption{}
\includegraphics[width=\columnwidth,height=4cm,clip=true,keepaspectratio]{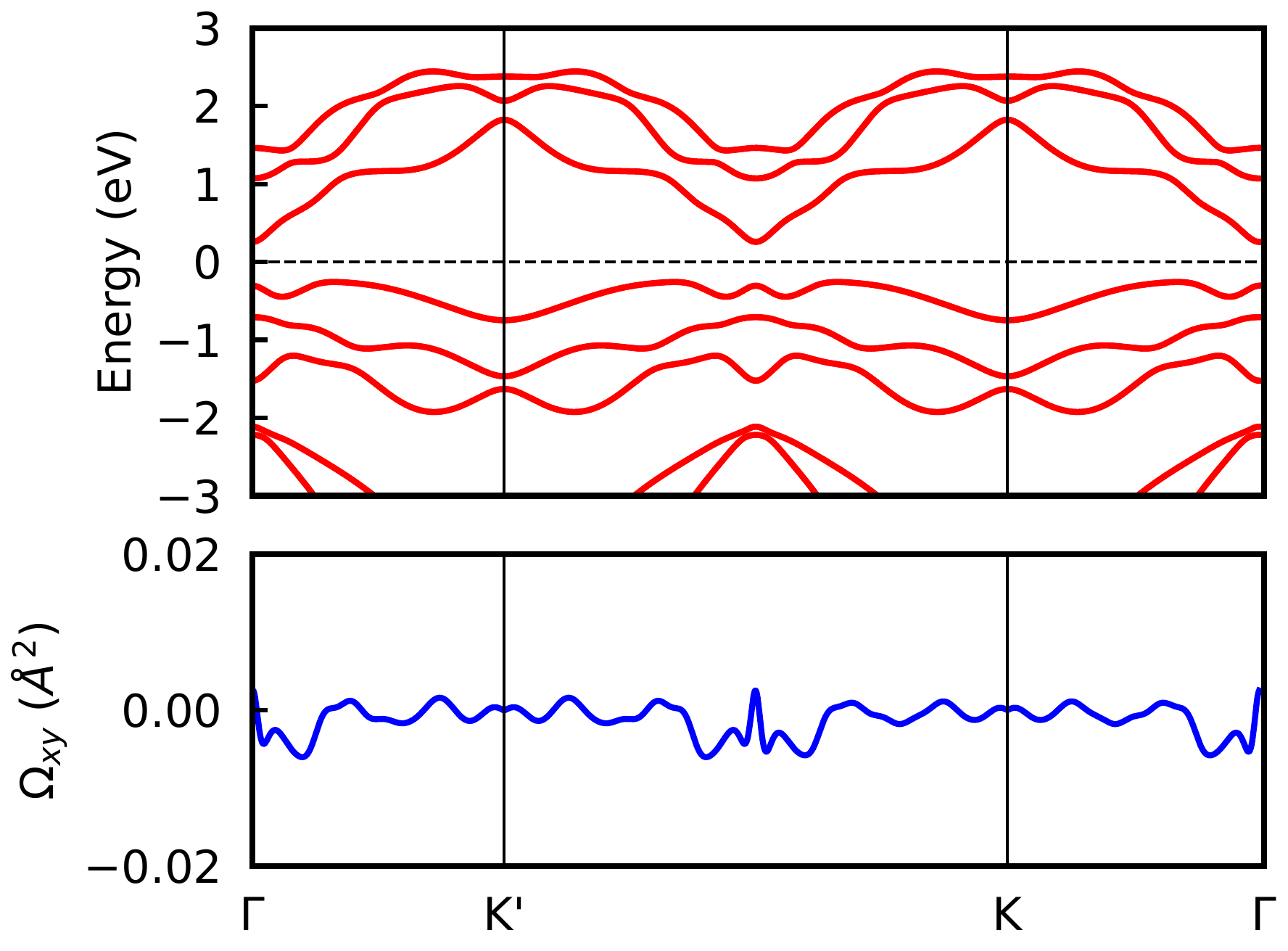}
\end{subfigure}
\begin{subfigure}[b]{0.4\columnwidth}
\subcaption{}
\includegraphics[width=\columnwidth,height=4cm,clip=true,keepaspectratio]{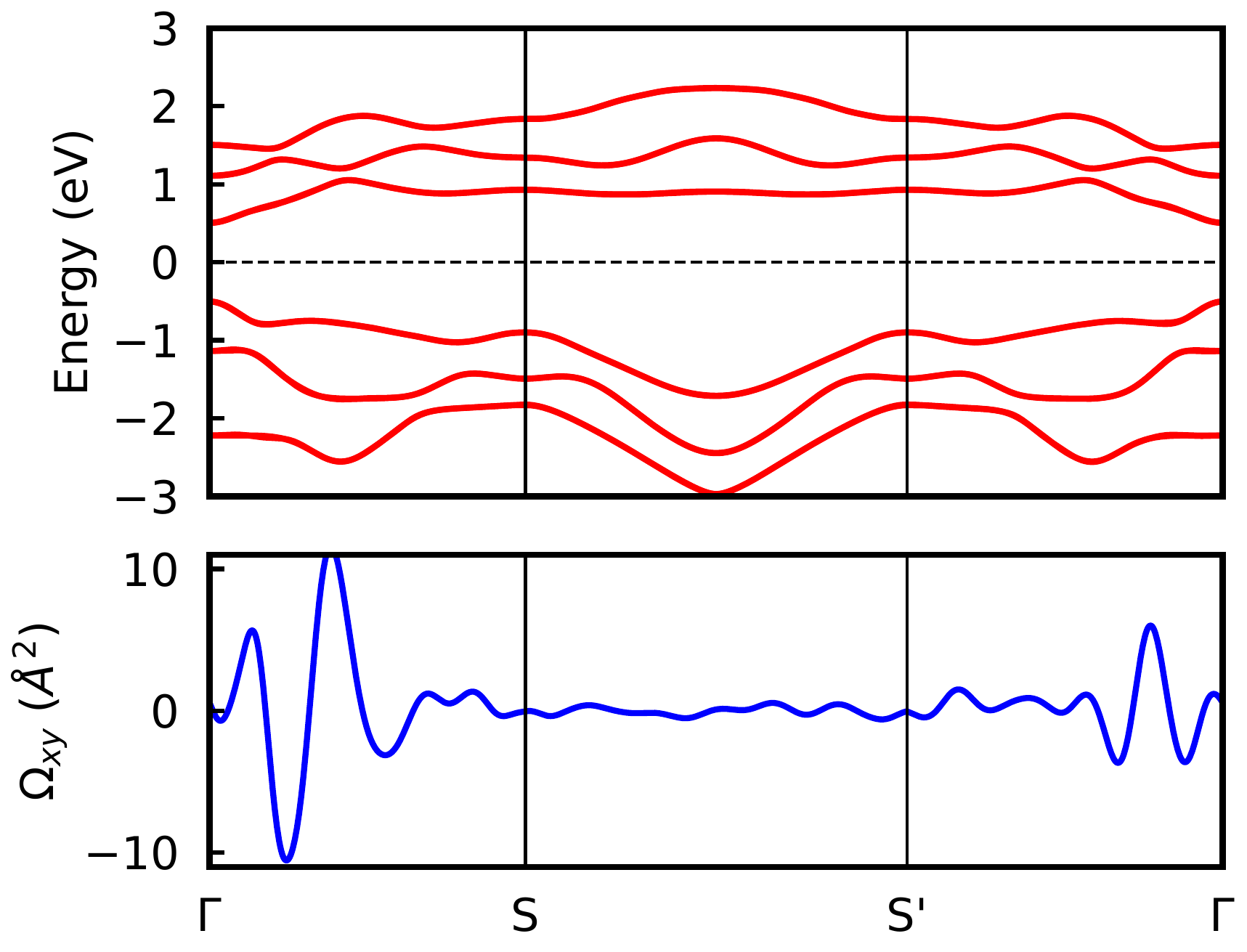}
\end{subfigure}
\begin{subfigure}[b]{0.4\columnwidth}
\subcaption{}
\includegraphics[width=\columnwidth,height=4cm,clip=true,keepaspectratio]{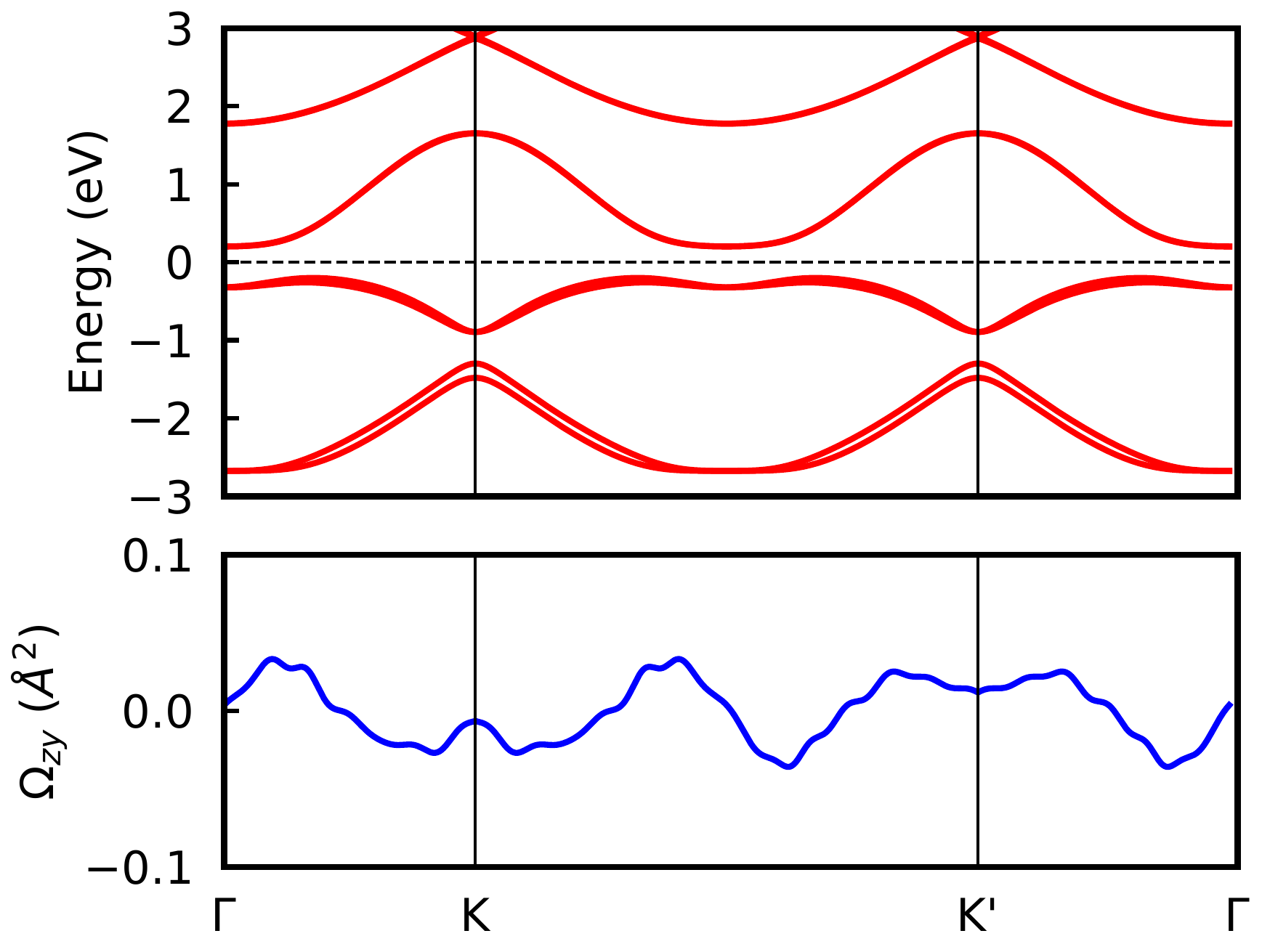}
\end{subfigure}
\begin{subfigure}[b]{0.4\columnwidth}
\subcaption{}
\includegraphics[width=\columnwidth,height=4cm,clip=true,keepaspectratio]{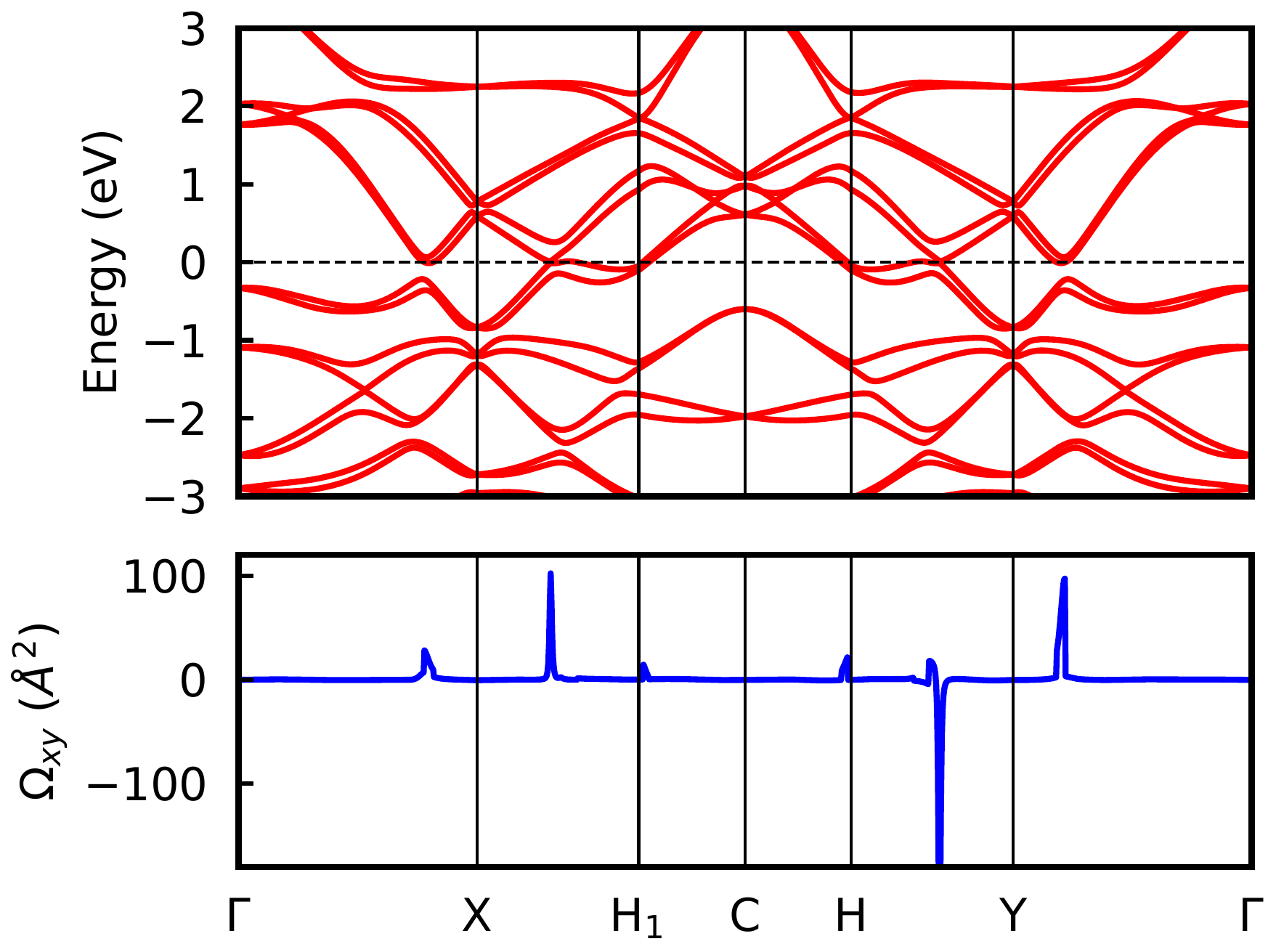}
\end{subfigure}
\begin{subfigure}[b]{0.4\columnwidth}
\subcaption{}
\includegraphics[width=\columnwidth,height = 4cm,clip=true,keepaspectratio]{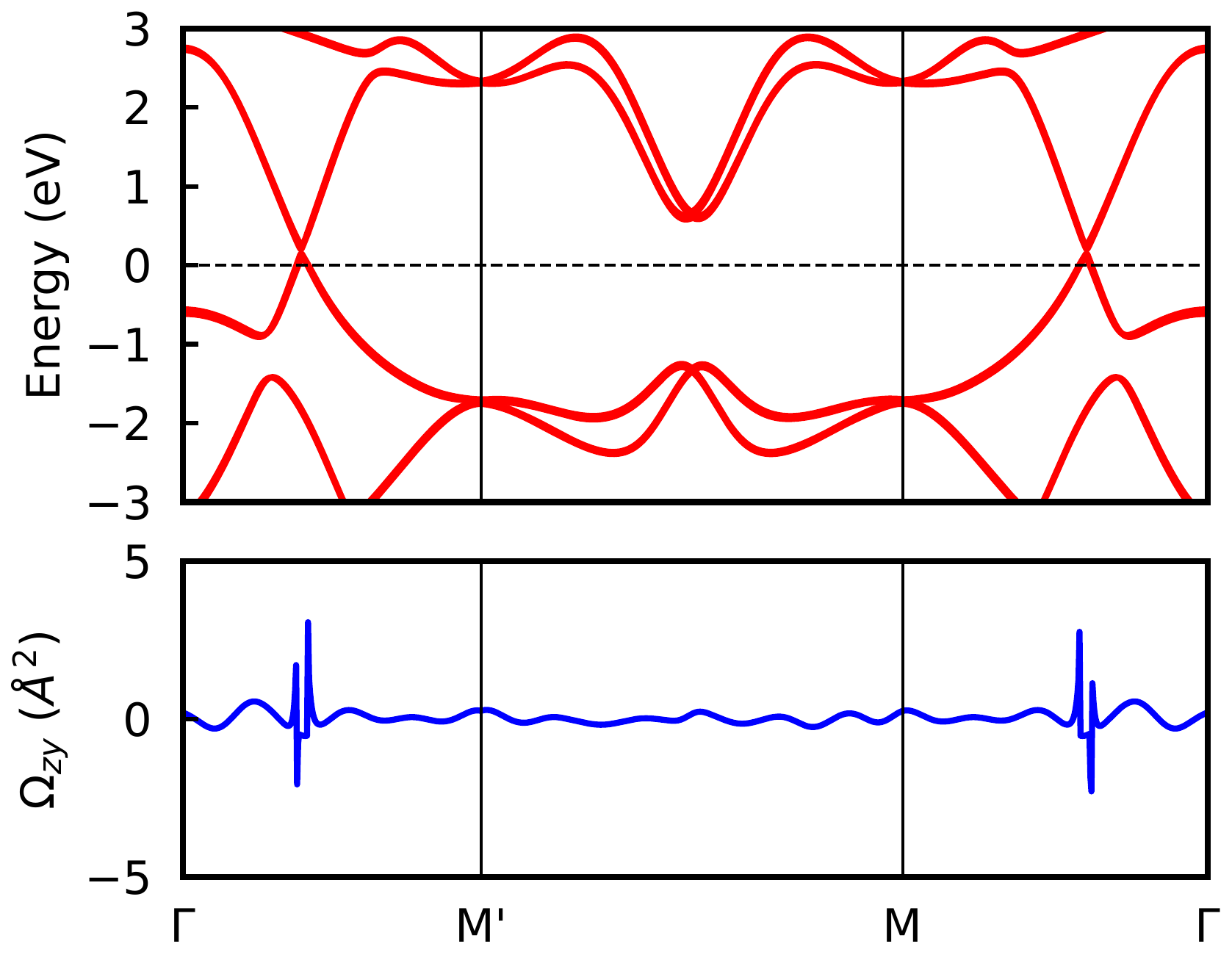}
\end{subfigure}
\begin{subfigure}[b]{0.4\columnwidth}
\subcaption{}
\includegraphics[width=\columnwidth,height = 4cm,clip=true,keepaspectratio]{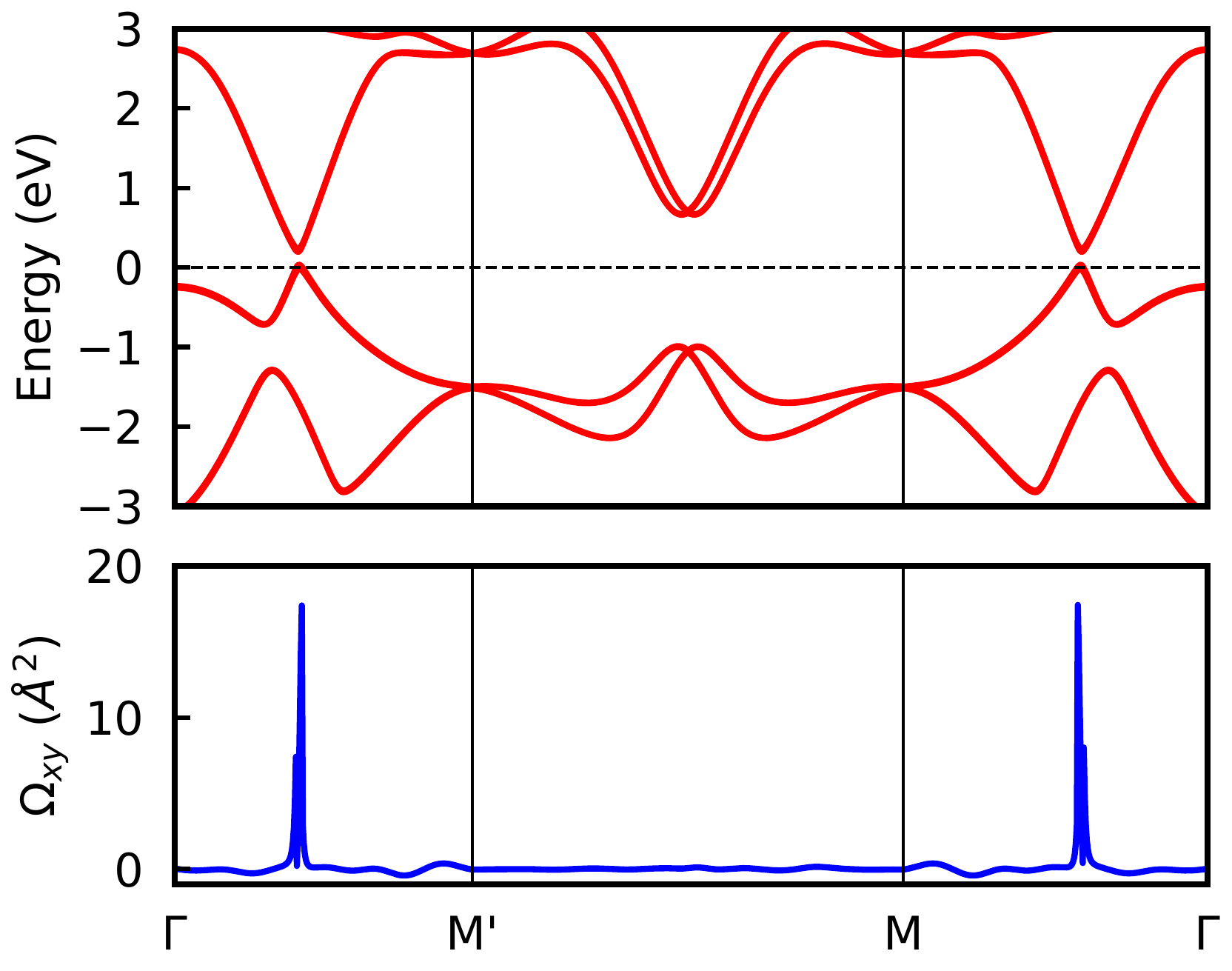}
\end{subfigure}
\caption{\label{fig:berry} Berry curvature distributions calculated within the two-dimensional Brillouin zone for different tellurium phases: (a) $\alpha$-tellurene, (b) $\beta$-tellurene, (c) one-side hydrogen-passivated hexagonal tellurene, (d) buckled kagome tellurene, (e) buckled square tellurene, and (f) buckled square tellurene under 5\% strain.}
\end{figure}

\begin{table}[H]
\centering
\caption{\label{tab:Z2} $\mathbb{Z}_2$ topological invariant calculated within MLWF-HSE06.}
\begin{tabular}{lccc}
\toprule
phase &	$\mathbb{Z}_2$ &	Topological class &	Berry-curvature character\\
\midrule
$\alpha$        & 0     & trivial       &                  \\
$\beta$         & 0	& trivial 	& \\
buckled kagome	& 1	& non-trivial	& strong SOC-induced hotspots\\
buckled square	& 1	& non-trivial	& moderate, symmetry-localized\\
passivated hexagonal	& 1	& non-trivial	& valley-localized \\
\bottomrule
\end{tabular}
\end{table}

The topological properties of the two-dimensional tellurene phases were characterized using the Chern number and the  $\mathbb{Z}_2$  invariant,
both evaluated from the Berry curvature and the evolution of the
occupied electronic subspace in momentum space. For systems that break time-reversal symmetry, the Chern number $C$ is
defined as the Brillouin-zone integral of the Berry curvature summed
over all occupied bands,

\begin{equation}
C = \frac{1}{2\pi} \int_{\mathrm{BZ}} \Omega_{xy}(\mathbf{k}) \, d^2k ,
\end{equation}

where $\Omega_{xy}(\mathbf{k})$ is the Berry curvature summed over the occupied
bands, $\Omega_{xy}(\mathbf{k})=\sum_{n\in\mathrm{occ}}\Omega_{n,xy}(\mathbf{k})$.
In our implementation, $\Omega_{n,xy}(\mathbf{k})$ is evaluated using the
Kubo (velocity-matrix) expression, as implemented in \textsc{WannierTools},

  \begin{equation}
\Omega_{n,xy}(\mathbf{k}) =
-2\,\mathrm{Im}\sum_{m\neq n}
\frac{
\langle u_{n\mathbf{k}}|\,\hat v_x\,|u_{m\mathbf{k}}\rangle\,
\langle u_{m\mathbf{k}}|\,\hat v_y\,|u_{n\mathbf{k}}\rangle
}{
\left(\varepsilon_{m\mathbf{k}}-\varepsilon_{n\mathbf{k}}\right)^2
},
\end{equation}
where $|u_{n\mathbf{k}}\rangle$ and $\varepsilon_{n\mathbf{k}}$ are the
cell-periodic eigenstates and eigenvalues of the Wannier-interpolated tight-binding
Hamiltonian $H(\mathbf{k})$, and the velocity operators are obtained from
$\hat v_{\alpha}=(1/\hbar)\,\partial H(\mathbf{k})/\partial k_{\alpha}$.
In time-reversal-symmetric systems, the Berry curvature satisfies
$\Omega_{xy}(\mathbf{k})=-\Omega_{xy}(-\mathbf{k})$, enforcing a vanishing total
Chern number, although sizable local Berry-curvature contributions may still occur.
For time-reversal-invariant two-dimensional systems, the relevant topological index
is the $\mathbb{Z}_2$ invariant, which distinguishes trivial insulators from quantum
spin Hall phases. The $\mathbb{Z}_2$ invariant was determined from the evolution of
the Wilson loop, or equivalently the Wannier charge centers, of the occupied bands
across the Brillouin zone. A nontrivial topological phase corresponds to an odd
winding of the Wannier charge centers and yields $\mathbb{Z}_2=1$, whereas the
absence of winding indicates a trivial phase with $\mathbb{Z}_2=0$.

The Berry curvature and Chern number were
computed using tight-binding Hamiltonians constructed from maximally
localized Wannier functions, ensuring an
accurate interpolation of the electronic structure on dense
$\mathbf{k}$-point meshes. Spin--orbit coupling was included in all
topological calculations, as it is essential for capturing the
SOC-driven band inversions and avoided crossings responsible for
nontrivial topology in tellurene\,\cite{wantool}.

The topological behavior of tellurene polymorphs is shown to be highly
sensitive to lattice geometry, buckling, and symmetry breaking,
leading to a rich hierarchy of quantum phases within the same chemical
composition. Results are shown in Table~\ref{tab:Z2}. Among the four phases examined here, $\alpha$ and $\beta$-tellurene are
topologically trivial, whereas buckled kagome, buckled square, and
one-side passivated hexagonal tellurene all realize nontrivial
two-dimensional  $\mathbb{Z}_2 = 1$ topology.

$\beta$-tellurene remains a trivial insulator ($\mathbb{Z}_2 = 0$) despite the presence
of spin–orbit coupling. Although SOC induces band splittings and
generates localized Berry-curvature features associated with avoided
crossings, it does not alter the global band connectivity of the
occupied manifold. No SOC-driven band inversion occurs, and the
Wilson-loop evolution exhibits no topological winding. This
demonstrates that Berry-curvature hotspots alone are insufficient to
guarantee nontrivial  $\mathbb{Z}_2 = 1$  topology. Rather, a change in the global
ordering of bands across the Brillouin zone is required.

In contrast, buckled kagome tellurene exhibits nontrivial $\mathbb{Z}_2$ topology of the occupied band. The kagome lattice naturally hosts a dense manifold of
near-degenerate bands, and buckling enhances SOC-induced hybridization
among them. In the absence of SOC, the system lies close to a
multi-band crossing regime, while SOC lifts these degeneracies and
opens a bulk gap. The resulting Berry curvature is highly localized
and unusually large in magnitude near SOC-gapped avoided crossings,
reflecting the strong interband mixing inherent to the kagome
geometry. Despite the extreme local curvature, time-reversal symmetry
enforces a vanishing Chern number, and the nontrivial topology is
instead encoded in the  $\mathbb{Z}_2 = 1$ invariant. This phase represents the most
pronounced manifestation of SOC-driven topology among the structures
considered.

Buckled square tellurene also realizes a  $\mathbb{Z}_2 = 1$-nontrivial  phase,
but through a comparatively simpler mechanism. Here, SOC gaps
symmetry-allowed near-crossings present in the non-SOC band structure,
yielding a well-defined insulating manifold with nontrivial
Wilson-loop winding. The Berry curvature is more moderate and
spatially confined than in the kagome case, indicating fewer competing
near-degeneracies. Importantly, the persistence of  $\mathbb{Z}_2 = 1$ under
moderate strain demonstrates that the square-lattice nontrivial  phase is not
accidental but instead represents a stable topological regime tunable
by lattice deformation.

One-side passivated hexagonal tellurene provides a complementary route
to QSH behavior through explicit breaking of out-of-plane
symmetry. Passivation shifts the system toward a near-crossing regime
in the absence of SOC, while SOC opens a bulk gap and stabilizes a
 $\mathbb{Z}_2 = 1$ phase. Unlike the kagome and square lattices, the Berry
curvature in this structure is predominantly valley-localized around
the K and K' points, reflecting the hexagonal symmetry and inversion
asymmetry of the lattice. The resulting curvature pattern cancels
globally under time-reversal symmetry but highlights the
valley-resolved nature of the topological response, suggesting
potential interplay between QSH and valley physics.

These results establish that nontrivial $\mathbb{Z}_2 = 1$
  topology in tellurene is not an isolated phenomenon but emerges
  systematically when lattice geometry and symmetry place the system
  near a SOC crossing regime. The kagome lattice maximizes this effect
  through band multiplicity, the square lattice offers a simpler and
  strain-robust realization, and one-side passivation enables topology
  through symmetry breaking and valley selectivity. In contrast,
  $\beta$-tellurene lacks the necessary band reordering and remains
  topologically trivial. This comparative analysis highlights
  tellurene as a versatile platform for engineering two-dimensional
  quantum spin Hall phases through structural design rather than
  chemical substitution.

\section{Conclusions}

In this work, we conducted a comprehensive first-principles
investigation of tellurium across its dimensional hierarchy—bulk Te-I,
2D monolayers, and 1D helical nanowires—addressing structural,
electronic, vibrational, and topological properties with full
inclusion of spin–orbit coupling (SOC).

We confirmed that bulk trigonal Te-I is dynamically and
thermodynamically stable and hosts Weyl nodes arising from broken
inversion symmetry and strong SOC. Two-dimensional $\alpha$-Te and $\beta$-Te monolayers are shown to be stable semiconductors and topologically trivial within our calculations ($\mathbb{Z}_2 = 0$), with no SOC-driven band inversion in the occupied manifold; however, their strong SOC suggests they are promising candidates for topological phase transitions driven by external strain, doping, or magnetic perturbations.

Buckled kagome, buckled square, and one-side passivated hexagonal
tellurene exhibit nontrivial two-dimensional $\mathbb{Z}_2 = 1$
topology of the occupied bands, with the latter realizing a fully
gapped quantum spin Hall phase and the former two representing
incipient QSH regimes accessible upon tuning the chemical potential.
The realization of a nontrivial $\mathbb{Z}_2 = 1$ topology in metallic kagome and square
tellurene highlights an experimentally realistic route to quantum spin
Hall phases, where topology is established at the band-structure level
and insulating behavior can be achieved through electrostatic or
substrate-induced tuning. Moreover, the associated SOC-induced
Berry-curvature hotspots suggest enhanced spin and charge transport
responses even in the metallic regime.  Together, these results
establish tellurium as a uniquely tunable platform for topology
engineering across dimensionality. By bridging 3D Weyl physics, 2D
symmetry-modulated topological phases, and termination-induced
edge-localized states in finite nanowires, our study provides a
unified framework for designing tellurium-based topological
matter. This work not only advances the fundamental understanding of
symmetry-driven phases in chalcogen systems but also highlights the
potential of Te nanostructures for next-generation topological
electronics, spintronics, and optoelectronic devices.

\section{Acknowledgements}

All authors acknowledge the financial support from the Brazilian
funding agency CNPq under grant numbers 305174/2023-1, 408144/2022-0
and 305952/2023-4. G.E.G.A. thanks CNPq for a undergraduate
scholarship. We thank computational resources from Supercomputers LaMCAD at UFG, Santos
Dumont at LNCC and CENAPAD-SP at Unicamp.

\clearpage
\newpage

\begin{center}
{\bf Supplementary Information\\
  Interplay between structural, electronic and topological properties in low-dimensional tellurium} 
\end{center}

\section{Thermal properties}

\setcounter{figure}{0}
\setcounter{table}{0}
\renewcommand{\thefigure}{S\arabic{figure}}
\renewcommand{\thetable}{S\arabic{table}}

\begin{figure}[H]
\centering
\begin{subfigure}[b]{0.48\textwidth}
        \subcaption{}
        \includegraphics[width=\linewidth]{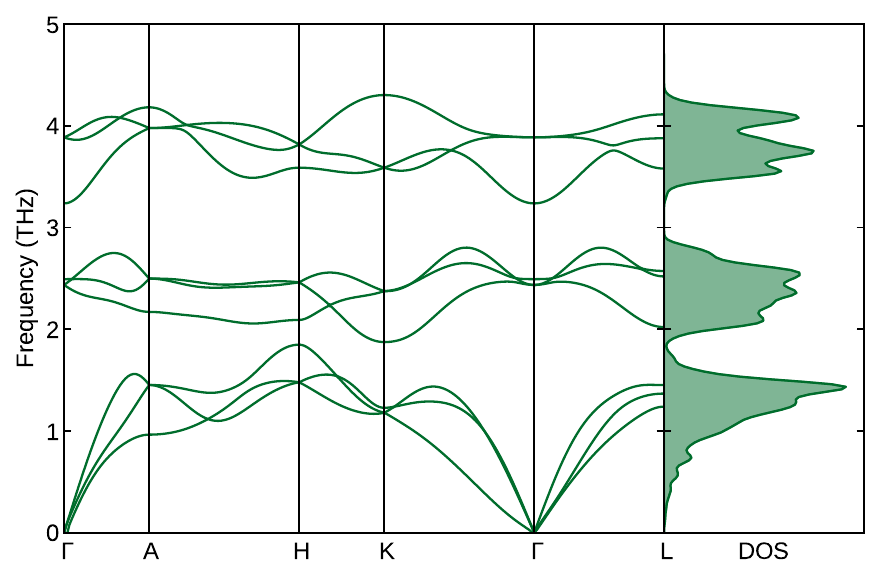}
    \end{subfigure}
    \hfill 
    \begin{subfigure}[b]{0.48\textwidth}
        \subcaption{}
        \includegraphics[width=\linewidth]{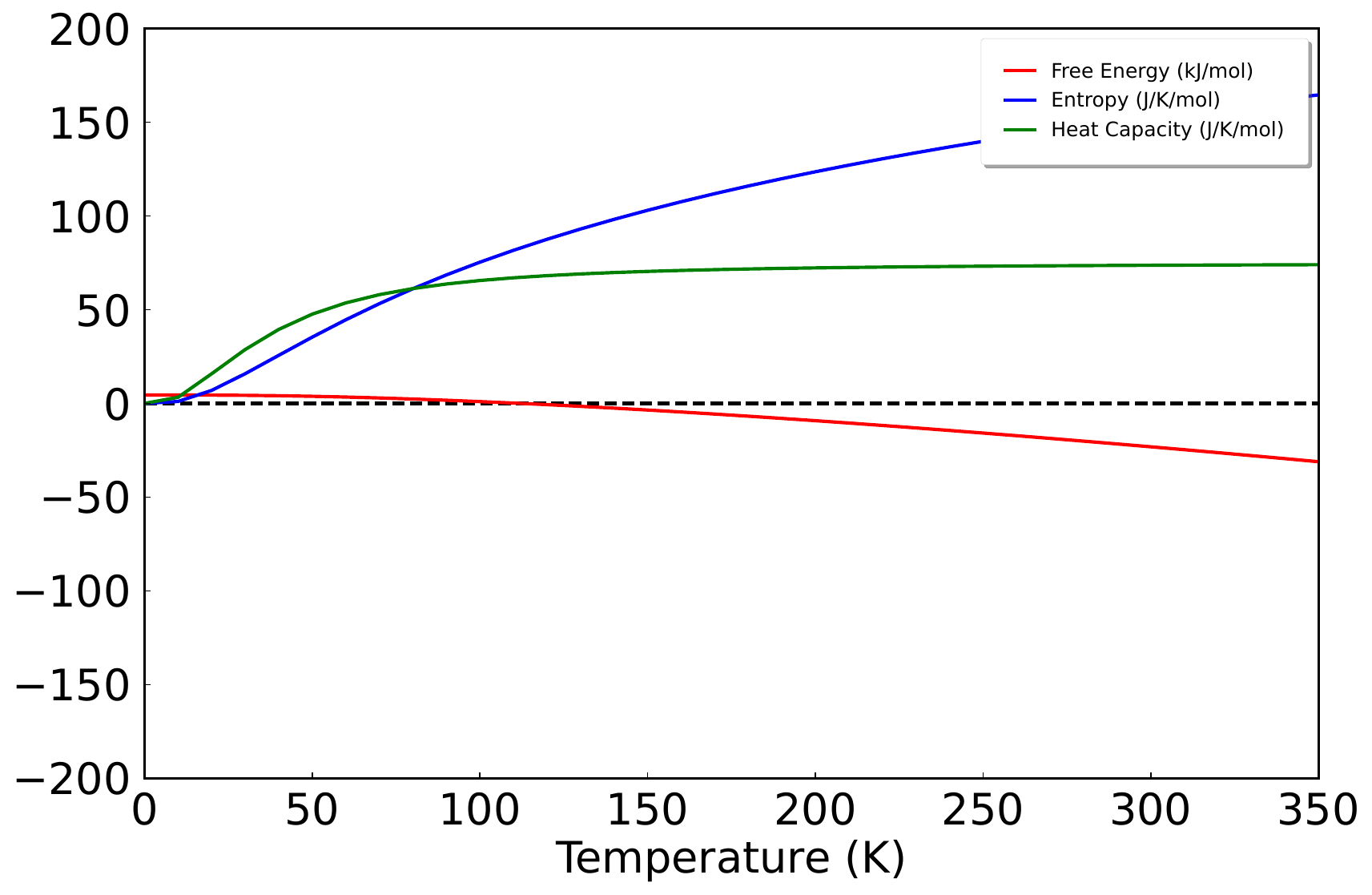}
    \end{subfigure}
    \captionsetup[subfigure]{labelformat=empty}
   \begin{subfigure}[b]{0.2\columnwidth}
    \subcaption{c) 2.44 THz}
\includegraphics[width=\columnwidth]{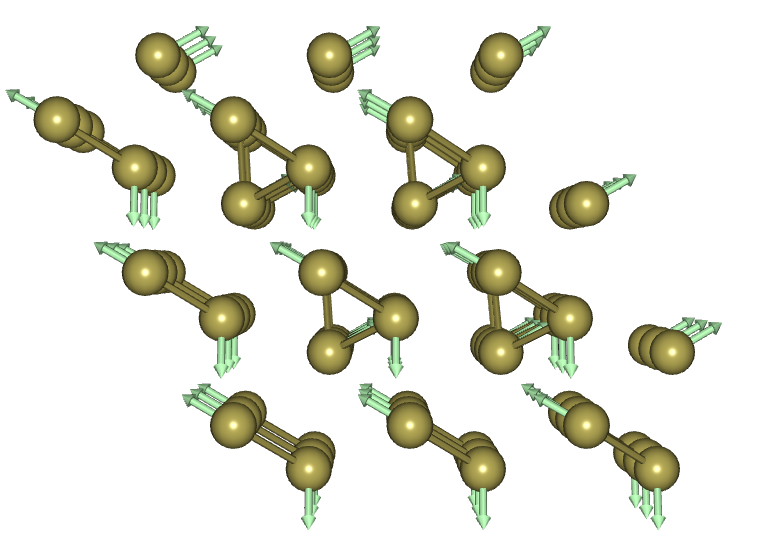}
\subcaption{view along $\vec{c}$}
   \end{subfigure}
	\hspace{0.1cm}
	\begin{subfigure}[b]{0.2\columnwidth}
    \subcaption{d) 2.49 THz}
		\includegraphics[width=\columnwidth]{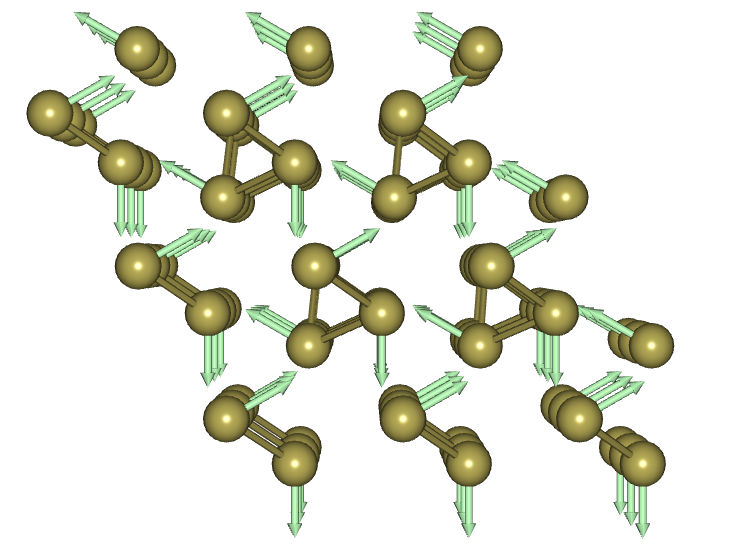}
\subcaption{view along $\vec{c}$}
        \end{subfigure}
	\hspace{0.1cm}
	\begin{subfigure}[b]{0.2\columnwidth}
    \subcaption{e) 3.24 THz}
\includegraphics[width=\columnwidth]{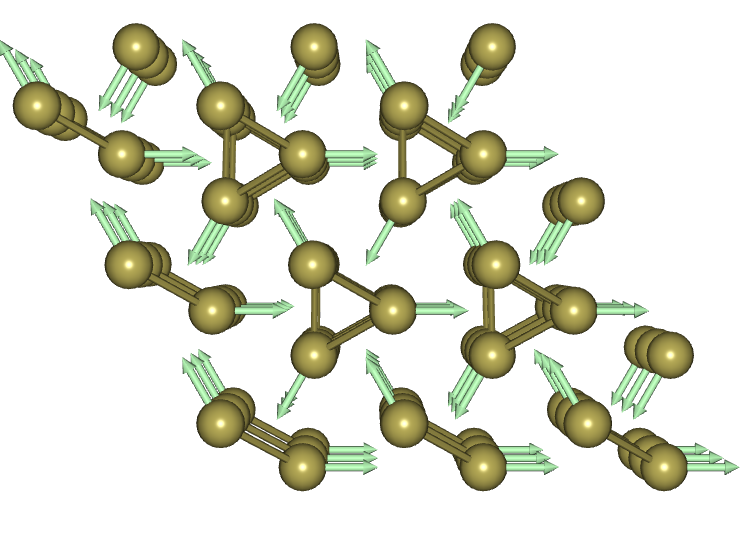}
\subcaption{view along $\vec{c}$}
	\end{subfigure}
	\hspace{0.1cm}
	\begin{subfigure}[b]{0.2\columnwidth}
    \subcaption{f) 3.89 THz}
		\includegraphics[width=\columnwidth,clip=true,keepaspectratio,height=3cm]{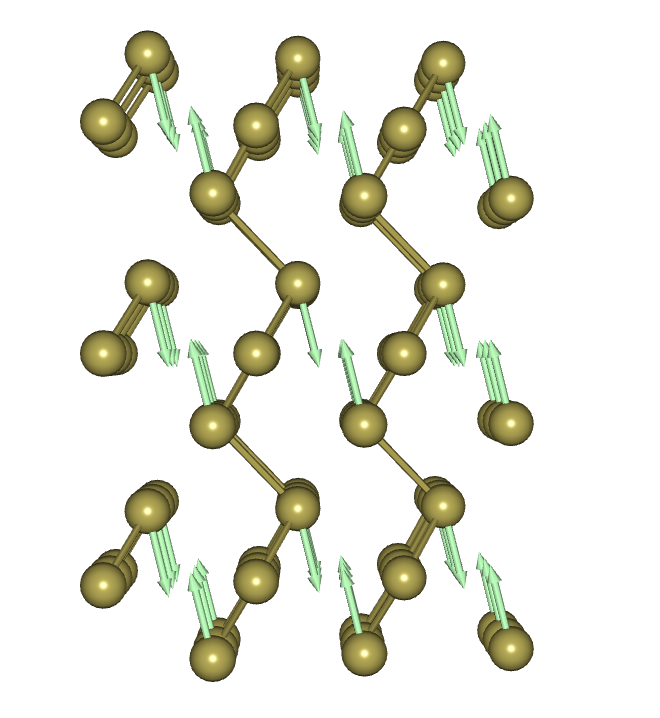}
 \subcaption{view along $\vec{a}$}
        \end{subfigure}
	\caption{\label{fig:S1}Vibrational and thermodynamic properties of trigonal tellurium (Te-I). a) Phonon dispersion curve and density of states and b) free energy, entropy and heat capacity. Selected phonon modes in Te-I: (c-e) projected along the  $\vec{c}$-axis and (f) is viewed along the $\vec{a}$ axis.}
\end{figure}

\begin{figure}[H]
\centering
\begin{subfigure}[b]{0.3\columnwidth}
\subcaption{}
\includegraphics[width=\columnwidth,clip=true,keepaspectratio]{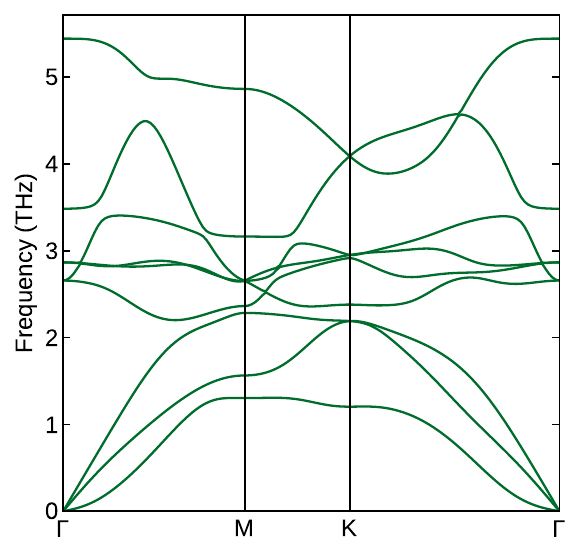}
\end{subfigure}
\begin{subfigure}[b]{0.3\columnwidth}
\subcaption{}
\includegraphics[width=\columnwidth,clip=true,keepaspectratio]{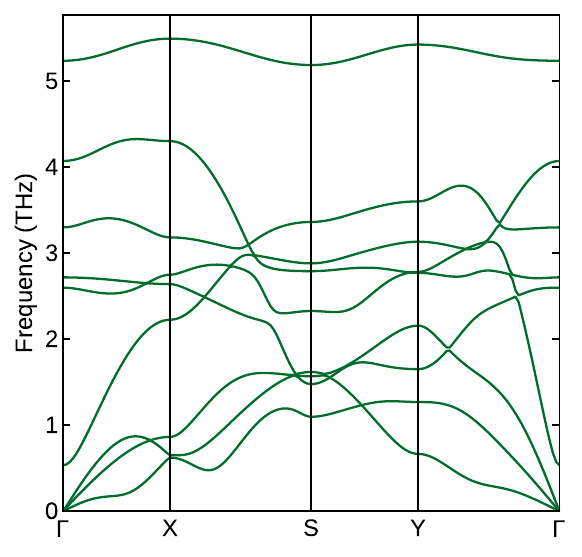}
\end{subfigure}
\begin{subfigure}[b]{0.3\columnwidth}
\subcaption{}
\includegraphics[width=\columnwidth,clip=true,keepaspectratio]{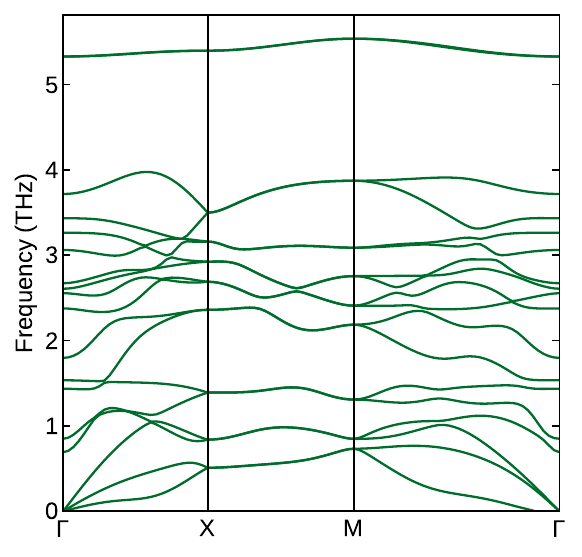}
\end{subfigure}\\
\begin{subfigure}[b]{0.3\columnwidth}
\subcaption{}
\includegraphics[width=\columnwidth,clip=true,keepaspectratio]{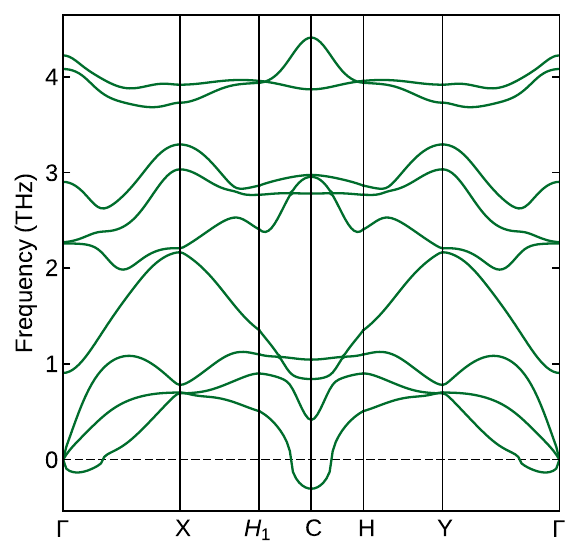}
\end{subfigure}
\begin{subfigure}[b]{0.3\columnwidth}
\subcaption{}
\includegraphics[width=\columnwidth,clip=true,keepaspectratio]{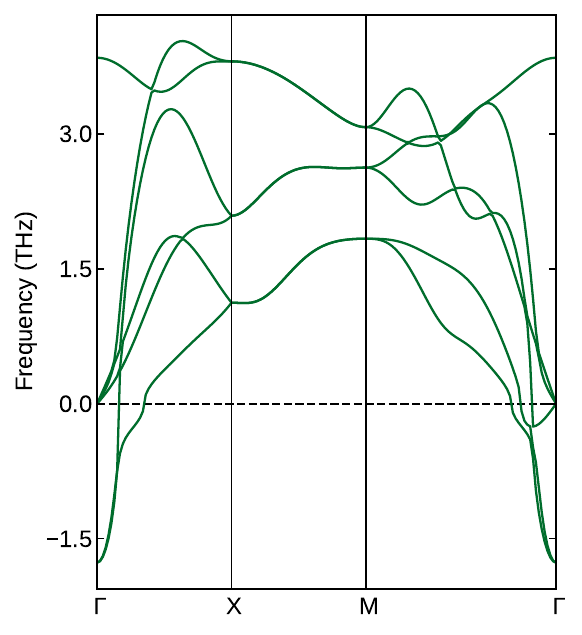}
\end{subfigure}
\caption{\label{fig:S2} Phonon dispersion curves of 2D tellurene phases: a) $\alpha$-Te, b) $\beta$-Te,  c) buckled pentagonal, d) buckled kagome and e) buckled square.}
\end{figure}

\begin{figure}[H]
	\centering
    \captionsetup[subfigure]{labelformat=empty}
	\begin{subfigure}[b]{0.2\columnwidth}
          \subcaption{a) 2.66 THz}
          \includegraphics[width=\columnwidth,clip=true]{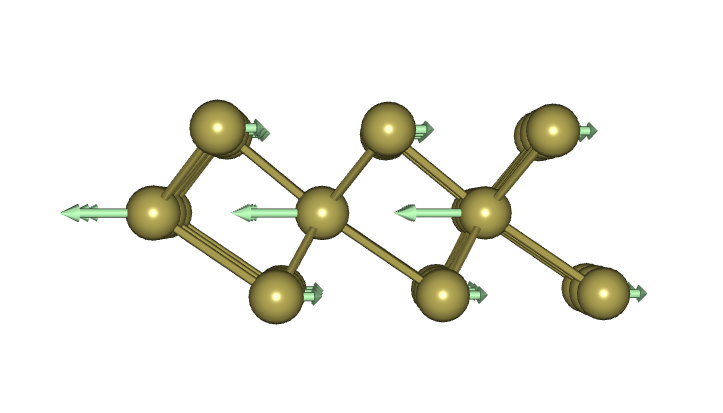}
         \subcaption{$\vec{b}$-axis}
		
	\end{subfigure}
	\begin{subfigure}[b]{0.2\columnwidth}
    \subcaption{b) 2.87 THz}	\includegraphics[width=\columnwidth,clip=true]{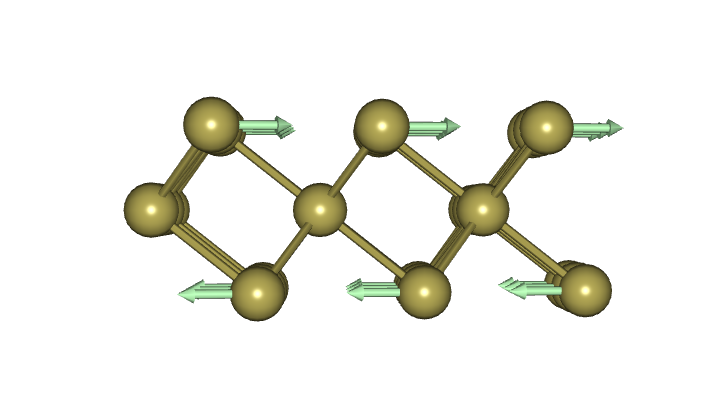}	
        \subcaption{$\vec{b}$-axis}
	\end{subfigure}
	\begin{subfigure}[b]{0.2\columnwidth}
    \subcaption{c) 3.48 THz}
		\includegraphics[width=\columnwidth,clip=true]{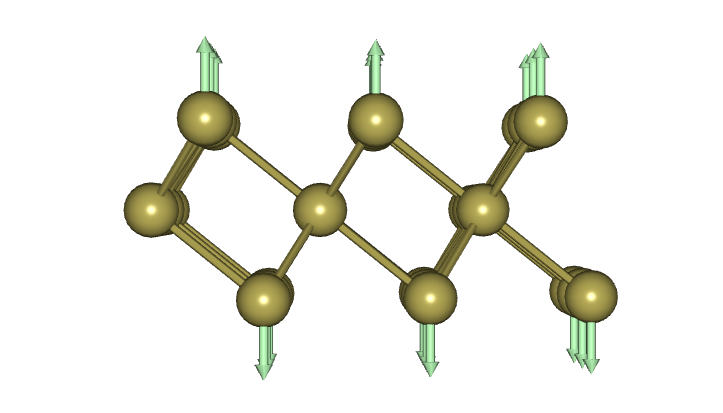}
        \subcaption{$\vec{b}$-axis}
	\end{subfigure}
	\begin{subfigure}[b]{0.18\columnwidth}
    \subcaption{d) 5.44 THz}
	\includegraphics[width=\columnwidth,clip=true]{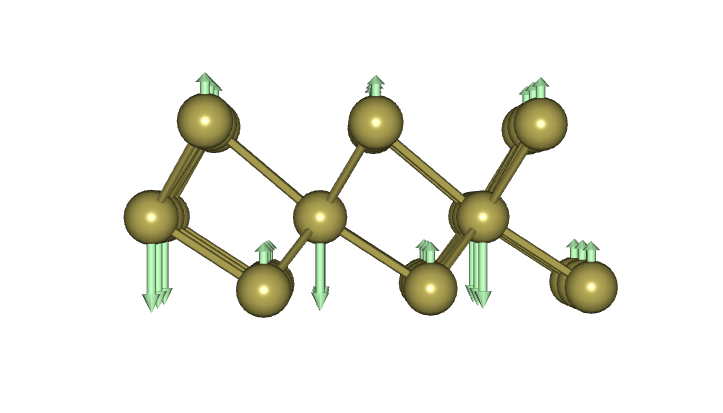}	
         \subcaption{$\vec{b}$-axis}
	\end{subfigure}
\vspace{1cm}
	\begin{subfigure}[b]{0.18\columnwidth}
    \subcaption{e) 0.54 THz}
\includegraphics[width=\columnwidth,clip=true]{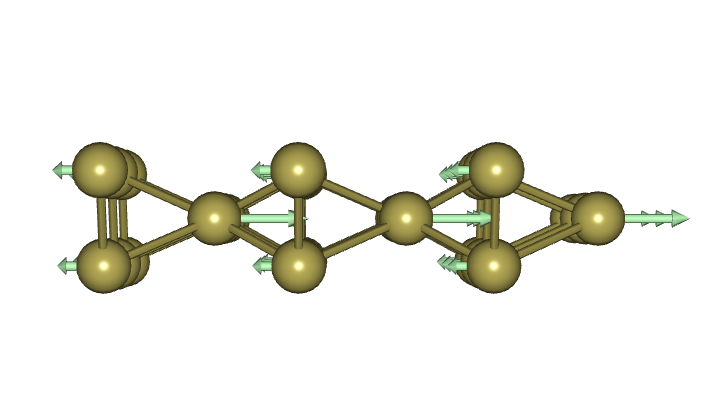}
		\subcaption{$\vec{b}$-axis}
	\end{subfigure}
	\begin{subfigure}[b]{0.18\columnwidth}
    \subcaption{f) 2.60 THz}	\includegraphics[width=\columnwidth,clip=true]{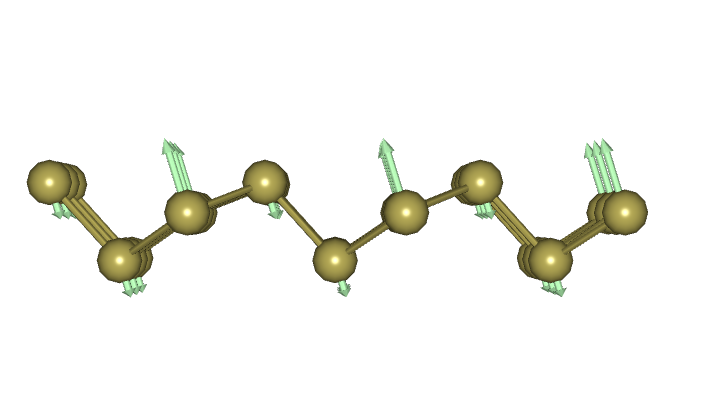}
         \subcaption{$\vec{a}$-axis}
		
	\end{subfigure}
	\begin{subfigure}[b]{0.18\columnwidth}
    \subcaption{g) 2.72 THz}
		\includegraphics[width=\columnwidth,clip=true]{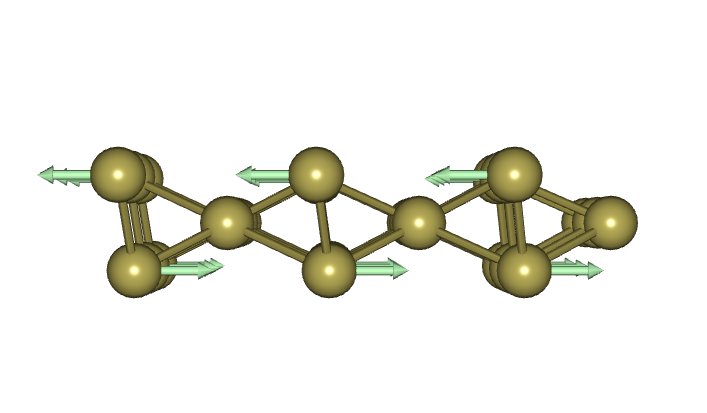}	
         \subcaption{$\vec{b}$-axis}		
	\end{subfigure}
	\begin{subfigure}[b]{0.18\columnwidth}
    \subcaption{h) 3.29 THz}
\includegraphics[width=\columnwidth,clip=true]{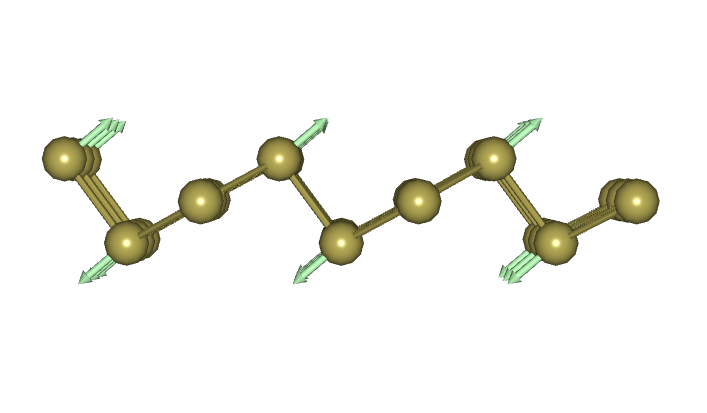}	
        \subcaption{$\vec{a}$-axis}
	\end{subfigure}	
	\begin{subfigure}[b]{0.18\columnwidth}
    \subcaption{i) 4.07 THz}
\includegraphics[width=\columnwidth,clip=true]{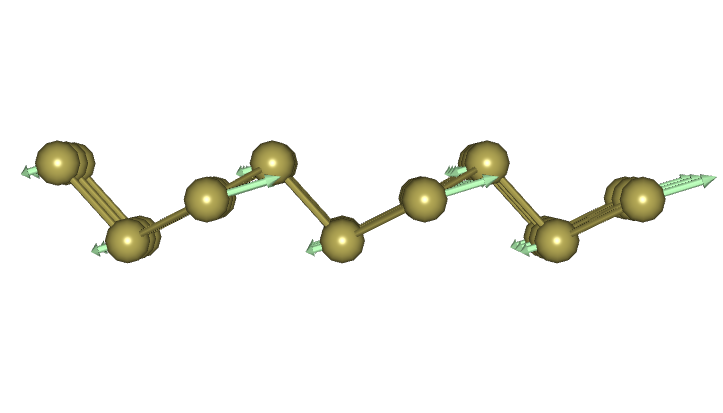}
        \subcaption{$\vec{a}$-axis}
		\end{subfigure}
	\begin{subfigure}[b]{0.18\columnwidth}
    \subcaption{j) 5.26 THz}
	\includegraphics[width=\columnwidth,clip=true]{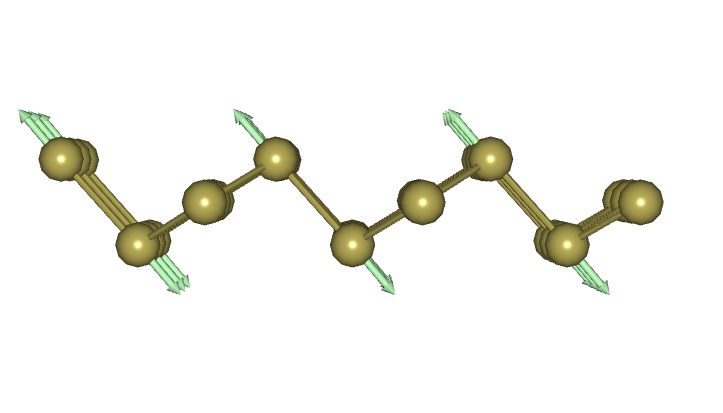}
         \subcaption{$\vec{a}$-axis}
	\end{subfigure}
	\caption{\label{fig:S3} Selected phonon modes at $\Gamma$-point calculated withon GGA: (a-d) $\alpha$-Te and (i-j) $\beta$-Te.}
\end{figure}

\begin{figure}[H]
\centering
\begin{subfigure}[b]{0.32\textwidth}
\subcaption{}
\includegraphics[width=\textwidth,clip=true,keepaspectratio]{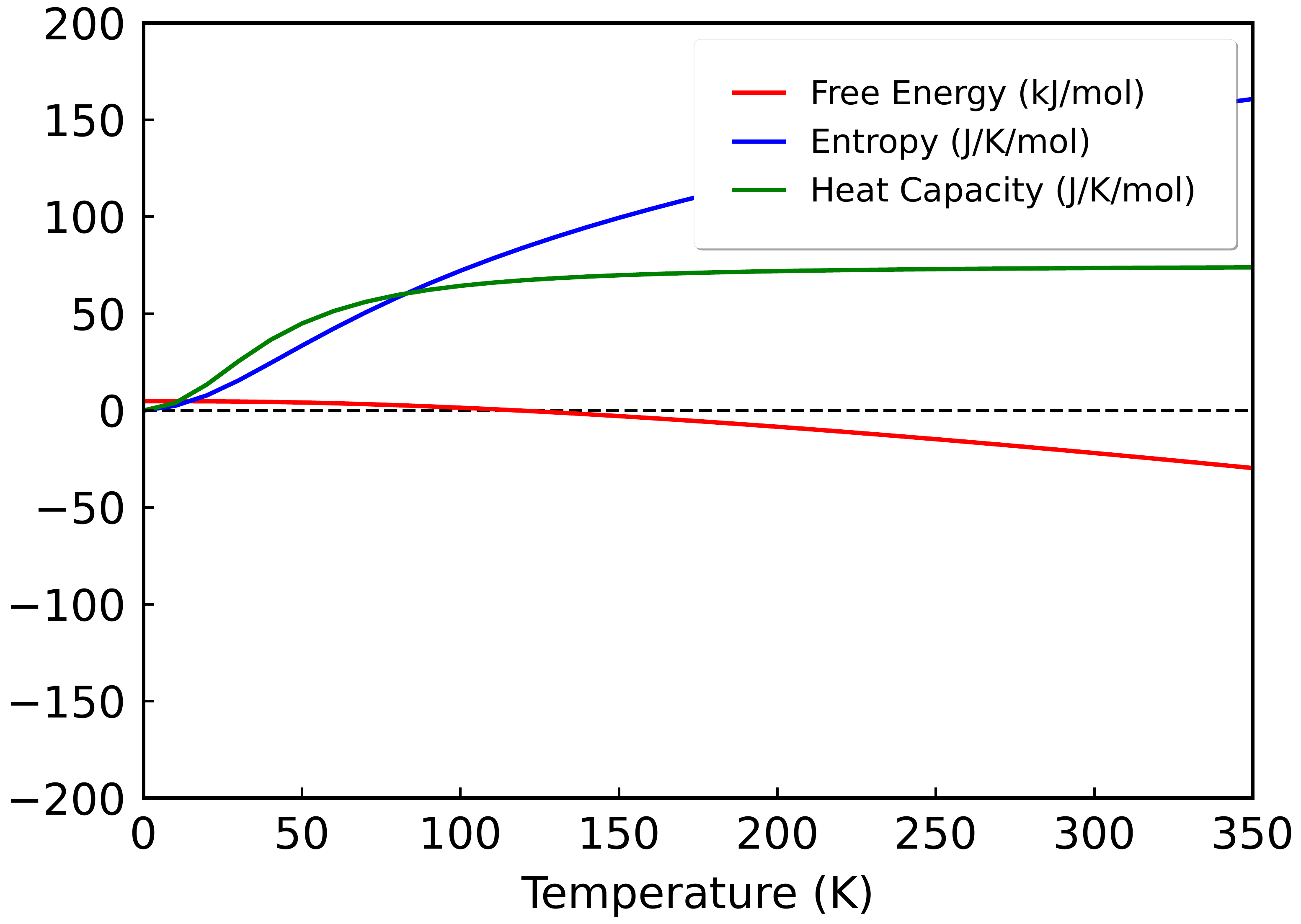}
\end{subfigure}
\begin{subfigure}[b]{0.32\textwidth}
\subcaption{}
\includegraphics[width=\textwidth,clip=true]{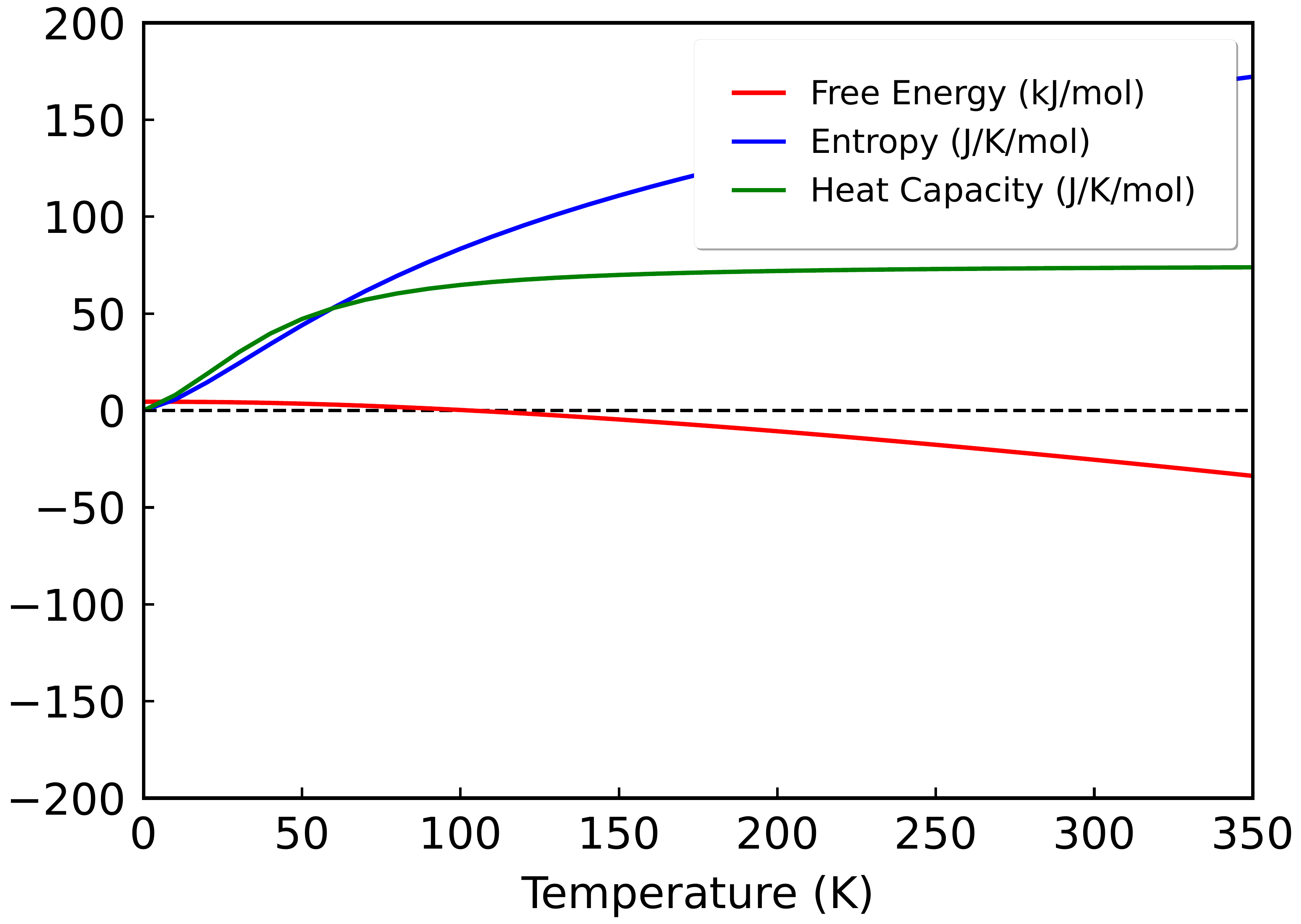}
\end{subfigure}
\begin{subfigure}[b]{0.32\textwidth}
\subcaption{}
\includegraphics[width=\textwidth,clip=true,keepaspectratio]{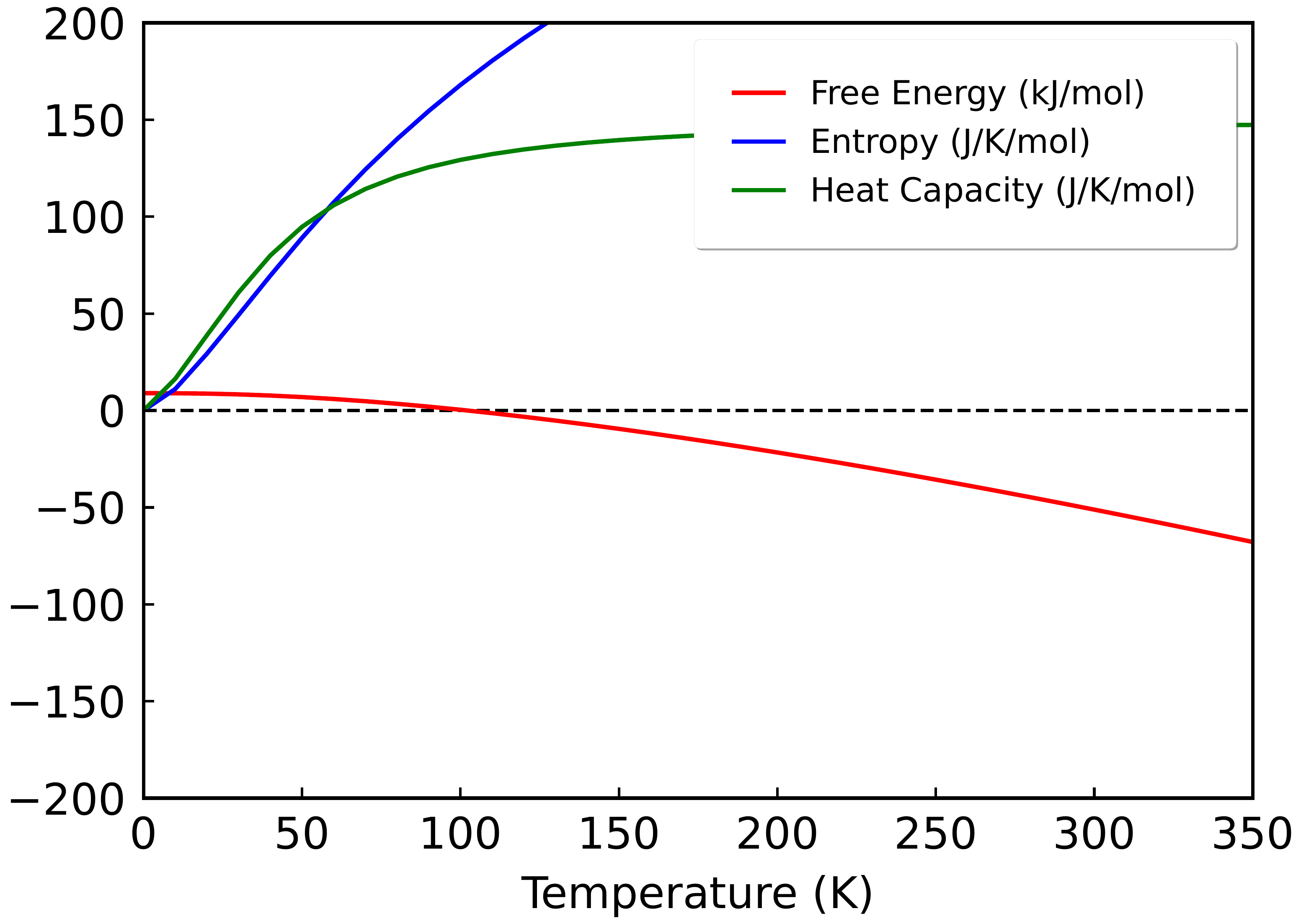}
\end{subfigure}
\begin{subfigure}[b]{0.32\textwidth}
\subcaption{}
\includegraphics[width=\textwidth,clip=true,keepaspectratio]{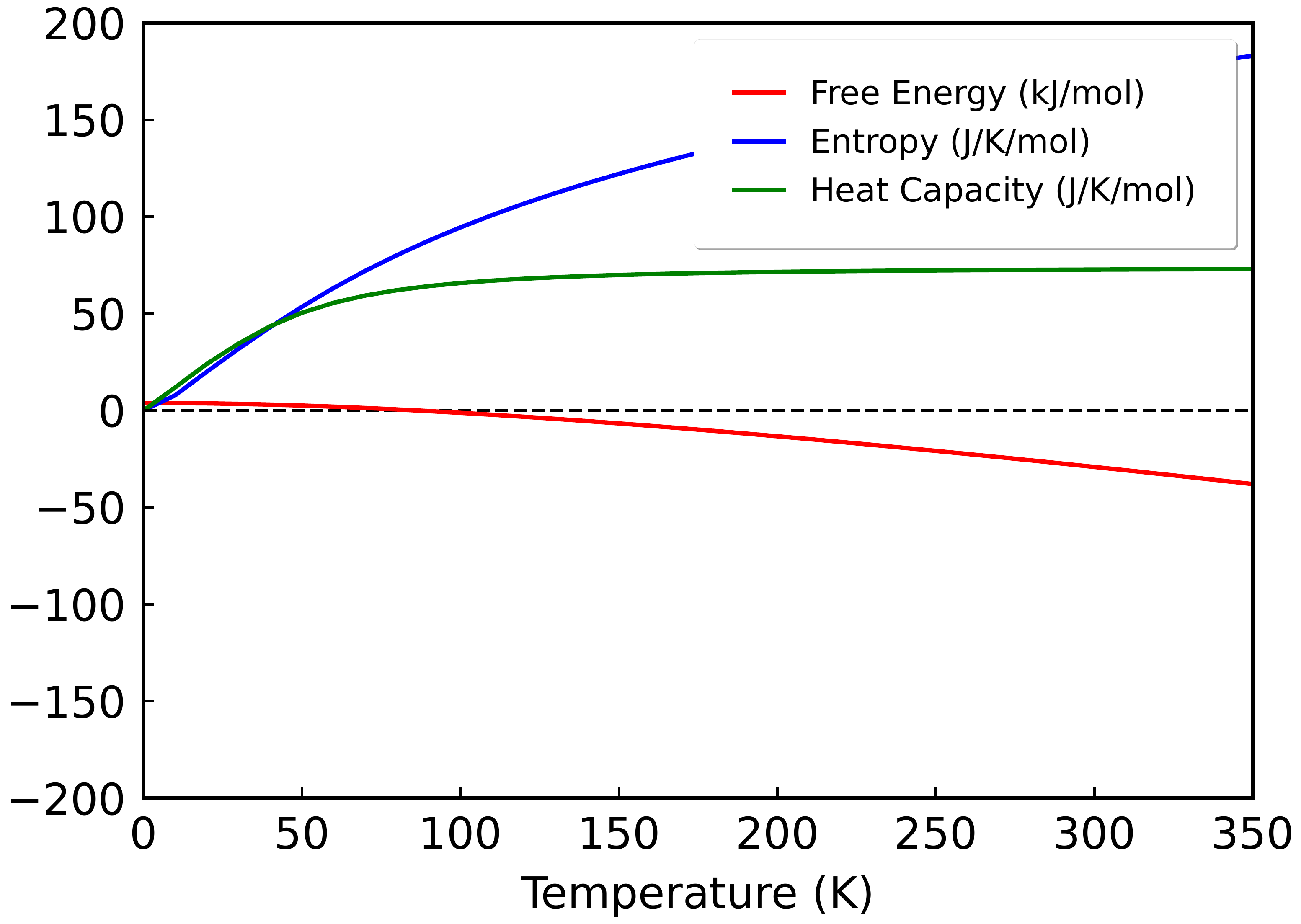}
\end{subfigure}
\begin{subfigure}[b]{0.32\textwidth}
\centering
\subcaption{}
\includegraphics[width=\textwidth,clip=true,keepaspectratio]{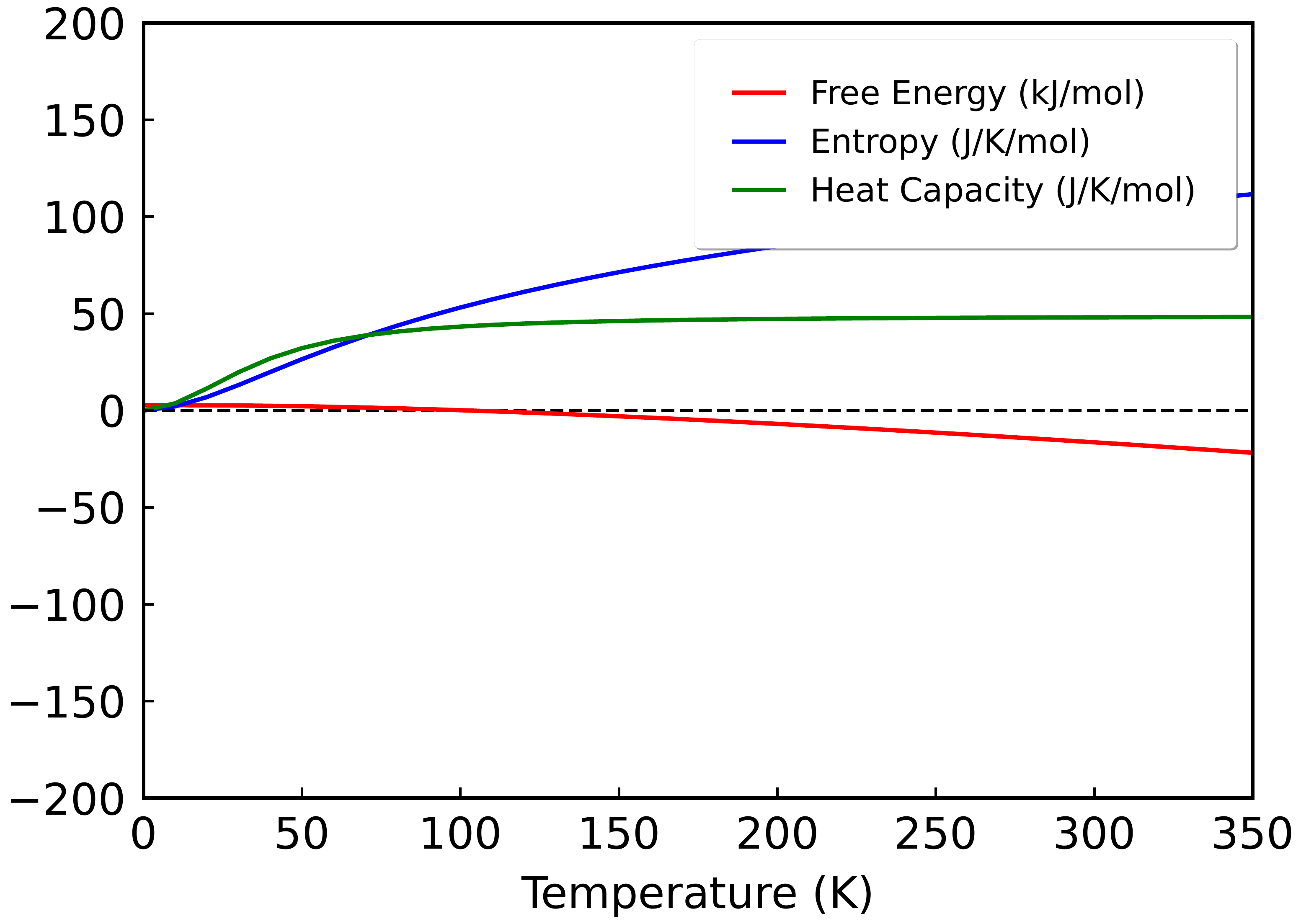}
\end{subfigure}
\caption{\label{fig:S4} Thermal properties of tellurium 2D lattices: heat capacity ($C_V$), Helmholtz free energy ($F$), and entropy ($S$) calculated within GGA. a) $\alpha$-Te, b) $\beta$-Te,  c) buckled pentagonal, d) buckled kagome and e) buckled square.}
\end{figure}

\begin{figure}[H]
\centering
\begin{subfigure}[b]{0.4\columnwidth}
    \subcaption{}
\includegraphics[width=\columnwidth,height=4cm,clip=true,keepaspectratio]{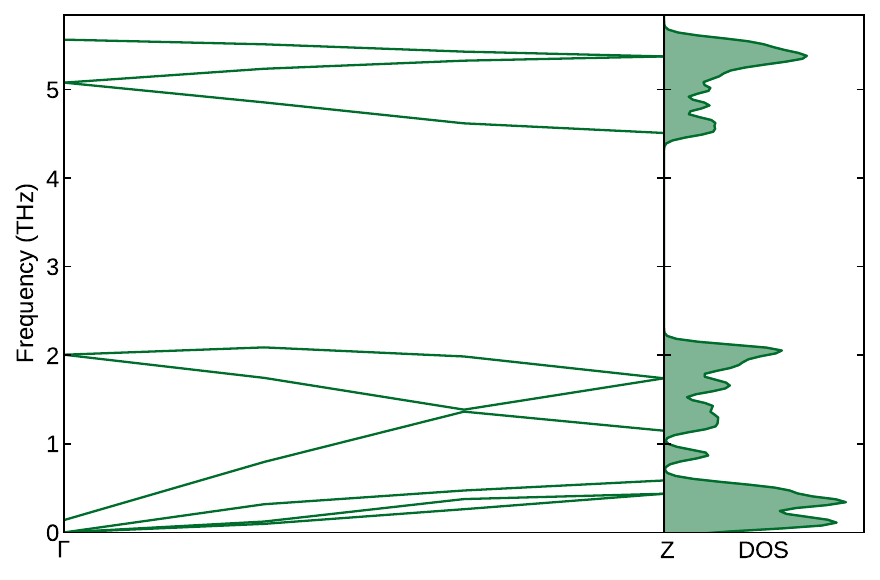}
\end{subfigure}
\begin{subfigure}[b]{0.4\columnwidth}
    \subcaption{}
\includegraphics[width=\columnwidth,height=4cm,clip=true,keepaspectratio]{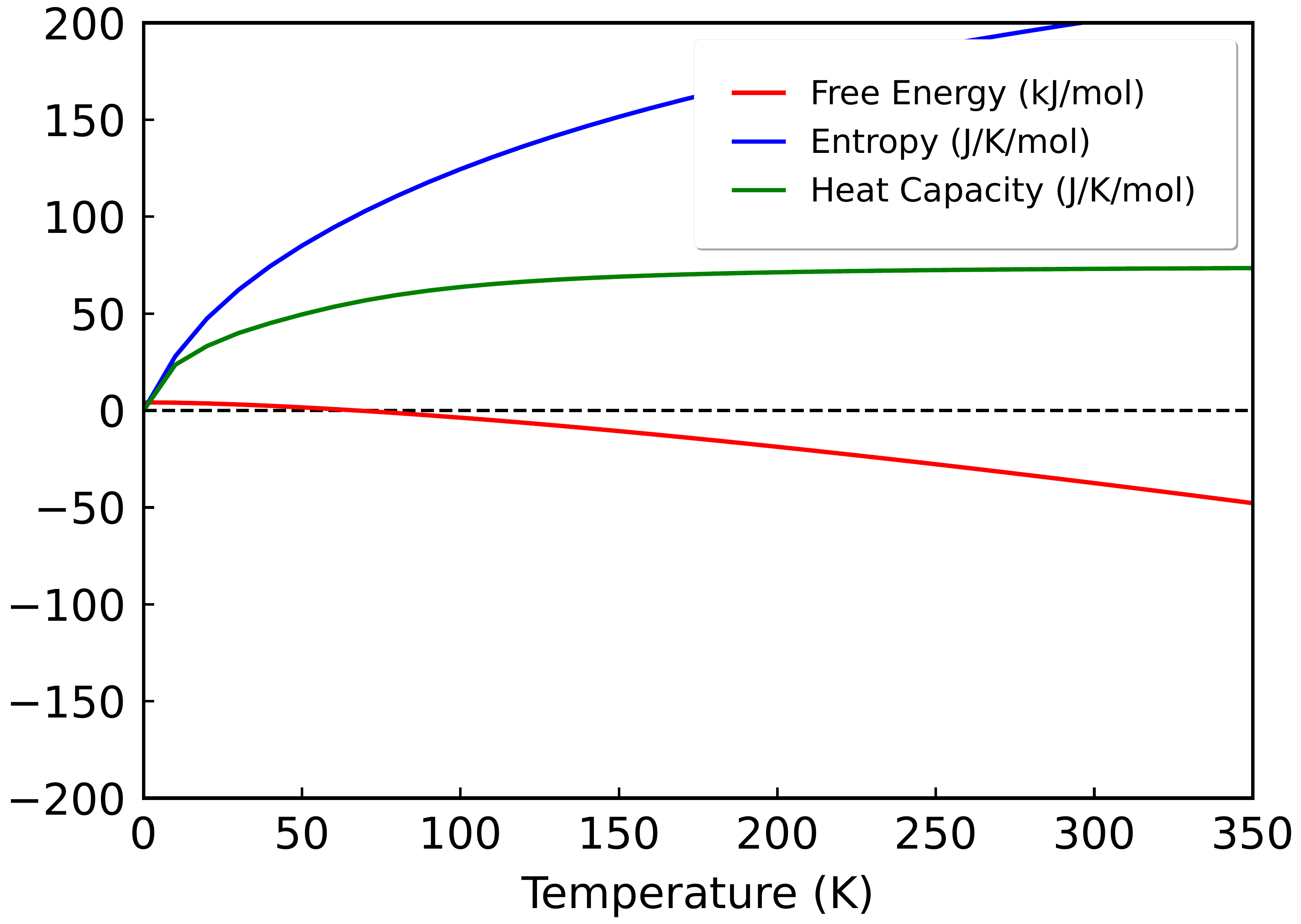}
\end{subfigure}
\begin{subfigure}[b]{0.3\columnwidth}
    \subcaption{0.14 THz}
\includegraphics[width=\columnwidth,clip=true,keepaspectratio]{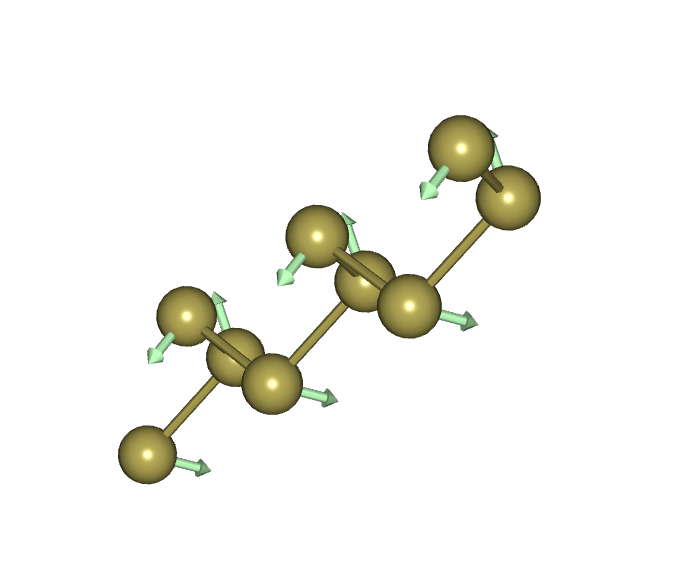}
	\end{subfigure}
	\hspace{0.1cm}
	\begin{subfigure}[b]{0.3\columnwidth}
    \subcaption{2.00 THz}
\includegraphics[width=\columnwidth,clip=true,keepaspectratio]{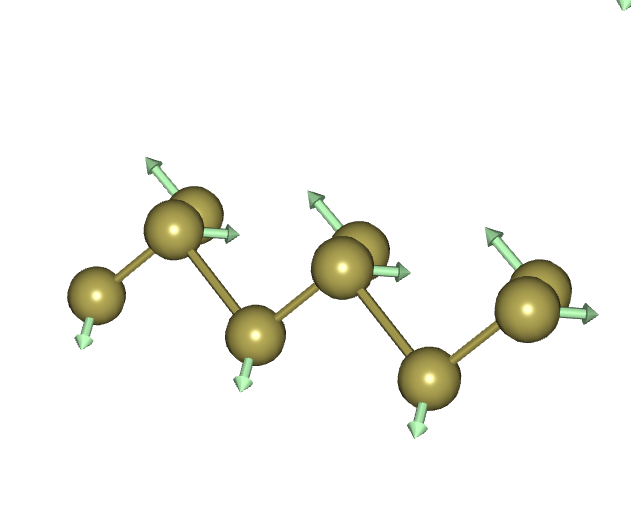}
	
	\end{subfigure}\\
	
	\begin{subfigure}[b]{0.3\columnwidth}
    \subcaption{5.07 THz}
		\includegraphics[width=\columnwidth,clip=true,keepaspectratio]{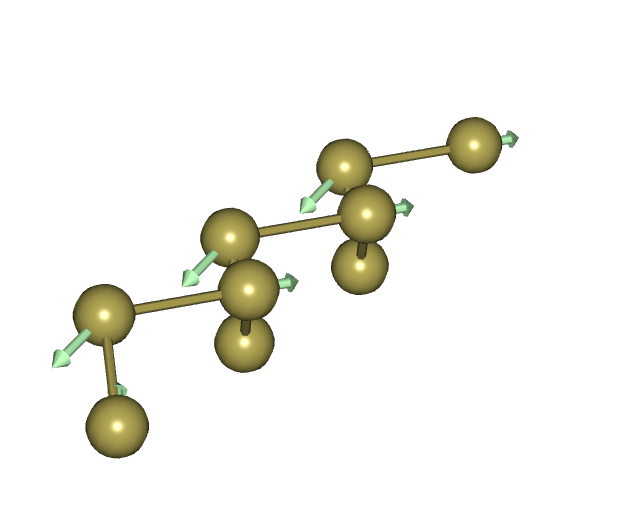}
	
	\end{subfigure}
	\hspace{0.1cm}
	\begin{subfigure}[b]{0.3\columnwidth}
    \subcaption{5.56 THz}
	\includegraphics[width=\columnwidth,clip=true,keepaspectratio]{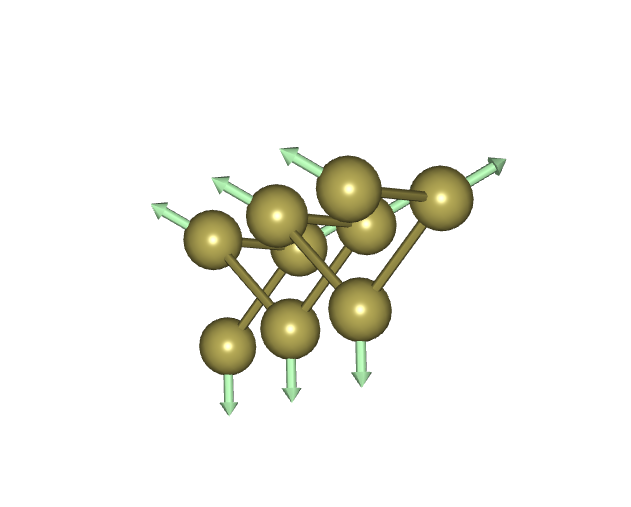}
		
	\end{subfigure}
	\caption{\label{fig:S5} a) Phonon dispersion curve with phonon DOS and b) thermal properties of Te-h calculated within GGA. (c)-(f) Selected phonon modes at specific frequencies at $\Gamma$-point in Te-h.}
\end{figure}

\begin{table}[H]
\caption{\label{tab:S1} Entropy, free energy and specific heat at constant volume C$_v$ at the dulong-Petit limit of tellurium phases.}
 \centering
\renewcommand{\arraystretch}{1.5}
\setlength{\tabcolsep}{10pt}
\begin{tabular}{ c c c c }
\toprule
phase & Free energy ($KJ/\text{mol}$) & Entropy ($J/K\cdot \text{mol}$) & \textbf{$C_v$} ($J/K\cdot \text{mol}$) \\
\midrule
Te-I          & -23.16 & 153.20 & 73.68 \\
$\alpha$-Te   & -21.95 & 149.34 & 73.50 \\
$\beta$-Te    & -25.42 & 160.87 & 73.53 \\
buckled pentagonal     & -25.96 & 160.43 & 72.07 \\
buckled kagome &  -29.11  & 171.72 & 72.70 \\
buckled square & -16.43 & 104.15 & 48.06 \\
Te-h          & -37.47 & 201.05 & 73.07 \\
\bottomrule
\end{tabular}
\end{table}

\section{Role of SOC in the band structure}

\begin{figure}[H]
\begin{subfigure}[b]{0.3\columnwidth}
\subcaption{}
\includegraphics[width=\columnwidth,clip=true,keepaspectratio]{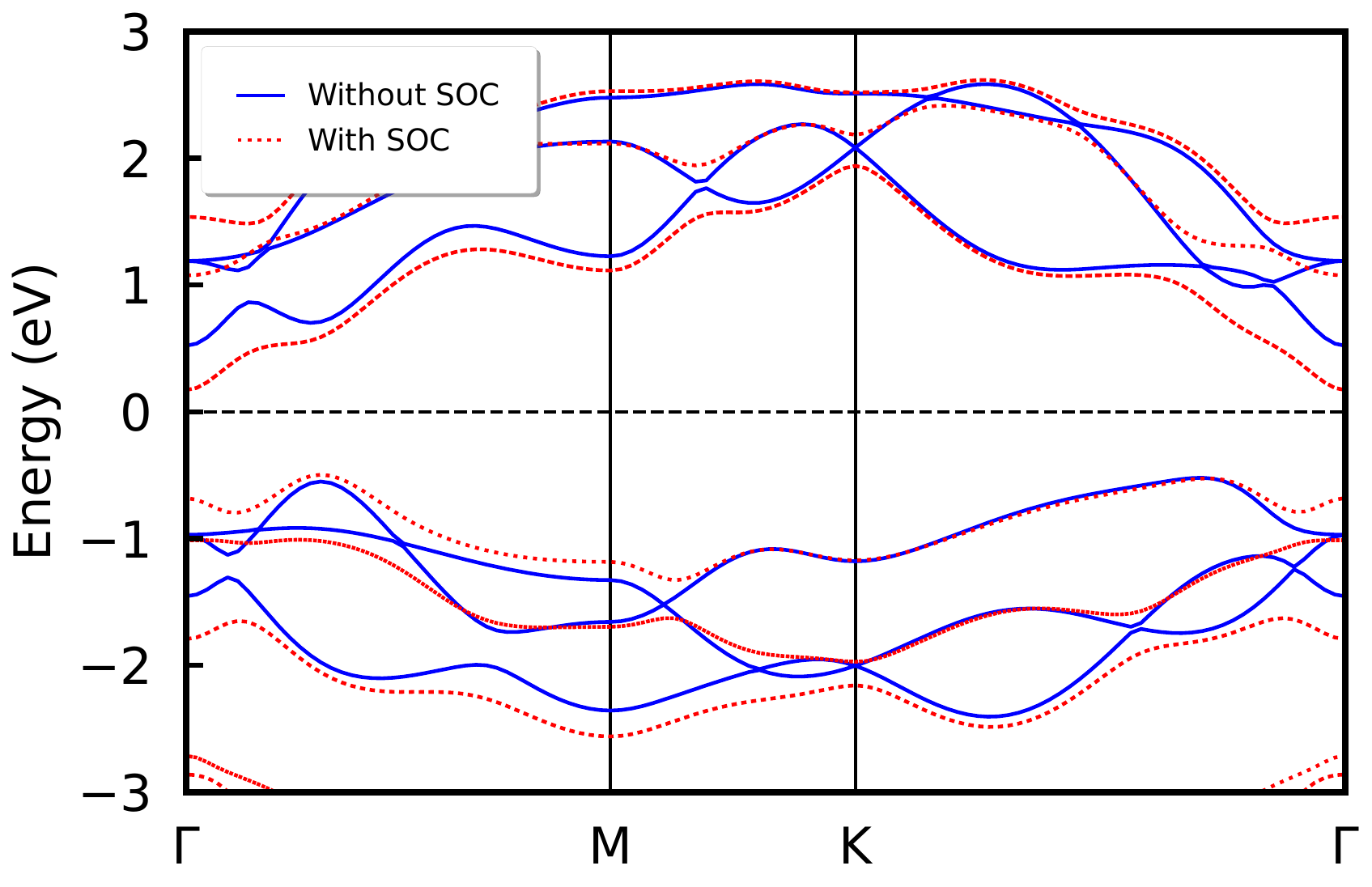}
\end{subfigure}
\begin{subfigure}[b]{0.3\columnwidth}
\subcaption{}
\includegraphics[width=\columnwidth,clip=true,keepaspectratio]{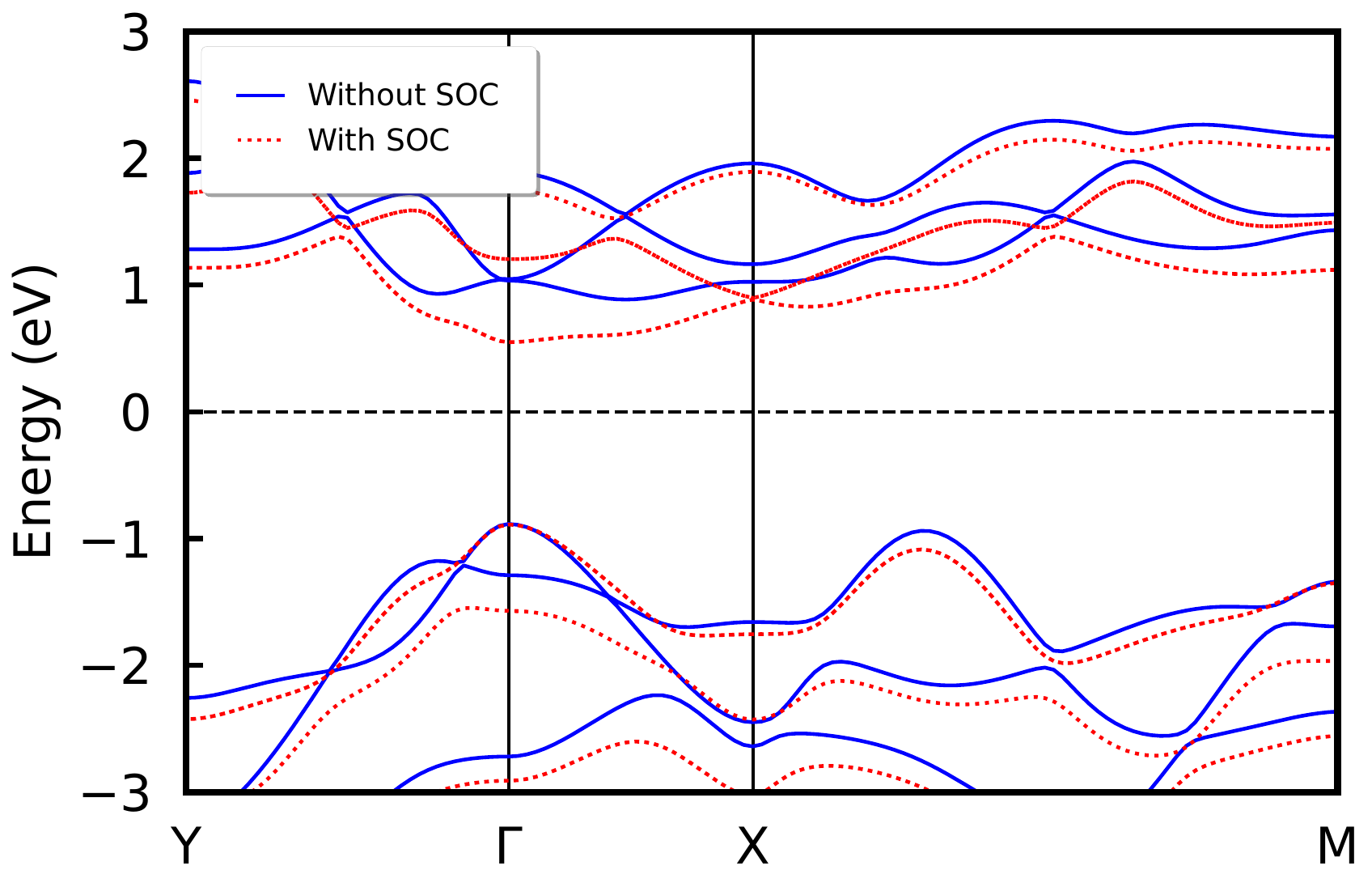}
\end{subfigure}
\begin{subfigure}[b]{0.3\columnwidth}
\subcaption{}
\includegraphics[width=\columnwidth,clip=true,keepaspectratio]{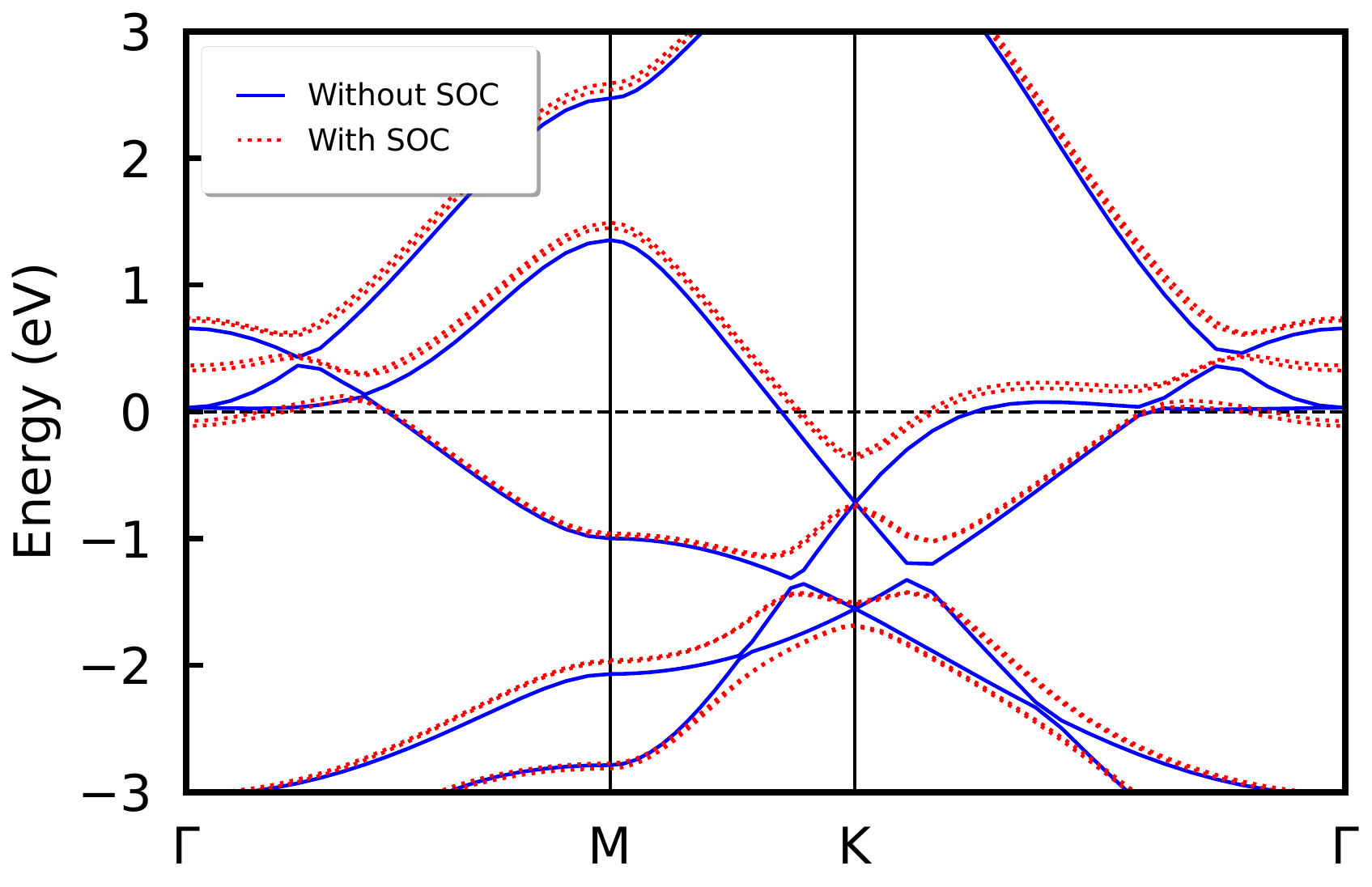}
\end{subfigure}
\begin{subfigure}[b]{0.3\columnwidth}
\subcaption{}
\includegraphics[width=\columnwidth,clip=true,keepaspectratio]{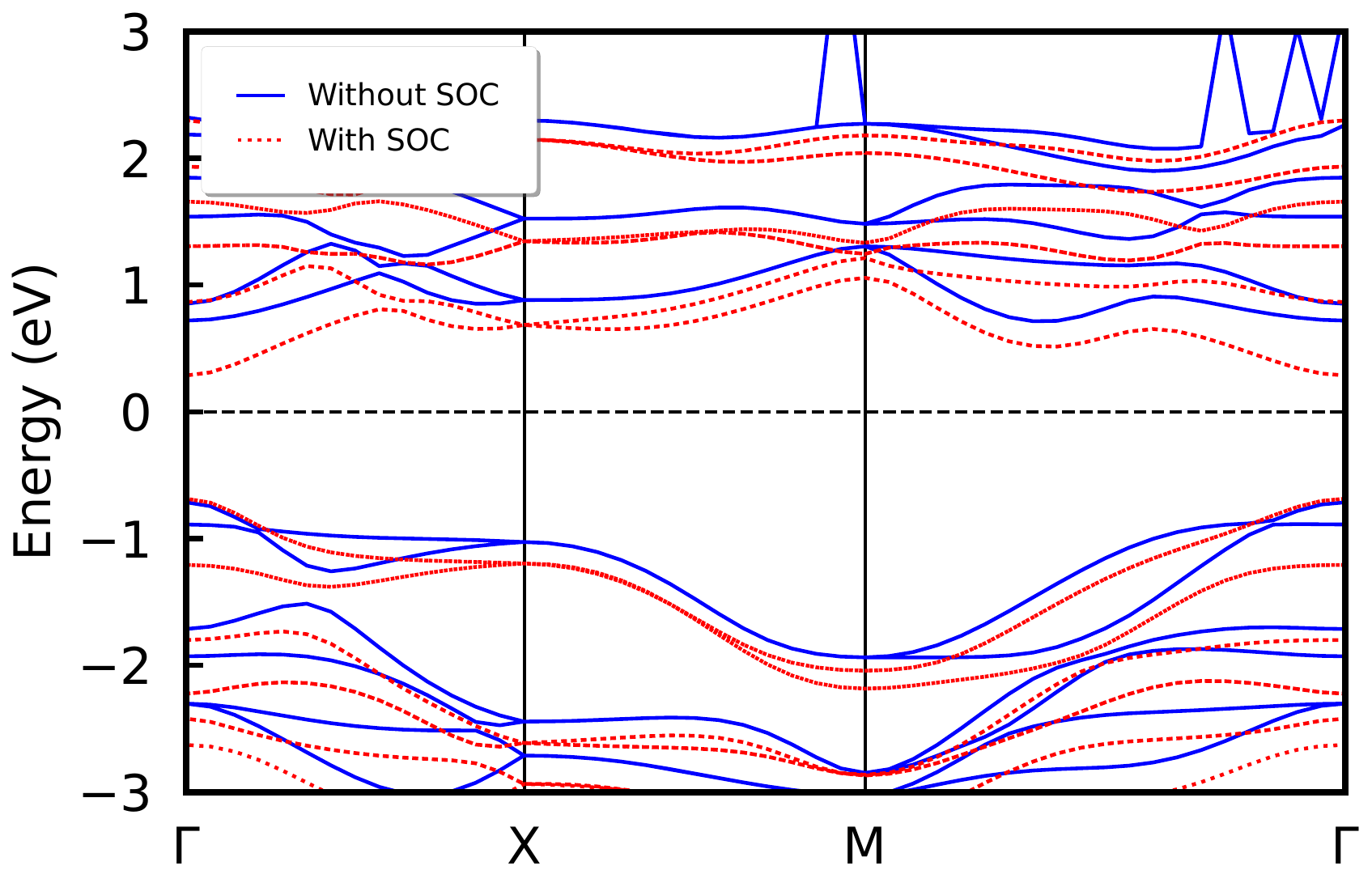}
\end{subfigure}
\begin{subfigure}[b]{0.3\columnwidth}
\subcaption{}
\includegraphics[width=\columnwidth,clip=true,keepaspectratio]{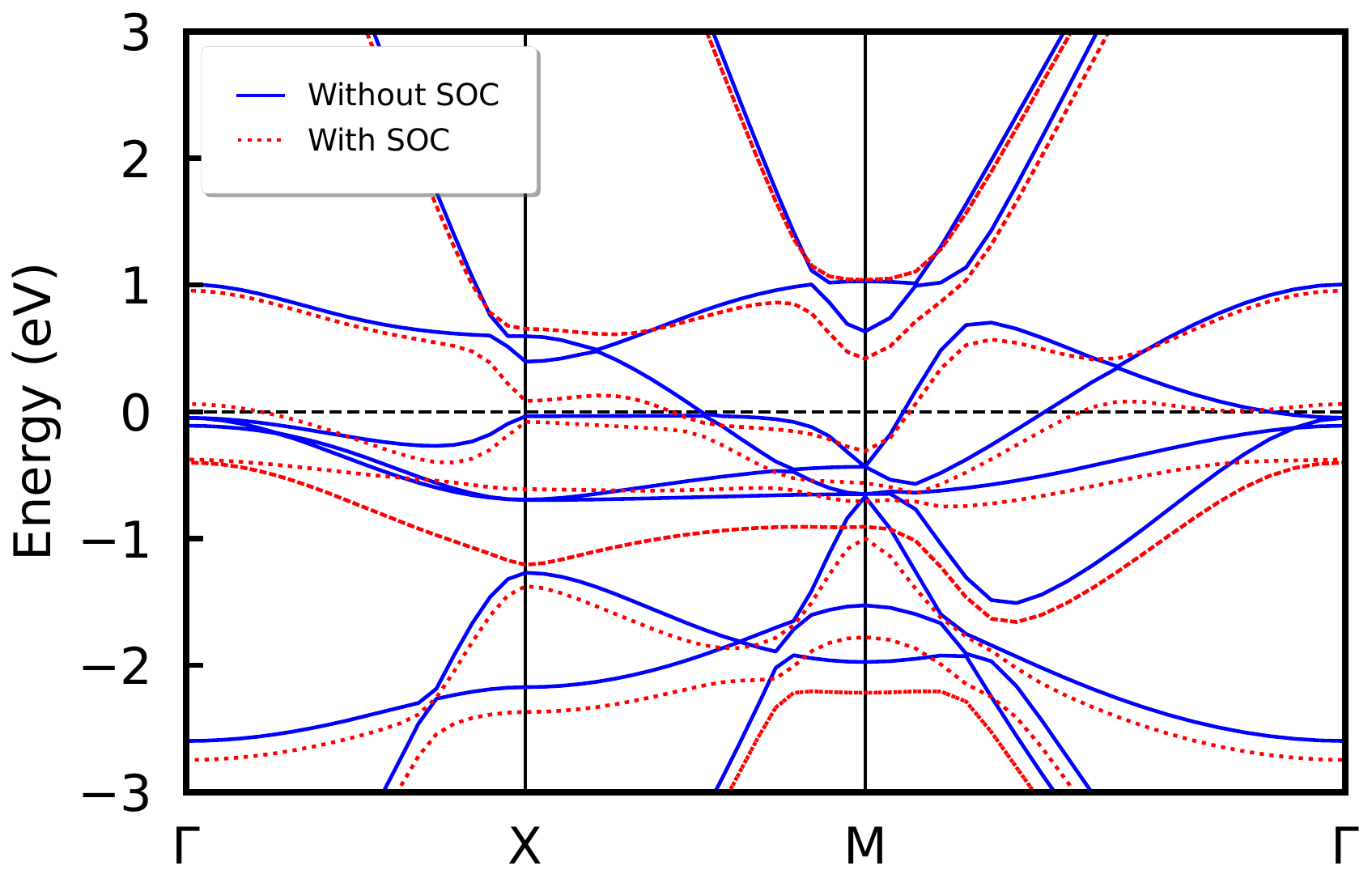}
\end{subfigure}
\begin{subfigure}[b]{0.3\columnwidth}
\subcaption{}
\includegraphics[width=\columnwidth,clip=true,keepaspectratio]{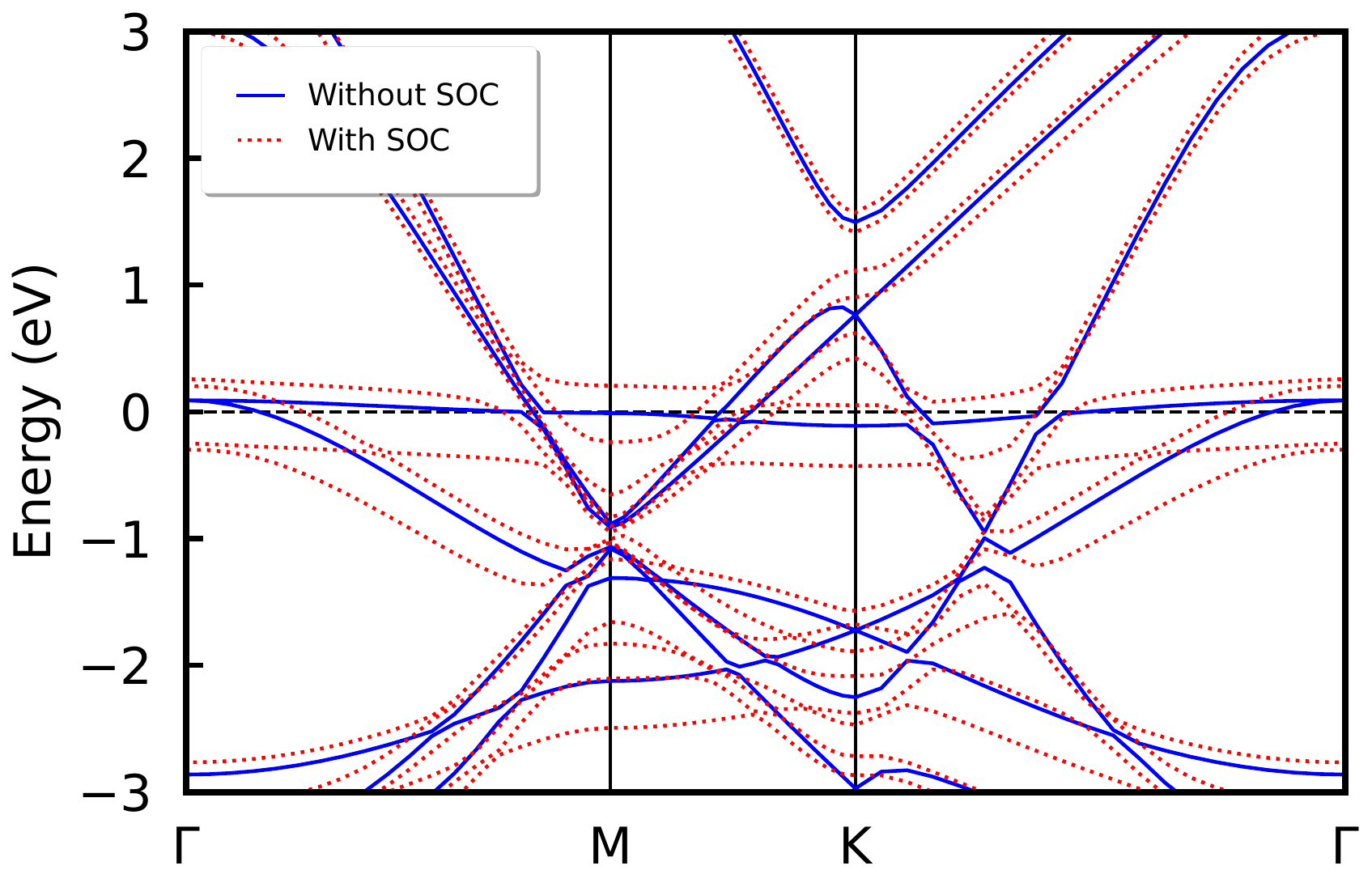}
\end{subfigure}\\
\begin{subfigure}[b]{0.3\columnwidth}
\subcaption{}
\includegraphics[width=\columnwidth,clip=true,keepaspectratio]{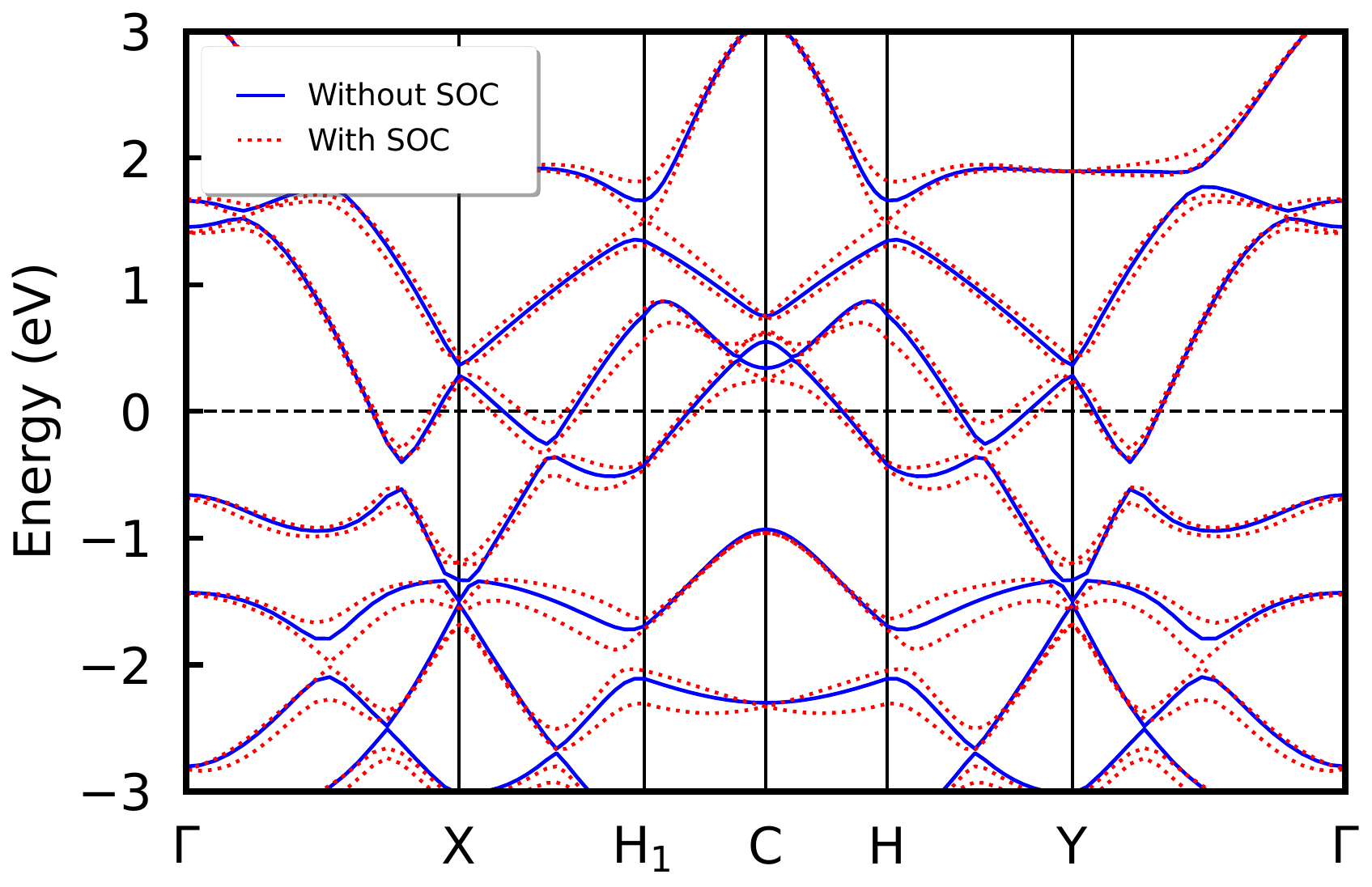}
\end{subfigure}
\begin{subfigure}[b]{0.3\columnwidth}
\subcaption{}
\includegraphics[width=\columnwidth,clip=true,keepaspectratio]{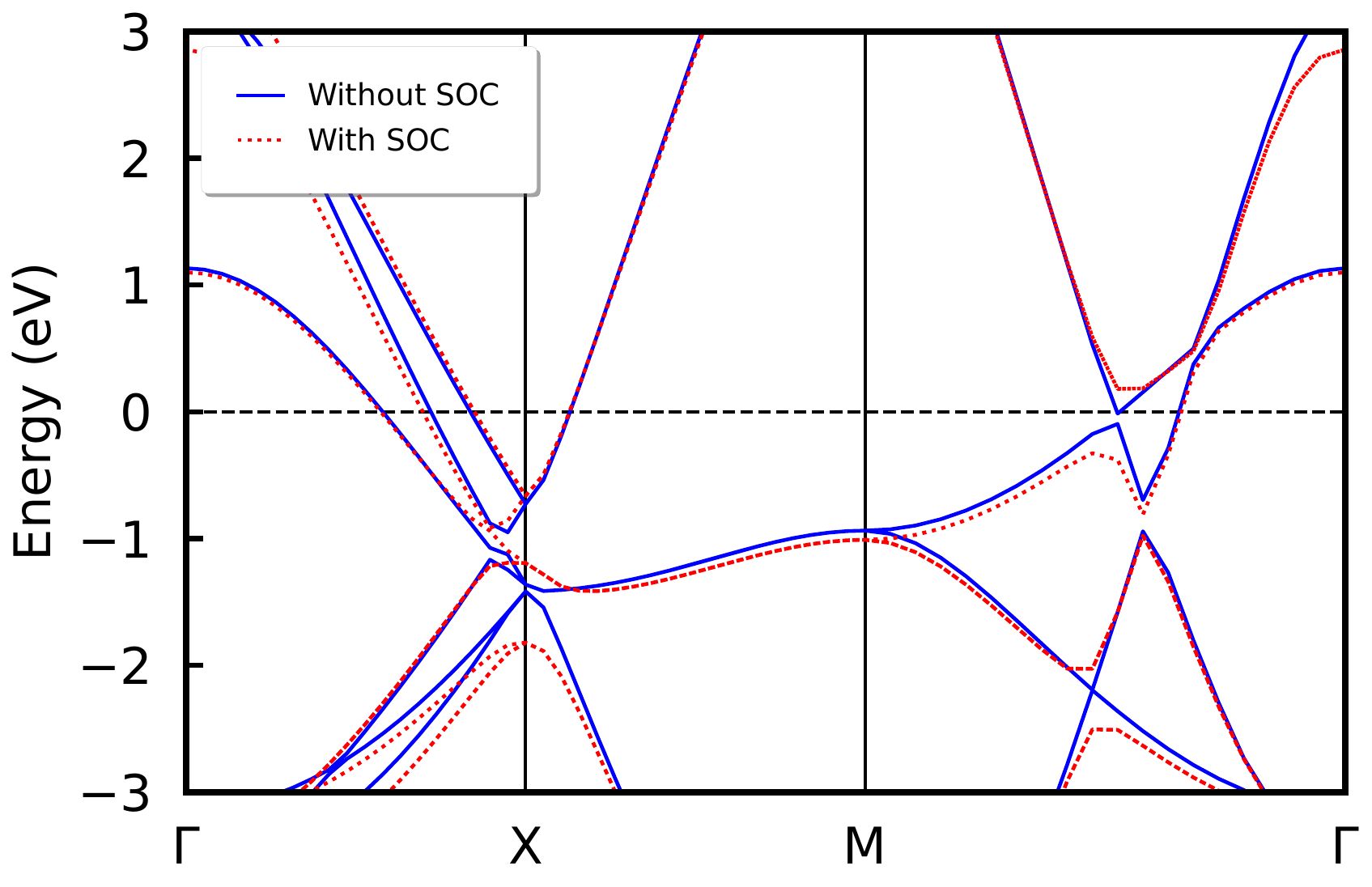}
\end{subfigure}
\begin{subfigure}[b]{0.3\columnwidth}
\subcaption{}
\includegraphics[width=\columnwidth,clip=true,keepaspectratio]{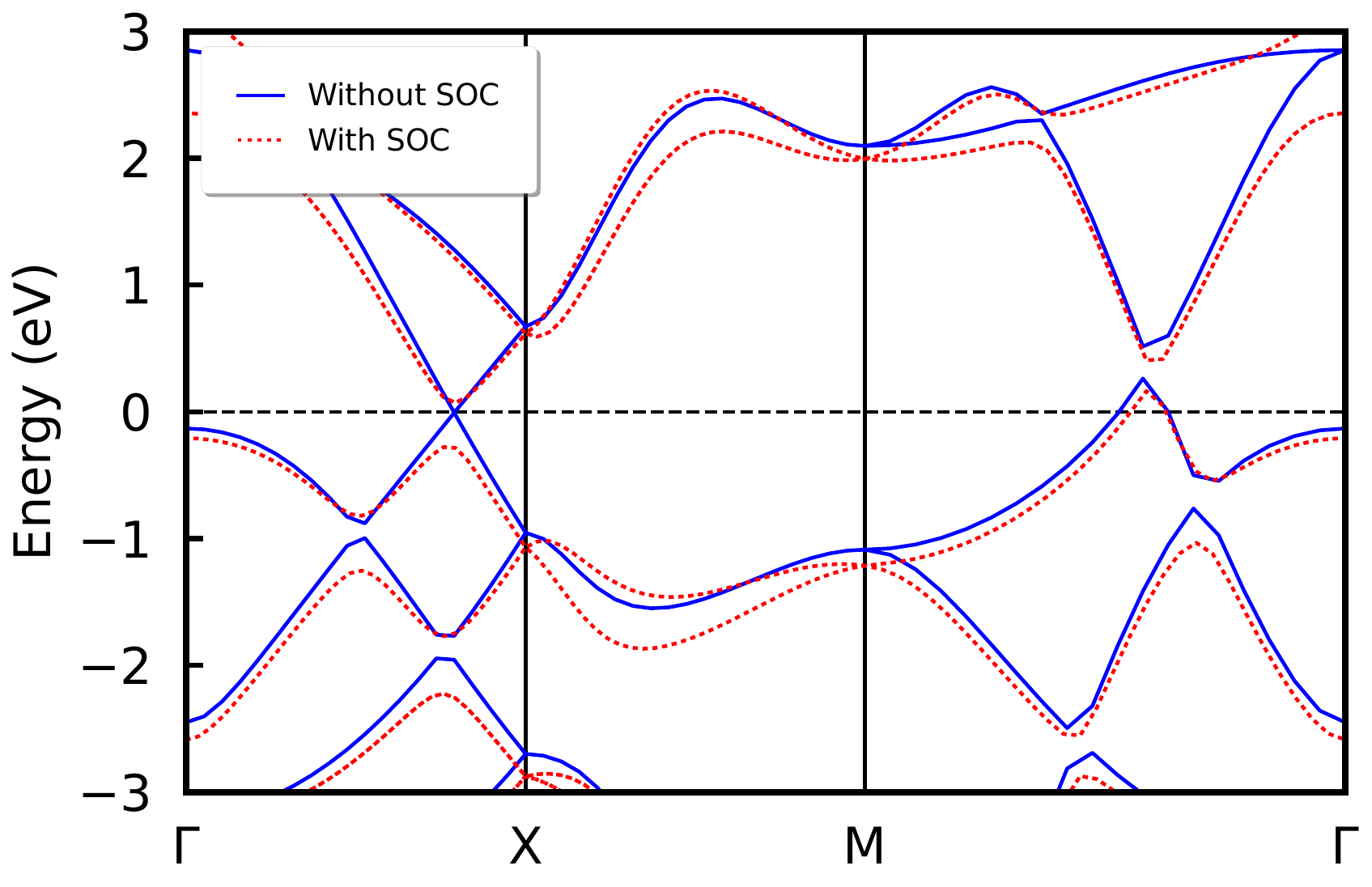}
\end{subfigure}\\
\begin{subfigure}[b]{0.3\columnwidth}
\subcaption{}
\includegraphics[width=\columnwidth,clip=true,keepaspectratio]{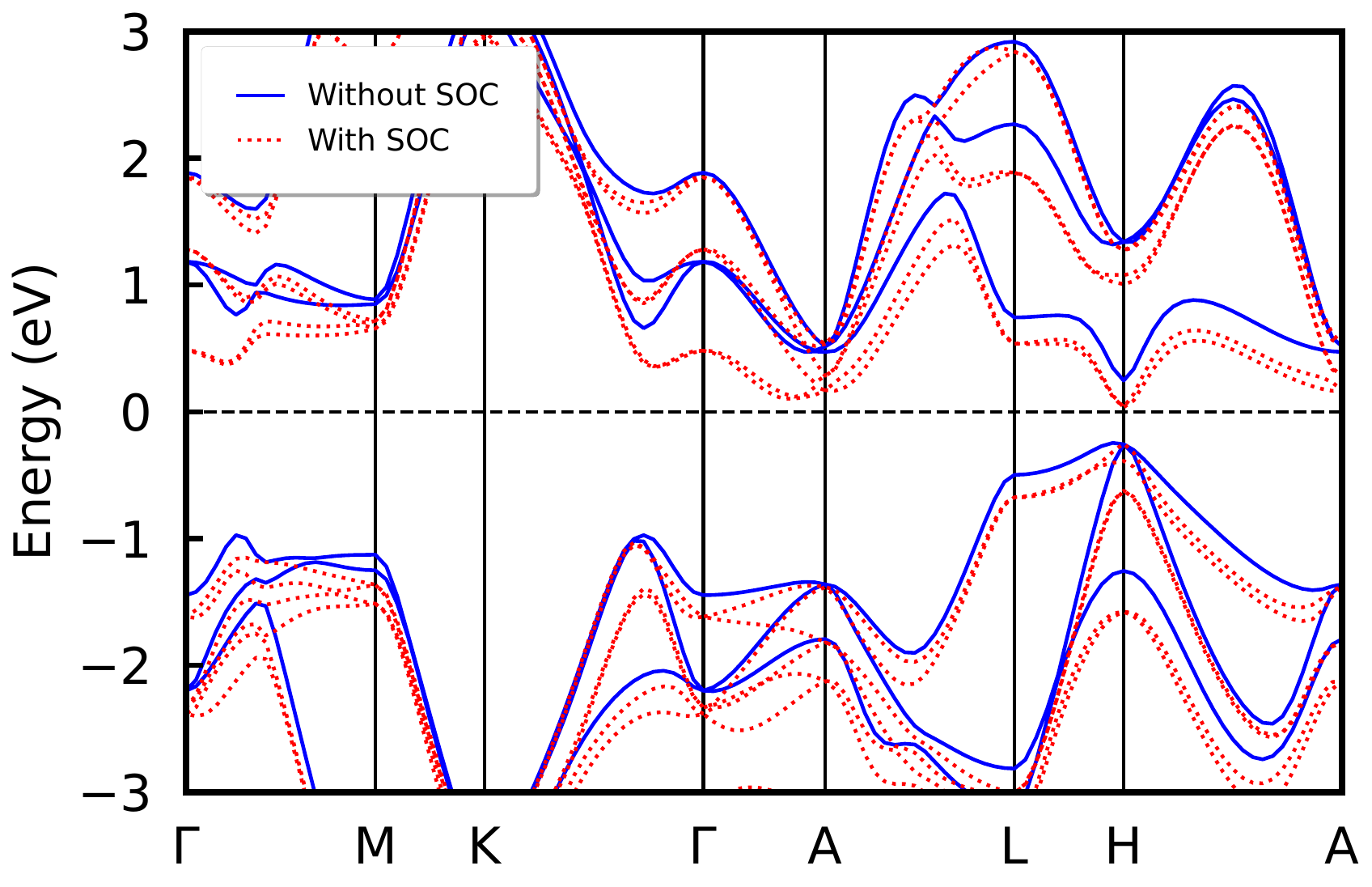}
\end{subfigure}
\begin{subfigure}[b]{0.3\columnwidth}
\subcaption{}
\includegraphics[width=\columnwidth,clip=true,keepaspectratio]{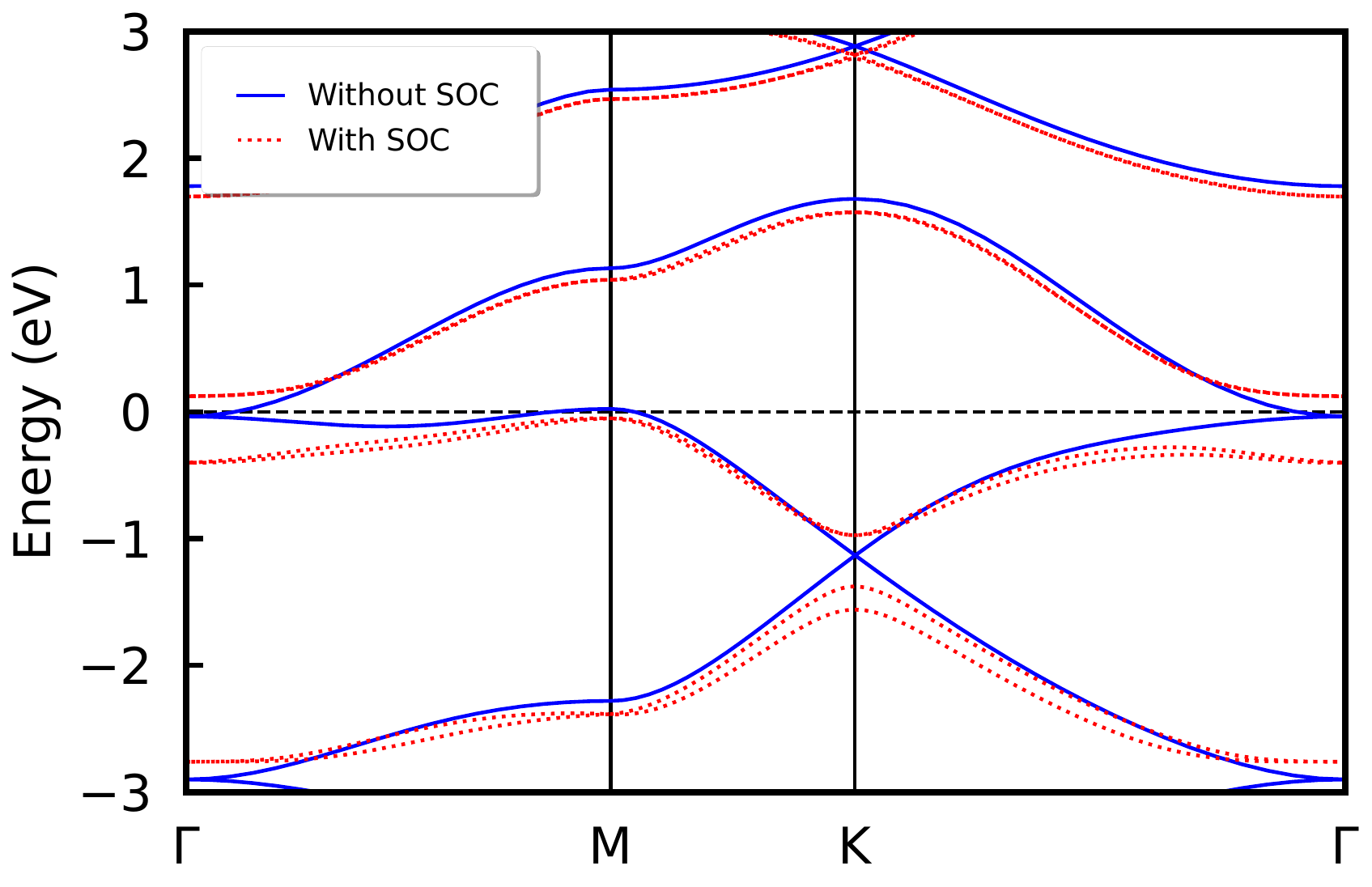}
\end{subfigure}
\begin{subfigure}[b]{0.3\columnwidth}
\subcaption{}
\includegraphics[width=\columnwidth,clip=true,keepaspectratio]{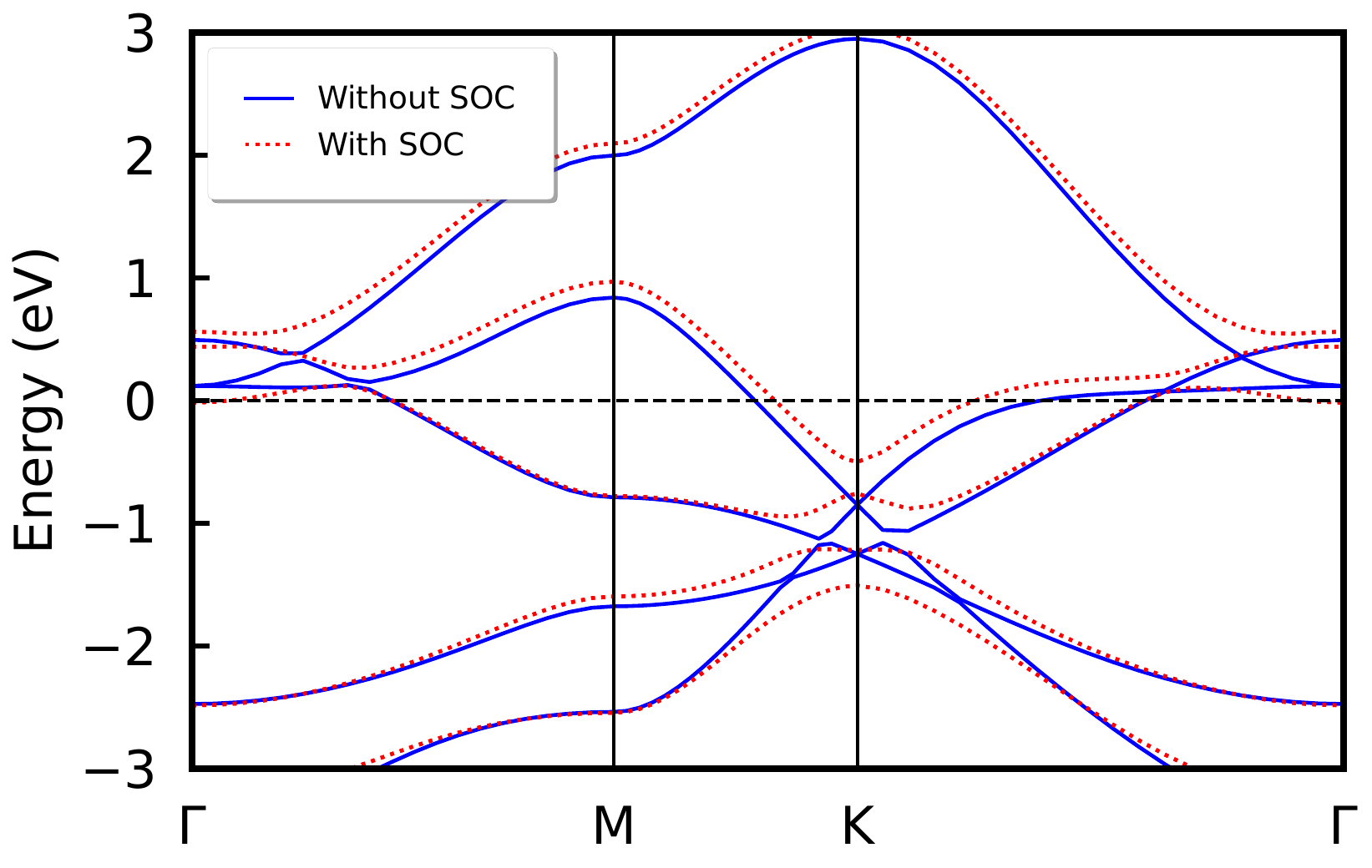}
\end{subfigure}
\caption{\label{fig:S6} Band structure calculated within MLWF-TB/HSE06 of tellurium phases:  a) $\alpha$, b) $\beta$, c) hexagonal planar, d) Lieb-like, e) pentagonal buckled, f) planar kagome, g) buckled kagome, h) planar square, i) buckled square, j) trigonal Te-I, k) one-side hydrogen passivated hexagonal and l) hexagonal planar strained (5\%). The highlighted region in k) reveals the emergence of Weyl nodes near the Fermi level, resulting from inversion symmetry breaking combined with strong SOC effects. Red (blue) lines are calculations with (without) spin-orbit coupling (SOC).}
\end{figure}

\section{Topological characterization of one-dimensional systems}

In one-dimensional crystalline systems, the topological classification of electronic bands differs fundamentally from that of two- and three-dimensional materials. For nonmagnetic systems with spin--orbit coupling that preserve time-reversal symmetry, the electronic structure belongs to symmetry class AII in the Altland--Zirnbauer classification\cite{Schnyder2008,Ryu2010}. In contrast to higher dimensions, this symmetry class does not support a nontrivial $\mathbb{Z}_2$ topological insulating phase in one dimension. Consequently, a quantum spin Hall--type invariant cannot be defined for strictly one-dimensional, time-reversal-invariant band insulators. Instead, the relevant bulk quantity in one dimension is the Berry phase accumulated along the Brillouin zone, commonly referred to as the Zak phase\cite{Zak1989}. This quantity is directly related to the electronic polarization and provides the appropriate framework for analyzing the topology of one-dimensional insulating systems.

For a one-dimensional periodic system, the Zak phase associated with the occupied electronic states is defined as

\begin{equation}
\gamma = i \sum_{n \in \mathrm{occ}} \int_{\mathrm{BZ}}
\langle u_{n k} | \partial_k u_{n k} \rangle \, dk ,
\end{equation}
where $|u_{n k}\rangle$ are the cell-periodic Bloch functions and the integral is performed over the one-dimensional Brillouin zone. 

Within the modern theory of polarization, the Zak phase is related to the bulk electronic polarization $P$ via  \cite{KingSmith1993,Resta1994}
\begin{equation}
P = \frac{e}{2\pi} \gamma \quad (\mathrm{mod}\ e),
\end{equation}
where $e$ is the elementary charge. Only the polarization modulo the electron charge is physically meaningful, reflecting the gauge freedom associated with the choice of unit cell.

The Zak phase is, in general, not a topological invariant, as its
value depends on the choice of real-space origin and gauge. However,
in the presence of spatial inversion symmetry or chiral symmetry, the
Zak phase becomes quantized to values of $0$ or $\pi$ (mod
$2\pi$)\cite{Resta2000,Hughes2011}. In such cases, the quantized Zak
phase serves as a symmetry-protected topological invariant and
enforces the existence of robust boundary phenomena, such as topological end states. 

Although Berry curvature plays a central role in the description of
topological phases in higher-dimensional systems, it does not
constitute a bulk topological invariant in strictly one-dimensional
insulators. Nevertheless, Berry curvature distributions can provide
useful qualitative information regarding band hybridization and
spin--orbit--coupling effects. In particular, the absence of
pronounced Berry-curvature peaks or singular features near the band
edges is consistent with the absence of band inversion and supports a
topologically trivial characterization of the electronic structure. In
one dimension, such Berry-curvature analyses should therefore be
regarded as complementary diagnostics rather than as indicators of a
topological phase.

\begin{figure}[H]
\begin{subfigure}[b]{0.3\columnwidth}
\subcaption{}
\includegraphics[width=\columnwidth,clip=true,keepaspectratio]{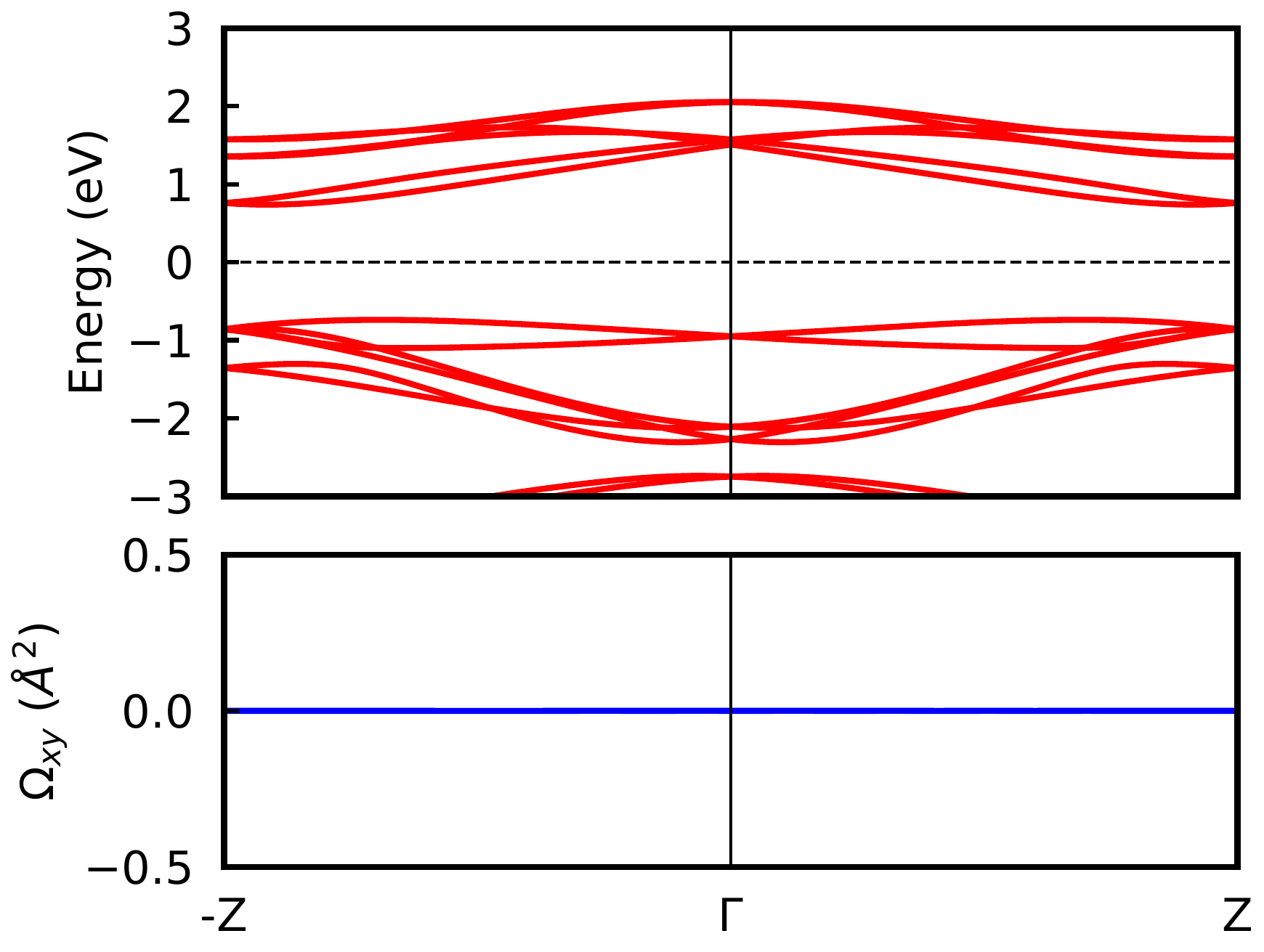}
\end{subfigure}
\caption{\label{fig:berry_zak_nw} Zak phase of the one-dimensional tellurium nanowire.
Berry phase accumulated by the occupied electronic states along the one-dimensional Brillouin zone. The Zak phase is found to be essentially zero, corresponding to a vanishing bulk polarization.}

\end{figure}

The tellurium nanowire considered in this work is a one-dimensional,
time-reversal-invariant system lacking inversion and chiral
symmetries. Consequently, its Zak phase is not symmetry-quantized. The
computed Zak phase is found to be essentially zero, indicating a
vanishing bulk polarization. This establishes that the periodic
nanowire is topologically trivial in the normal (non-superconducting)
state. Accordingly, any localized states observed at the ends of
finite nanowires are not protected by bulk topology and arise from
termination-specific effects. The result for the Zak phase is shown in
Fig.\ref{fig:berry_zak_nw}, calculated using VASP and WannierTools.

\section{Effective masses}

\begin{table}[H]
 \caption{\label{tab:S2} Directional effective masses ($m^*/m_e$). For indirect band gap materials, paths are shown from the Valence Band Maximum (VBM) or Conduction Band Minimum (CBM). The fractional coordinates for these extrema are: Square buckled (VBM at [0.21, 0.21, 0.00]; CBM at [0.39, 0.00, 0.00]) and Hexagonal passivated (VBM at [0.49, 0.00, 0.00]; CBM at $\Gamma$).}
    \label{effmass_materials}
    \centering
    \renewcommand{\arraystretch}{1.5}
    \setlength{\tabcolsep}{10pt}
    \begin{tabular}{c c c}
        \toprule 
        & \multicolumn{2}{c}{m$^*$ (m$_{e}$)} \\
        \cmidrule(lr){2-3}
        Phases & Electron & Hole \\
        \midrule
        $\alpha$-Te  &  0.108 & 0.135  \\
        $\beta$-Te   &  \begin{tabular}{@{}c@{}}1.009 ($\Gamma\rightarrow$X) \\ 0.203 ($\Gamma\rightarrow$Y)\end{tabular} & \begin{tabular}{@{}c@{}} 0.368 ($\Gamma\rightarrow$X) \\ 0.127 ($\Gamma\rightarrow$Y)\end{tabular} \\
        Pentagonal buckled & 0.220 & 0.172 \\
        Square buckled & \begin{tabular}{@{}c@{}}0.100 (CBM$\rightarrow\Gamma$) \\ 0.148 (CBM$\rightarrow$X)\end{tabular} & \begin{tabular}{@{}c@{}}0.459 (VBM$\rightarrow\Gamma$) \\ 0.239 (VBM$\rightarrow$M)\end{tabular} \\
        Hexagonal passivated & \begin{tabular}{@{}c@{}}2.400 ($\Gamma\rightarrow$M) \\ 2.310 ($\Gamma\rightarrow$K)\end{tabular} & 1.184 (VBM$\rightarrow\Gamma$) \\        
        \bottomrule
    \end{tabular}
\end{table}

\bibliography{referencias}

\end{document}